\documentclass[aps,pre,10pt,longbibliography,twocolumn,nofootinbib]{revtex4-1}

\usepackage[colorlinks, allcolors=blue, linktoc=all]{hyperref}
\usepackage{newtxtext}
\usepackage{newtxmath}
\usepackage{inconsolata}

\usepackage[english]{babel}
\usepackage{amsmath}
\usepackage{amssymb}
\usepackage{mathtools}
\usepackage{graphicx}
\usepackage{pythonhighlight}

\newcommand{\bigO}{\mathcal{O}}

\newcommand{\ceil}[1]{\mathrm{ceil}\left(#1\right)}
\newcommand{\modulo}[2]{#1 \, \mathrm{mod} \, #2}
\newcommand{\sgn}{\mathrm{sgn}}
\renewcommand{\vec}[1]{\mathbf{#1}}

\newcommand{\sfrac}[2]{#1/#2}

\newcommand{\diff}[1]{\mathrm{d} #1}
\newcommand{\pdiff}[1]{\partial_{#1}}

\newcommand{\spderiv}[2]{\pdiff{#2} #1}
\newcommand{\pderiv}[2]{\frac{\pdiff{} #1}{\pdiff{} #2}}
\newcommand{\npderiv}[3]{\frac{\pdiff{}^{#3} #1}{\pdiff{} #2^{#3}}}

\newcommand{\lap}[1]{\nabla^2 #1}

\renewcommand{\eqref}[1]{Eq.~(\ref{#1})}
\newcommand{\secref}[1]{\S\ref{#1}}
\newcommand{\figref}[1]{Fig.~\ref{#1}}

\newcommand{\appref}[1]{Appendix~\ref{#1}}

\newcommand{\chref}[2]{\href{#1}{#2}\footnote{\url{#1}}}
\graphicspath{{./figures_dedalus/}{./figures_examples/}}

\newif\iftoc
\toctrue

\begin{document}

\title{Dedalus: A Flexible Framework for Numerical Simulations with Spectral Methods}

\author{Keaton~J.~Burns}
\affiliation{Massachusetts Institute of Technology Departments of Mathematics and Physics, Cambridge, MA 02139}
\affiliation{Center for Computational Astrophysics, Flatiron Institute, New York, NY 10010}

\author{Geoffrey~M.~Vasil}
\affiliation{University of Sydney School of Mathematics and Statistics, Sydney, NSW, Australia}

\author{Jeffrey~S.~Oishi}
\affiliation{Bates College Department of Physics and Astronomy, Lewiston, ME 04240}

\author{Daniel~Lecoanet}
\affiliation{Princeton Center for Theoretical Science and Princeton University Department of Astrophysical Sciences, Princeton, NJ 08544}

\author{Benjamin~P.~Brown}
\affiliation{University of Colorado Laboratory for Atmospheric and Space Physics and Department of Astrophysical and Planetary Sciences, Boulder, CO 80309}

\begin{abstract}

Numerical solutions of partial differential equations enable a broad range of scientific research.
The Dedalus Project is a flexible, open-source, parallelized computational framework for solving general partial differential equations using spectral methods.
Dedalus translates plain-text strings describing partial differential equations into efficient solvers.
This paper details the numerical method that enables this translation, describes the design and implementation of the codebase, and illustrates its capabilities with a variety of example problems.
The numerical method is a first-order generalized tau formulation that discretizes equations into banded matrices.
This method is implemented with an object-oriented design.
Classes for spectral bases and domains manage the discretization and automatic parallel distribution of variables.
Discretized fields and mathematical operators are symbolically manipulated with a basic computer algebra system.
Initial value, boundary value, and eigenvalue problems are efficiently solved using high-performance linear algebra, transform, and parallel communication libraries.
Custom analysis outputs can also be specified in plain text and stored in self-describing portable formats.
The performance of the code is evaluated with a parallel scaling benchmark and a comparison to a finite-volume code.
The features and flexibility of the codebase are illustrated by solving several examples: the nonlinear Schr{\"o}dinger equation on a graph, a supersonic magnetohydrodynamic vortex, quasigeostrophic flow, Stokes flow in a cylindrical annulus, normal modes of a radiative atmosphere, and diamagnetic levitation.
The Dedalus code and the example problems are available online at \url{http://dedalus-project.org/}.

\end{abstract}
\maketitle

\iftoc
\makeatletter
\def\l@subsubsection#1#2{}
\makeatother
\tableofcontents
\onecolumngrid
\phantom{space}
\vspace{1em}
\twocolumngrid
\fi

\section{Introduction}

Partial differential equations (PDEs) describe continuum processes.
The continuous independent variables typically represent space and time, but can also represent more abstract quantities such as momentum, energy, age of a population, or currency.
The ability to equate the infinitesimal rates of change of different quantities produces endless possible applications.
Important examples include wave propagation, heat transfer, fluid flow, quantum mechanical probability flux, chemical \& nuclear reactions, biological phenomena, and even financial markets \cite{Toth:2011gh} or social/population dynamics \citep{Burridge:2017kd,Keyfitz:1997gg}.
Even more intriguing are possible combinations of several of the above \citep{Franks:2002fj,Lemmerer:2019cg}.

Apart from a small handful of closed-form solutions, the vast majority of PDEs require serious numerical and computational intervention.
A wide variety of numerical algorithms solve PDEs through the general approach of discretizing its continuous variables and operators to produce a finite-sized algebraic system yielding an approximate solution.
Finite element, finite volume, and finite difference methods are common schemes that discretize the domain of the PDE into cells or points and derive algebraic relations between the values at neighboring cells or points from the governing equations.
These methods can accommodate complex geometries (such as the flow around an aircraft), but can be difficult to implement for complex equations and typically converge relatively slowly as additional cells or points are added.

In contrast, spectral methods discretize variables by expanding them in a finite set of basis functions and derive equations for the coefficients of these functions.
These methods are well-suited to many equation types and provide rapidly converging solutions (e.g.\ exponential for smooth functions) as additional modes are included.
However, spectral methods are typically limited to simple geometries (such as boxes, cylinders, and spheres).
Recent literature has developed sparse representations of equations that are substantially better conditioned and faster than traditional dense collocation techniques \citep{Greengard:1991kg,Julien:2009kb,Muite:2010kr,Olver:2013ed,channelflow,Viswanath:2015jk,Miquel:2017hq}.
These features make spectral methods an attractive choice for scientists seeking to study a wide variety of physical processes with high precision.

While computing capacities have grown exponentially over the past few decades, the progression of software development has been more gradual.
Many software packages have chosen one or a few closely related PDEs and focused on creating highly optimized implementations of algorithms that are well-suited to those choices.
These solvers usually hardcode not only the PDE but also the dynamical variables, choice of input control parameters, integration scheme, and analysis output.
While many world-class simulation codes have been developed this way, often scientific questions lead beyond what a dedicated code can do.
This is not always because of a lack of computational power or efficiency, but often because continued progress requires an alternative model, dynamical variable reformulation, or more exotic forms of analysis.

Simulation packages with flexible model specification also address an underserved scientific niche.
It is often straightforward to write serial codes to solve simple one-dimensional equations for particular scientific questions.
It is also worthwhile to invest multiple person-years building codes that solve well-known equations.
However, it can be difficult to justify spending significant time developing codes for novel models that are initially studied by only a few researchers.
We believe this leaves many interesting questions unaddressed simply from a local cost-benefit analysis.
Flexible toolkits can lower the barrier to entry for a large number of interesting scientific applications.

The \chref{https://fenicsproject.org}{FEniCS} and \chref{https://www.firedrakeproject.org}{Firedrake} packages both allow users to symbolically enter their equations in variational form and produce finite-element discretizations suitable for forward-modeling and optimization calculations.
These are very powerful tools for solving wide ranges of PDEs in complicated geometries, however they remain less efficient than spectral methods for many PDEs in simple geometries.
\chref{http://channelflow.org}{Channelflow} uses sparse Chebyshev methods to simulate the Navier-Stokes equations and allows users to find and analyze invariant solutions using dynamical systems techniques.
However, the code is restricted to solving incompressible flow in a periodic channel geometry.
The \chref{http://www.chebfun.org}{Chebfun} and \chref{https://github.com/JuliaApproximation/ApproxFun.jl}{ApproxFun} packages are highly flexible toolkits for performing function approximation using spectral methods.
They include a wide variety of features including sparse, well-conditioned, and adaptive methods for efficiently solving differential equations to machine precision.
However, these packages are not optimized for the solution of multidimensional PDEs on parallel architectures.

The goal of the Dedalus Project is to bridge this gap and provide a framework applying modern, sparse spectral techniques to highly parallelized simulations of custom PDEs.
The codebase allows users to discretize domains using the direct products of spectral series and symbolically specify systems of PDEs on those domains.
The code then produces a sparse discretization of the equations and automatically parallelizes the solution of the resulting model.
The Dedalus codebase is open-source, highly modular, and easy to use.
While its development has been motivated by the study of turbulent flows in astrophysics and geophysics, Dedalus is capable of solving a much broader range of PDEs.
To date, it has been used for applications and publications in
applied mathematics \citep{Lecoanet:2018ie,Tobias:2016kp,Tobias:2018iq,Michel:2019kx,Marcotte:2019ua},
astrophysics \citep{Lecoanet:2014ij,Lecoanet:2015jx,Vasil:2015cw,Lecoanet:2016ew,Lecoanet:2017ep,Anders:2017fp,Clark:2017er,Clark:2017dy,Currie:2017cb,Seligman:2017ho,Currie:2018cr,Quataert:2019ku,Anders:2019gt,Clarke:2019up,Anders:2019dq,Lecoanet:2019hm},
atmospheric science \citep{Tarshish:2018jx,Couston:2018iu,Vallis:2019ez,Perrot:2018vq,Lecoanet:2018wi,McKim:2019te},
biology \citep{Mickelin:2018fv,Mussel:2019ia,Mussel:2018vv},
condensed matter physics \citep{Marciani:2019tk,Heinonen:2019gq},
fluid dynamics \citep{Lecoanet:2015dh,Couston:2017eq,Anders:2018ej,Couston:2018eg,Balci:2018kd,Lepot:2018ih,Michel:2019eu,Burns:2019wm,Foldes:2017vv,Olsthoorn:2019db,Saranraj:2018vc,Rocha:2020ch},
glaciology \citep{Kim:2018hj},
limnology \citep{Olsthoorn:2019bw},
numerical analysis \citep{Vasil:2016kb,Vasil:2019ir,Lecoanet:2019jn,Hester:2019vk},
oceanography \citep{Wenegrat:2018hg,Callies:2018in,Tauber:2019ej,Kar:2018ub,Holmes:2019gy},
planetary science \citep{Bordwell:2018bw,Parker:2019bg}, and
plasma physics \citep{Davidovits:2016jw,Davidovits:2016fy,Fraser:2018he,Davidovits:2019dg,Zhu:2019vp,Parker:2019ua,Zhou:2019vu,Parker:2019ws}.

We begin this paper with a review of the fundamental theory of spectral methods and a description of the specific numerical method employed by Dedalus (\secref{sec.numerics}).
We then provide an overview of the project and codebase using a simple example problem (\secref{sec.overview}).
Sections \secref{sec.bases}--\secref{sec.analysis} detail the implementations of the fundamental modules of the codebase, with a particular emphasis on its systems for symbolic equation entry and automatic distributed-memory parallelization.
Although these sections describe essential details of the code, a careful reading is not necessary to begin using Dedalus.
Finally, \secref{sec.examples} demonstrates the features and performance of the codebase with a parallel scaling analysis, a comparison to a finite volume code, and example simulations of nonlinear waves on graphs, compressible magnetohydrodynamic flows, quasi-geostrophic flow in the ocean, Stokes flow in cylindrical geometry, atmospheric normal modes, and diamagnetic levitation.

\section{Sparse spectral methods}
\label{sec.numerics}

\subsection{Fundamentals of spectral methods}

\subsubsection{Spectral representations of functions}

A spectral method discritizes functions by expanding them over a set of basis functions.
These methods find broad application in numerical analysis and give highly accurate and efficient algorithms for manipulating functions and solving differential equations.
The classic reference \citet{Boyd:2001wu} covers the material in this section in great detail.

Consider a complete orthogonal basis $\{\phi_n(x)\}$ and the associated inner product $(\phi_n, \phi_m)_\phi \propto \delta_{n,m}$.
The spectral representation of a function $f(x)$ comprises the coefficients $\{f^\phi_n\}$ appearing in the expansion of $f(x)$ as
\begin{equation}
    f(x) = \sum_{n=0}^\infty f^\phi_n \phi_n(x),
\end{equation}
\noindent with
\begin{equation}
    f^\phi_n = \frac{(\phi_n, f)_\phi}{(\phi_n, \phi_n)_\phi} \equiv \langle \phi_n | f \rangle.
\end{equation}
\noindent We will use bra-ket notation to denote such normalized bra-family inner products.
Formally, an exact representation requires an infinite number of nonzero coefficients.
Numerical spectral methods approximate functions (e.g.\ PDE solutions) using expansions that are truncated after $N$ modes.
The truncated coefficients $\tilde{f}^\phi_n$ are computed using quadrature rules of the form
\begin{equation}
    \tilde{f}^\phi_n = \sum_{i=0}^{N-1} w_i f(x_i)
\end{equation}
where the weights, $w_i$, and collocation points, $x_i$, depend on the underlying inner-product space.
The quadrature scheme constitutes a discrete \emph{spectral transform} for translating between the spectral coefficients and $N$ samples of the function.

The error in the truncated approximation is often of the same order as the last retained coefficient.
The spectral coefficients of smooth functions typically decay exponentially with $n$, resulting in highly accurate representations.
The spectral coefficients of non-smooth functions typically decay algebraically as $n^{-\alpha}$, where $\alpha$ depends on the order of differentiability of $f(x)$.
Exact differentiation and integration on the underlying basis functions provides accurate calculus for general functions.
For PDEs with highly differentiable solutions, spectral methods therefore give significantly more accurate results than fixed-order schemes.

\subsubsection{Common spectral series}
\label{sec.spectral_pde_methods}

\begin{figure}
\centering
\includegraphics[width=\linewidth]{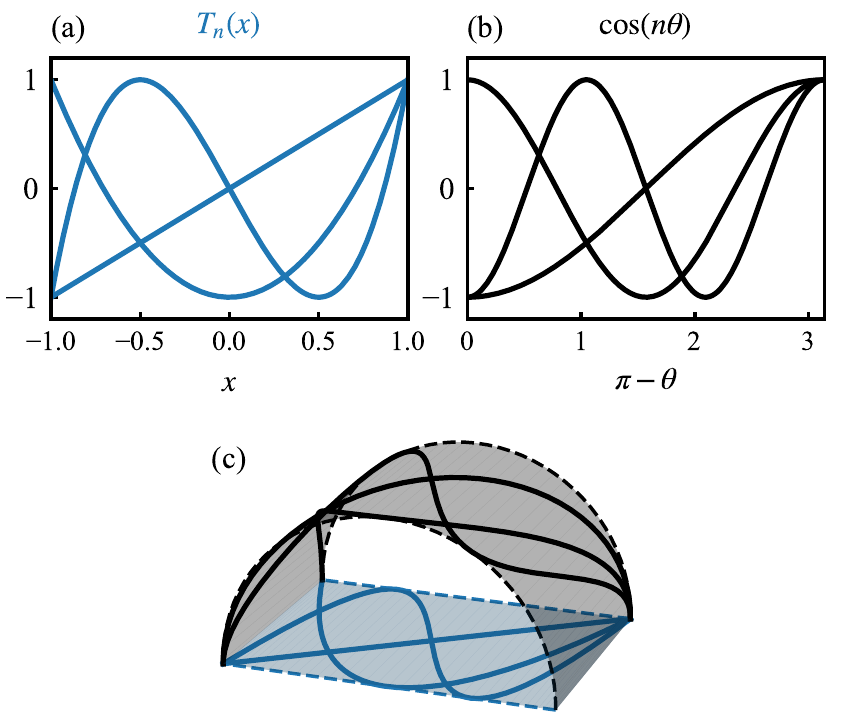}
\caption{Chebyshev polynomials (a) can be viewed as cosine functions (b) drawn on a cylinder and projected onto the bisecting \mbox{plane (c)}.}
\label{fig.chebyshev_polys}
\end{figure}

Trigonometric polynomials are the archetypal spectral bases: sine series, cosine series, and complex exponential Fourier series.
These bases provide exponentially converging approximations to smooth functions on periodic intervals.
The Fast Fourier Transform (FFT) can compute the series coefficients in $\bigO(N \log N)$ time, enabling computations requiring both the coefficients and grid values to be performed efficiently.

The classical orthogonal polynomials also frequently appear as spectral bases.
Most common are the Chebyshev polynomials $\{T_n(x)\}$, which provide exponentially converging approximations to smooth functions on the interval $[-1, 1]$.
A simple change of variables relates the Chebyshev polynomials to cosine functions,
\begin{equation}
    T_n(x) = \cos(n \cos^{-1}(x)).
\end{equation}
\noindent Geometrically, $T_n(x)$ is the projection of $\cos(nx)$ from the cylinder to the plane; see \figref{fig.chebyshev_polys}.
The relation to cosines enables transforming between Chebyshev coefficients and values on collocation points using the fast discrete cosine transform (DCT).
The fast transform often makes Chebyshev series preferable to other polynomials on finite intervals.

\subsubsection{Solving differential equations with spectral methods}

Spectral methods solve PDEs by creating algebraic equations for the coefficients of the truncated solution.
Different approaches for constructing and solving these systems each come with advantages and disadvantages.
In examining a few approaches, we consider a simple linear PDE of the form $L u(x) = f(x)$, where $L$ is a $x$-differential operator.

The \emph{collocation} approach is perhaps the most common polynomial spectral method.
In this case, the differential equation is enforced at the interior collocation points.
The solution is written in terms of the values at these points:
\begin{equation}
    L u(x_i) = f(x_i), \quad \quad i = 0,...,N-1.
\end{equation}
\noindent The boundary conditions typically replace the DE at the collocation endpoints.
The collocation method works well in a broad range of applications.
The primary advantages are that many boundary conditions are easily enforced and the solution occurs on the grid.
The primary disadvantages are that the method produces dense matrices ($L$) and more complicated boundary conditions require more care to implement \citep{Driscoll:2015ea}.

An alternative is the \emph{Galerkin} method, where the solution is written in terms of ``trial'' functions $\{\phi_n\}$ that satisfy the boundary conditions.
The differential equation is then projected against a set of ``test'' functions $\{\psi_n\}$:
\begin{equation}
   \sum_{n} \langle \psi_i | L \phi_n \rangle u^\phi_n  = \sum_{n} \langle \psi_i | \phi_n \rangle f^\phi_n, \quad  i = 0, ..., N-1.
\end{equation}
\noindent For periodic boundary conditions, the Galerkin method using Fourier series produces diagonal derivative operators, allowing constant-coefficient problems to be solved trivially.
Galerkin bases can be constructed from Chebyshev polynomials for simple boundary conditions, with the caveat that the series coefficients must be converted back to Chebyshev coefficients to apply fast transforms.

The \emph{tau} method generalizes the Galerkin method by solving the perturbed equation
\begin{equation}
    L u(x) + \tau P(x) = f(x),
\end{equation}
\noindent where $P(x)$ is specified.
The parameter $\tau$ adjusts to accommodate the boundary conditions, which are enforced simultaneously.
The tau method provides a conceptually straightforward way of applying general boundary conditions without needing a specialized basis.
The classical tau method \citep{lanczos1938trigonometric,Clenshaw:1957hi} uses the same test and trial functions and assumes $P(x) = \phi_{n-1}(x)$, making it equivalent to dropping the last row of the discrete $L$ matrix and replacing it with the boundary condition.
This classical formulation with Chebyshev series results in dense matrices, but (as described below) the tau method can be modified to produce sparse and banded matrices for many equations.

\subsection{A general sparse tau method}

This section describes the spectral method employed in Dedalus.
We use a tau method with different trial and test bases, which produces sparse matrices for general equations.
Formulating problems as first-order systems and using Dirichlet preconditioning (basis recombination) renders matrices fully banded.

The method is fundamentally one-dimensional, but generalizes trivially to $D$-dimensions with all but one separable dimension (e.g.\ Fourier-Galerkin).
Multidimensional problems then reduce to an uncoupled set of 1D problems that are solved individually.

Other approaches based on the ultraspherical method \citep{Olver:2013ed} or integral formulations \citep{Julien:2009kb} similarly result in sparse and banded matrices for many equations.
Our particular approach is designed to accommodate general systems of equations and boundary conditions automatically.

\subsubsection{Sparse differential operators}
\label{sec.sparse_diff_theory}

\begin{figure}
\centering
\includegraphics[width=\linewidth]{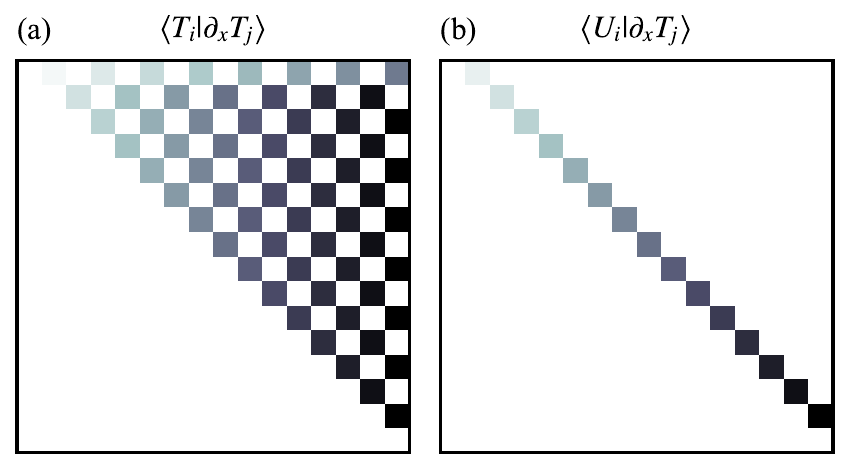}
\caption{Derivative matrices using (a) Chebyshev-T and (b) Chebyshev-U polynomials as test functions.
Using different families of test and trial functions allows general differential operators to be represented with sparse matrices.}
\label{fig.chebyshev_deriv_matrices}
\end{figure}

Traditional polynomial spectral formulations often result in dense derivative matrices.
In particular, the derivatives of Chebyshev polynomials are dense when expanded back in Chebyshev polynomials:
\begin{equation}
    \spderiv{T_j}{x}(x) = \sum_{i=0}^{j-1} T_i(x) \frac{2 j (\modulo{(j-i)}{2})}{1 + \delta_{i,0}}. \label{T-to-T-deriv}
\end{equation}
\figref{fig.chebyshev_deriv_matrices} shows the matrix version of \eqref{T-to-T-deriv}, referred to as the ``T-to-T'' form.
However, the derivatives of Chebyshev polynomials (of the first kind) are proportional to the Chebyshev polynomials of the second kind, $U_n(x)$:
\begin{equation}
    \spderiv{T_n}{x}(x) = n U_{n-1}(x).
\end{equation}
\noindent Defined trigonometrically,
\begin{equation}
    U_n(x) = \frac{\sin((n+1)\cos^{-1}(x))}{\sin(\cos^{-1}(x))}.
\end{equation}

Using test functions $\psi_n=U_{n}$ and trial functions $\phi_{n}=T_{n}$ produces a single-band derivative matrix, called the ``T-to-U'' form (right side in \figref{fig.chebyshev_deriv_matrices}).
We convert non-differential terms from T-to-U via the sparse conversion relation
\begin{equation}
   2T_n(x) = U_n(x) - U_{n-2}(x). \label{T-to-U-conv}.
\end{equation}
Together, these relations render first-order differential equations sparse.
Higher-order equations can be handled by utilizing ultraspherical polynomials for higher derivatives \citep{Olver:2013ed}, but our approach is to simply reduce all equations to first-order systems.
The sparse-$\tau$ method extends to other orthogonal polynomial series.
\appref{sec.diff_conv_matrices} lists the full set of derivative and conversion relations implemented in Dedalus.

\subsubsection{Banded boundary conditions}
\label{sec.sparse_bc_theory}

Choosing the  tau polynomial $P(x) = U_{N-1}(x)$ in a T-to-U method allows dropping the last matrix row and finding $u(x)$ without finding $\tau$.
The system then consists of a banded interior matrix, bordered with a dense boundary-condition row.
Applying a right-preconditioner renders the boundary row sparse and the system fully banded.
For Dirichlet boundary conditions, using the adjoint relation of \eqref{T-to-U-conv} gives the non-orthogonal polynomials,
\begin{equation}
    D_n(x) =
    \begin{cases}
    T_n(x) & n = 0,1 \\
    T_n(x) - T_{n-2}(x) & n \geq 2,
    \end{cases}
\end{equation}
where
\begin{equation}
    D_n(\pm 1) =
    \begin{cases}
    (\pm1)^n & n = 0,1 \\
    0 & n \geq 2.
    \end{cases}
\end{equation}
In this basis, Dirichlet boundary conditions only involve the first two expansion coefficients.
This technique is known as ``Dirichlet preconditioning'' or ``basis recombination''.

In summary, for first-order systems with Dirichlet boundary conditions, choosing $\phi_n = D_n$, $\psi_n = U_n$, and $P = U_{N-1}$ produces fully banded matrices.
The resulting matrices are efficiently sovled using sparse/banded algorithms.
With a first-order system, it is possible to reformulate any boundary condition (e.g.\ Neumann or global integral conditions) in terms of a Dirichlet condition on the first-order variables.
This formulation extends to other orthogonal polynomial bases.
\appref{sec.dirichlet_matrices} lists the full set of Dirichlet recombinations implemented in Dedalus.

\subsubsection{Non-constant coefficients}
\label{sec.sparse_ncc_theory}

Many physical problems require multiplication by spatially non-constant coefficients (NCCs) that vary slowly compared to the unknown solution.
\citet{Olver:2013ed}, in the context of Chebyshev polynomials, observed that multiplication by such NCCs corresponds to band-limited spectral operators.

Multiplication by a general NCC $g(x)$ acts linearly on $u(x)$ via
\begin{equation}
	\sum_{j} \langle \psi_i | g \phi_j \rangle u^\phi_j = \sum_{j} G^{\psi \phi}_{i, j} u^\phi_j.
\end{equation}
\noindent Given an expansion of the NCC in some basis as $g(x) = \sum_{n=0}^{\infty} g^{\xi}_n \xi_n(x)$, the NCC matrix is
\begin{equation}
	G^{\psi \phi}_{i, j} = \sum_{n=0}^{\infty} g^{\xi}_n  \langle \psi_i | \xi_n \phi_j \rangle.
\end{equation}

For all orthogonal polynomials, $\langle \psi_i | \xi_n \phi_j \rangle = 0$ if $|i - j| > n$.
We truncate NCC expansions by dropping all terms where $g^\xi_n$ is smaller than some threshold amplitude.
The overall bandwidth of $G^{\psi \phi}$ is therefore $N'$, the number of terms that are retained in the expansion of the NCC.
For smooth NCCs, $N' \ll N$ and the multiplication matrix has low bandwidth.
\appref{sec.ncc_matrices} lists the full set of multiplication matrices implemented in Dedalus.

\subsubsection{Solving systems of equations}
\label{sec.sparse_system_theory}

For coupled systems of equations with $S$ variables, $\sum_{j=1}^{S} L^{i,j} u_j(x) = f_i(x)$ for $i = 1, ..., S$.
In block-operator form
\begin{equation}
    \mathcal{L} \cdot \mathcal{X} = \mathcal{F},
\end{equation}
\noindent where $\mathcal{X}$ is the state-vector of variables, $u_j$, and $\mathcal{L}$ is a matrix of the operators.
We discretize the system by replacing each $L^{i,j}$ with its sparse-tau matrix representation described above.
In block-banded form, using Kronecker products and the placement matrix $(E^{i,j})_{k,l} = \delta_{i,k} \delta_{k,l}$,
\begin{equation}
    L_\mathrm{sys} = \sum_{i,j=1}^S E^{i,j} \otimes L^{i,j}.
\end{equation}
The system matrix $L_\mathrm{sys}$ acts on the concatenation of the variable coefficients and has bandwidth $\bigO(S N)$.

If the operator matrices are interleaved rather than block-concatenated, the resulting system matrix will act on the interleaved variable coefficients and takes the form
\begin{equation}
    L_\mathrm{sys} = \sum_{i,j=1}^S L^{i,j} \otimes E^{i,j}.
\end{equation}
This matrix will have bandwidth $\bigO(S)$, making it practical to simultaneously solve coupled systems of equations with large $N$.

\subsubsection{Summary \& example}

\begin{figure*}
\centering
\includegraphics[width=\linewidth]{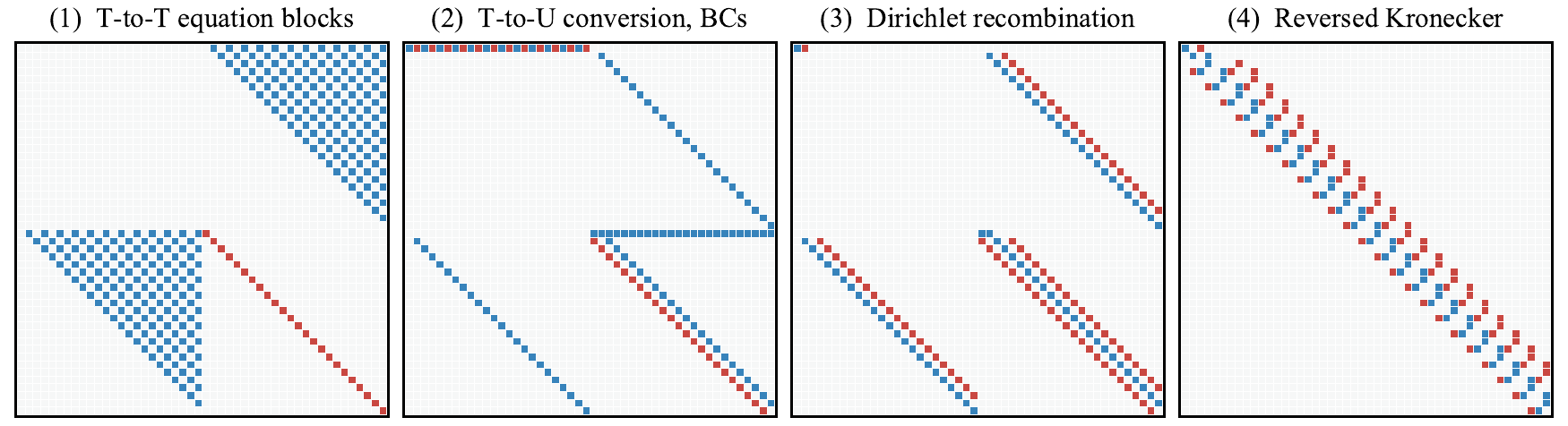}
\caption{Conceptual stages in matrix construction for the Chebyshev discretization of Poisson's equation with Dirichlet and Neumann boundary conditions (matrix entries colored by sign).
(1) Original (T-to-T) matrices with block columns corresponding to $u$ and $u_x$, and block rows corresponding to the LHS of $\spderiv{u_x}{x} = 0$ and $\spderiv{u}{x} - u_x = f(x)$.
The T-to-T differentiation matrices are dense upper triangular.
(2) After T-to-U conversion and the addition of boundary conditions via the tau method.
The derivatives are sparse and the identity-block bandwidths increase slightly.
The boundary conditions involve all coefficients of $u$ and $u_x$.
(3) After Dirichlet recombination.  Boundary rows are sparse and the equation-block bandwidths increase slightly.
(4) After reversing the Kronecker product to group by modes rather than variables.
This final matrix is highly sparse and completely banded.}
\label{fig.matrix_steps}
\end{figure*}

Dedalus uses a modified tau method that produces banded and well-conditioned matrices.
Carefully chosen test-trial basis pairs render derivatives sparse.
The first-order formulation makes all boundary conditions equivalent to Dirichlet conditions, which become banded under basis recombination.
Truncated NCC expansions retain bandedness for smooth NCCs.
Finally, matrices and coefficients are interleaved to keep systems of equations banded.

\figref{fig.matrix_steps} shows the matrices at various conceptual stages for a Chebyshev discretization of Poisson's equation in 1D with Dirichlet and Neumann boundary conditions:
\begin{equation}
    \npderiv{u}{x}{2} = f(x), \quad u(-1) = 0, \quad \spderiv{u}{x}(1) = 0.
\end{equation}
The combination of writing equations as first-order systems, T-to-U derivative mapping, Dirichlet preconditioning, and grouping modes before variables produces a banded pencil matrix.

We also note that coupled systems easily allow constraint equations (those without temporal derivatives) to be imposed alongside evolution equations, avoiding variable reformulations and/or splitting methods.
For instance, the divergence condition in incompressible hydrodynamics can be imposed directly (determining the pressure).
The momentum equation can then be integrated without splitting or derived pressure boundary conditions.

\section{Project overview and design}
\label{sec.overview}

\subsection{Codebase structure}

Dedalus makes extensive use of object-oriented programming to provide a simple interface for the parallel solution of general systems of PDEs.
The basic class structure reflects the mathematical objects that are encountered when posing and solving a PDE.
As an illustrative example, we consider solving the Fisher-KPP equation, a reaction-diffusion equation that first arose in ecology:
\begin{equation}
    \pderiv{u}{t} - D \lap{u} = R \, u (1 - u),
\end{equation}
\noindent with unknown variable $u(\vec{x},t)$, diffusion coefficient $D$, and reaction rate $R$ \cite{doi:10.1111/j.1469-1809.1937.tb02153.x, KPP1937}.

Properly posing the PDE first requires specifying its spatial domain.
This is done by creating a \pyth{Basis} object discretizing each dimension over a specified interval and forming a \pyth{Domain} object as the direct-product of these bases.
Here we construct a 2D channel domain, periodic in $x$ and with Neumann boundary conditions on the boundaries in $y$, as the direct product of a Fourier basis and a Chebyshev basis:
\begin{python}
import numpy as np
import dedalus.public as de

# Bases: names, modes, intervals, dealiasing
x_basis = de.Fourier('x', 128,
    interval=(0, 2*np.pi), dealias=3/2)
y_basis = de.Chebyshev('y', 64,
    interval=(0, 1), dealias=3/2)
# Domain: bases, datatype
domain = de.Domain([x_basis, y_basis], float)
\end{python}

Next, we define an initial value problem on this domain consisting of the PDE in first-order form (for temporal and Chebyshev derivatives) along with the boundary conditions.
This is done by creating an \pyth{Problem} object representing the problem type (here an initial value problem, \pyth{IVP}).
Problem parameters, string-substitutions (to simplify equation entry), and plain-text equations and boundary conditions are then added to the problem.
Under the hood, Dedalus constructs \pyth{Field} objects to represent the variables and \pyth{Operator} objects that symbolically represent the mathematical expressions in the PDE.
\begin{python}
# Problem: domain and variables
prob = de.IVP(domain, variables=['u','uy'])
# Fixed parameters
prob.parameters['D'] = 0.1
prob.parameters['R'] = 1
# Parsing substitutions
prob.substitutions['Lu'] = "dx(dx(u)) + dy(uy)"
# First-order reduction
prob.add_equation("uy - dy(u) = 0")
# Fisher-KPP equation
prob.add_equation("dt(u) - D*Lu = R*u*(1-u)")
# Neumann boundary conditions
prob.add_bc("left(uy) = 0")
prob.add_bc("right(uy) = 0")
\end{python}

To finish posing the \pyth{IVP}, we need to specify a temporal integration scheme, the temporal integration limits, and the initial values of the variables.
This is done through the \pyth{Solver} object built by each \pyth{Problem}.
The initial data of the state fields can be easily accessed from the solver and set in grid (\pyth{'g'}) or coefficient (\pyth{'c'}) space.
\begin{python}
# Pick a timestepper
ts = de.timesteppers.RK222
# Build solver
solver = prob.build_solver(ts)
# Set integration limits
solver.stop_sim_time = 10
# Set initial conditions
u = solver.state['u']
u['g'] = np.random.rand(*u['g'].shape)
\end{python}

Finally, the problem is solved by iteratively applying the temporal integration scheme to advance the solution in time.
In Dedalus, this main loop is directly written by the user, allowing for arbitrary data interactions as the integration progresses.
Although not included in this example, Dedalus also provides extensive analysis tools for evaluating and saving quantities during the integration.
\begin{python}
dt = 0.01
# Main loop chceking stopping criteria
while solver.ok:
    # Step forward
    solver.step(dt)
    # Perform some analysis
    print(np.mean(u['g']), np.std(u['g']))
\end{python}

While this simple example covers the core user-facing classes, several other classes control the automatic MPI parallelization and efficient solution of the solvers.
All together, the fundamental class hierarchy consists of the following:

\begin{itemize}
    \item \pyth{Basis}: A one-dimensional spectral basis.
    \item \pyth{Domain}: The direct product of multiple bases, forming the spatial domain of dependence of a PDE.
    \item \pyth{Distributor}: Directs the parallel decomposition of a domain and spectral transformations of distributed data between different states.
    \item \pyth{Layout}: A distributed transformation state, e.g.\ grid-space or coefficient-space.
    \item \pyth{Field}: A scalar-valued field over a given domain. The fundamental data unit in Dedalus.
    \item \pyth{Operator}: Mathematical operations on sets of fields, composed to form mathematical expressions.
    \item \pyth{Handler}: Captures the outputs of multiple operators to store in memory or write to disk.
    \item \pyth{Evaluator}: Efficiently coordinates the simultaneous evaluation of the tasks from multiple handlers.
    \item \pyth{Problem}: User-defined PDEs (initial, boundary, and eigen-value problems).
    \item \pyth{Timestepper}: ODE integration schemes that are used to advance initial value problems.
    \item \pyth{Solver}: Coordinates the actual solution of a problem by evaluating the underlying operators and performing time integration, linear solves, or eigenvalue solves.
\end{itemize}

\noindent The following sections detail the functionality and implementation of these classes.

\subsection{Dependencies}

Dedalus is provided as an open-source Python3 package.
We choose to develop the code in Python because it is an open-source, high-level language with a vast ecosystem of libraries for numerical analysis, system interaction, input/output, and data visualization.
While numerical algorithms written directly in Python sometimes suffer from poor performance, it is quite easy to wrap optimized C libraries into high-level interfaces with Python.
A typical high-resolution Dedalus simulation will spend a majority of its time in optimized C libraries.

The primary dependencies of Dedalus include:
\begin{itemize}
    \item The Numpy, Scipy, and Cython packages for Python \citep{scipybib,Behnel:2011gs}.
    \item The FFTW C-library for fast Fourier transforms \citep{Frigo:cp}.
    \item An implementation of the MPI communication interface and its Python wrapper mpi4py \citep{Dalcin:2008iq}.
    \item The HDF5 C-library for reading and writing HDF5 files and its Python wrapper h5py \citep{hdf5,collette_python_hdf5_2014}.
\end{itemize}
A wide range of standard-library Python packages are used to build logging, configuration, and testing interfaces following standard practices.

Additionally, and perhaps counter-intuitively, we have found that creating algorithms to accommodate a broad range of equations and domains has resulted in a compact and maintainable codebase.
Currently, the Dedalus package consists of roughly 10,000 lines of Python.
By producing sufficiently generalized algorithms, it is possible to compactly and robustly provide a great deal of functionality.

\subsection{Documentation}

Dedalus has been publicly available under the open-source GPL3 license since its creation, and is developed under distributed version control.
The online documentation includes a series of tutorials and example problems demonstrating the code's capabilities and walking new users through the basics of constructing and running a simulation.
Links to the source code repository and the documentation are available through the project website, \url{http://dedalus-project.org/}.

\subsection{Community}

The Dedalus collaboration uses open-source code development and strongly supports open scientific practices.
The benefits of the philosophy include distributed contributions to the codebase, a low barrier-to-entry (especially for students), and detailed scientific reproducibility.
We outline these ideas in detail in \citet{Oishi:2018th}.

Code development occurs through a public system of pull requests and reviews on the source code repository.
Periodic releases are issued to the Python Package Index (PyPI), and a variety of full-stack installation channels are supported, including single-machine and cluster install scripts and a conda-based build procedure.

The core developers maintain mailing lists for the growing Dedalus user and developer communities.
The mailing list is publicly archived and searchable, allowing new users to find previous solutions to common problems with installation and model development.
The Dedalus user list currently has over 150 members.
A list of publications using the code is \chref{http://dedalus-project.org/citations/}{maintained online}.

\section{Spectral bases}
\label{sec.bases}

Dedalus currently represents multidimensional fields using the direct product of one-dimensional spectral bases.
This direct product structure generally precludes domains including coordinate singularities, such as full disks or spheres.
However, curvilinear domains without coordinate singularities, such as cylindrical annuli can still be represented using this direct product structure (for an example, see \ref{sec.stokes_example}).
Basis implementations form the lowest level of the program's class hierarchy.
The primary responsibilities of the basis classes are to define their collocation points and to provide an interface for transforming between the spectral coefficients of a function and the values of the function on their collocation points.

An instance of a basis class represents a series of its respective type truncated to a given number of modes $N_c$, and remapped from \emph{native} to \emph{problem} coordinates with an affine map.
A basis object is instantiated with a \pyth{name} string defining the coordinate name, a \pyth{base_grid_size} integer setting $N_c$, parameters fixing the affine coordinate map (a \emph{problem} coordinate interval for bases on finite intervals, or stretching and offset parameters for bases on infinite intervals), and a \pyth{dealias} factor.
Each basis class is defined with respect to a \emph{native} coordinate interval, and contains a method for producing a collocation grid of $N_g$ points on this interval, called a \emph{native grid} of \emph{scale} $s = N_g / N_c$.
Conversions between the \emph{native coordinates} $x_n$ and \emph{problem coordinates} $x_p$ are done via an affine map of the form $x_p = a x_n + b$, which is applied to the native grid to produce the \emph{basis grid}.

Each basis class defines methods for \emph{forward transforming} (moving from grid values to spectral coefficients) and \emph{backward transforming} (vice versa) data arrays along a single axis.
Using objects to represent bases allows the transform methods to easily cache plans or matrices that are costly to construct.
The basis classes also present a unified interface for implementing identical transforms using multiple libraries with different performance and build requirements.
We now define the basis functions, grids, and transform methods for the currently implemented spectral bases.

\begin{figure*}
\centering
\includegraphics[width=\textwidth]{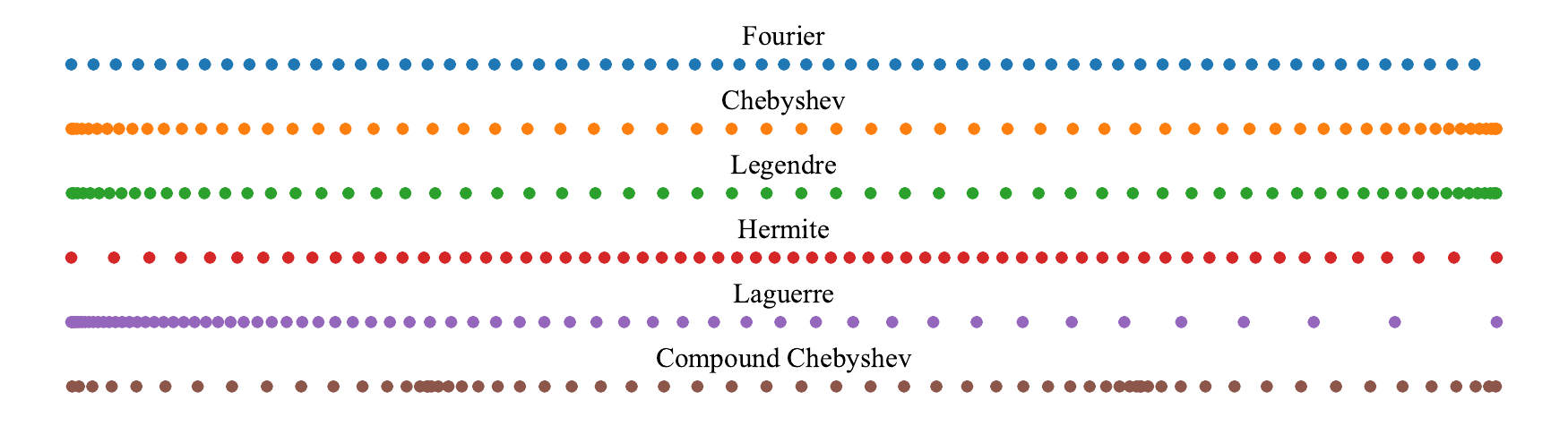}
\caption{Example collocation grids for the implemented bases.
The \pyth{Fourier} basis grid is evenly spaced on a periodic interval.
The \pyth{Chebyshev} and \pyth{Legendre} basis grids cluster near the ends of a finite interval.
The \pyth{Hermite} basis grid spreads out over the real line.
The \pyth{Laguerre} basis grid spreads one way over the half line.
The \pyth{Compound} basis concatenates other bases on adjascent segments; e.g.\ three Chebyshev segments.}
\label{fig.basis_grids}
\end{figure*}

\subsection{Fourier basis}

For periodic dimensions, we implement a \pyth{Fourier} basis consisting of complex exponential modes on the native interval $[0, 2\pi]$:
\begin{equation}
    \phi^F_k(x) = \exp(i k x)
\end{equation}
and a native grid consisting of evenly-spaced points beginning at the left side of the interval:
\begin{equation}
    x_i = \frac{2 \pi i}{N_g}, \quad \quad i = 0, ..., N_g - 1.
\end{equation}

A function is represented as a symmetric sum over positive and negative wavenumbers
\begin{equation}
    f(x) = \sum_{-k_{m}}^{k_{m}} f_k \phi^F_k(x)
\end{equation}
where $k_{m} = \mathrm{floor}((N_c - 1) / 2)$ is the maximum resolved wavenumber, excluding the Nyquist mode $k_{N} = N_c / 2$ when $N_c$ is even.
When $f$ is a real function, we store only the complex coefficients corresponding to $k \geq 0$; the $k < 0$ coefficients are determined by conjugate symmetry.
We discard the Nyquist mode since it is only marginally resolved: for real functions, the Nyquist mode captures $\cos(k_N x)$, but not $\sin(k_N x)$, which vanishes on the grid when $N_g = N_c$.

The expansion coefficients are given explicitly by
\begin{align}
    f_k &= \frac{1}{2 \pi} \int_0^{2 \pi} f(x) \phi^{F*}_k(x) \diff{x} \\
    &= \frac{1}{N_g} \sum_{i=0}^{N_g-1} f(x_i) \phi^{F*}_k(x_i)
\end{align}
and are computed with the fast Fourier transform (FFT).
We implement FFTs from both the Scipy and FFTW libraries, and rescale the results to match the above normalizations, i.e.\  the coefficients directly represent mode amplitudes.
The coefficients are stored in the traditional FFT output format, starting from $k=0$ and increasing to $k_m$, then following with $-k_m$ and increasing to $-1$.

\subsection{Sine/Cosine basis}

For periodic dimensions possessing definite symmetry with respect to the interval endpoints, we implement a \pyth{SinCos} basis consisting of either sine waves or cosine waves on the native interval $[0, \pi]$:
\begin{equation}
    \phi^c_k(x) = \cos(k x),
\end{equation}
\begin{equation}
    \phi^s_k(x) = \sin(k x)
\end{equation}
and a native grid consisting of evenly-spaced interior points:
\begin{equation}
    x_i = \frac{\pi (i + \sfrac{1}{2})}{N_g}, \quad \quad i = 0, ..., N_g - 1.
\end{equation}

Functions with even parity are represented with cosine series as
\begin{equation}
    f(x) = \sum_{k=0}^{N_c-1} f_k \phi^c_k(x)
\end{equation}
while functions with odd parity are represented with sine series as
\begin{equation}
    g(x) = \sum_{k=1}^{N_c-1} g_k \phi^s_k(x).
\end{equation}
The Nyquist mode $k_N = N_c$ is dropped from the sine series, since the corresponding cosine mode vanishes on the grid when $N_g = N_c$.

The expansion coefficients are given explicitly by
\begin{align}
    f_k &= \frac{2 - \delta_{k,0}}{\pi} \int_0^{\pi} f(x) \phi^{c}_k(x) \diff{x} \\
    &= \frac{2 - \delta_{k,0}}{N_g} \sum_{i=0}^{N_g-1} f(x_i) \phi^{c}_k(x_i),
\end{align}
\begin{align}
    g_k &= \frac{2}{\pi} \int_0^{\pi} g(x) \phi^{s}_k(x) \diff{x} \\
    &= \frac{2}{N_g} \sum_{i=0}^{N_g-1} g(x_i) \phi^{s}_k(x_i)
\end{align}
and are computed using the fast discrete cosine transform (DCT) and discrete sine transform (DST).
The same grid is used for both series, corresponding to type-II DCT/DSTs for the forward transforms, and type-III DCT/DSTs for the backward transforms.
We implement transforms from both the Scipy and FFTW libraries, and rescale the results to match the above normalizations, i.e.\ the coefficients directly represent mode amplitudes.

These transforms are defined to act on real arrays, but since they preserve the data-type of their inputs, they can be applied simultaneously to the real and imaginary parts of a complex array.
The spectral coefficients for complex functions are therefore also complex, with their real and imaginary parts representing the coefficients of the real and imaginary parts of the function.

\subsection{Chebyshev basis}

For finite non-periodic dimensions, we implement a \pyth{Chebyshev} basis consisting of the Chebyshev-T polynomials on the native interval $[-1, 1]$:
\begin{equation}
    T_n(x) = \cos(n \cos^{-1}(x)).
\end{equation}
The native Chebyshev grid uses the Gauss-Chebyshev quadrature nodes (a.k.a.\ the \emph{roots} or \emph{interior} grid):
\begin{equation}
    x_i = -\cos\left(\frac{\pi (i + \sfrac{1}{2})}{N_g}\right), \quad \quad i = 0, ..., N_g - 1.
\end{equation}
Near the center of the interval, the grid approaches an even distribution where $\Delta x \approx \pi / N_g$.
Near the ends of the interval, the grid clusters quadratically and allows very small structures to be resolved (\figref{fig.basis_grids}).

A function is represented as
\begin{equation}
    f(x) = \sum_{n=0}^{N_c-1} f_n T_n(x).
\end{equation}
\noindent The expansion coefficients are given explicitly by
\begin{align}
    f_n &= \frac{2 - \delta_{n,0}}{\pi} \int_{-1}^{1} \frac{f(x) T_n(x)}{\sqrt{1 - x^2}} \diff{x} \\
    &= \frac{2 - \delta_{n,0}}{N_g} \sum_{i=0}^{N_g-1} f(x_i) T_n(x_i)
\end{align}
and are computed using the fast discrete cosine transform (DCT) via the change of variables $x = \cos(\theta)$.
The Chebyshev basis uses the same Scipy and FFTW DCT functions as the cosine basis, wrapped to handle the sign difference in the change-of-variables and preserve the ordering of the Chebyshev grid points.
It also behaves similarly for complex functions, preserving the data type and producing complex coefficients for complex functions.

Chebyshev rational functions can also be used to discretize the half line and the entire real line \citep{Miquel:2017hq}.
These functions are not implemented explicitly in Dedalus, but can be utilized with the Chebyshev basis by manually including changes of variables in the equations.

\subsection{Legendre basis}

For finite non-periodic dimensions, we also implement a \pyth{Legendre} basis consisting of the Legendre polynomials $P_n(x)$ on the native interval $[-1, 1]$.
The Legendre polynomials are orthogonal on this interval as
\begin{equation}
    \int_{1}^{1} P_n(x) P_m(x) \diff{x} = \delta_{n,m} N_n^2,
   \end{equation}
\noindent where $N_n^2 = 2 / (2 n + 1)$.

The native grid points $x_i$ are the Gauss-Legendre quadrature nodes, calculated using \pyth{scipy.special.roots_legendre}.

A function is represented as
\begin{equation}
	f(x) = \sum_{n=0}^{N_c-1} f_n P_n(x).
\end{equation}
\noindent The expansion coefficients are given by
\begin{align}
	f_n &= \frac{1}{N_n^2} \int_{-\infty}^{\infty} f(x) P_n(x) \diff{x} \\
	&= \frac{1}{N_n^2} \sum_{i=0}^{N_g-1} w_i f(x_i) P_n(x_i)
\end{align}
\noindent where $w_i$ are the Gauss-Legendre quadrature weights, also computed using \pyth{scipy.special.roots_legendre}.
We synthesize the basis functions with the standard recursion relations in extended precision, which prevents underflows or overflows in problems with large numbers of modes.
Quadrature-based Matrix-Multiply Transforms (MMTs) convert between grid values and coefficients.

\subsection{Hermite basis}

For problems on the whole real line $(-\infty, \infty)$, we implement a \pyth{Hermite} basis consisting of the physicists' Hermite polynomials $H_n(x)$ and the normalized Hermite functions
\begin{equation}
	\phi^H_{n}(x) = e^{-x^2/2} H_n(x) / N_n,
\end{equation}
\noindent where $N_n^2 = \pi^{1/2} 2^n n!$.
The Hermite polynomials are orthogonal under the Gaussian weight as
\begin{equation}
	\int_{-\infty}^{\infty} H_n(x) H_m(x) e^{-x^2} \diff{x} = \delta_{n,m} N_n^2.
\end{equation}
The ``enveloped'' Hermite functions incorporate the weight and normalizations so that
\begin{equation}
	\int_{-\infty}^{\infty} \phi^H_n(x) \phi^H_m(x) \diff{x} = \delta_{n,m}.
\end{equation}

Since the Hermite functions exist over the entire real line, the affine map from native to problem coordinates is fixed by specifying \pyth{center} and \pyth{stretch} parameters, rather than specifying a problem interval.
The native grid points $x_i$ are the Gauss-Hermite quadrature nodes, calculated using \pyth{scipy.special.roots_hermite}.

Polynomial functions are represented in the standard basis as
\begin{equation}
	f(x) = \sum_{n=0}^{N_c-1} f_n H_n(x),
\end{equation}
\noindent while functions that decay towards infinity are represented in the enveloped basis as
\begin{equation}
	g(x) = \sum_{n=0}^{N_c-1} g_n \phi^H_n(x).
\end{equation}

The expansion coefficients are
\begin{align}
	f_n &= \frac{1}{N_n^2} \int_{-\infty}^{\infty} f(x) H_n(x) e^{-x^2} \diff{x} \\
	&= \frac{1}{N_n^2} \sum_{i=0}^{N_g-1} w_i f(x_i) H_n(x_i)
\end{align}
\begin{align}
	g_n &= \int_{-\infty}^{\infty} g(x) \phi^H_n(x) \diff{x} \\
	&= \sum_{i=0}^{N_g-1} w_i e^{x_i^2} g(x_i) \phi^H_n(x_i)
\end{align}
\noindent where $w_i$ are the Gauss-Hermite quadrature weights, also computed using \pyth{scipy.special.roots_hermite}.
We synthesize the basis functions with the standard recursion relations in extended precision, which prevents underflows or overflows in problems with hundreds of modes.
Quadrature-based Matrix-Multiply Transforms (MMTs) convert between grid values and coefficients.

\subsection{Laguerre basis}

For problems on the half real line $(0, \infty)$, we implement a \pyth{Laguerre} basis consisting of the standard Laguerre polynomials $L_n(x)$ and the normalized Laguerre functions
\begin{equation}
	\phi^L_{n}(x) = e^{-x/2} L_n(x).
\end{equation}
The Laguerre polynomials are orthonormal under the exponential weight:
\begin{equation}
	\int_{0}^{\infty} L_n(x) L_m(x) e^{-x} \diff{x} = \delta_{n,m}.
\end{equation}
The enveloped functions incorporate the weight so that
\begin{equation}
	\int_{0}^{\infty} \phi^L_n(x) \phi^L_m(x) \diff{x} = \delta_{n,m}.
\end{equation}

Since the Laguerre functions exist over the positive half line, the affine map from native to problem coordinates is fixed by specifying \pyth{edge} and \pyth{stretch} parameters, rather than specifying a problem interval.
A negative value of the \pyth{stretch} parameters can be used to create a basis spanning the negative half line.
The native grid points $x_i$ are the Gauss-Laguerre quadrature nodes, calculated using \pyth{scipy.special.roots_laguerre}.

Polynomial functions are represented in the standard basis as
\begin{equation}
	f(x) = \sum_{n=0}^{N_c-1} f_n L_n(x)
\end{equation}
\noindent while functions that decay towards infinity are represented in the enveloped basis as
\begin{equation}
	g(x) = \sum_{n=0}^{N_c-1} g_n \phi^L_n(x).
\end{equation}

The expansion coefficients are
\begin{align}
	f_n &= \int_{0}^{\infty} f(x) L_n(x) e^{-x} \diff{x} \\
	&= \sum_{i=0}^{N_g-1} w_i f(x_i) L_n(x_i)
\end{align}
\begin{align}
	g_n &= \int_{0}^{\infty} g(x) \phi^L_n(x) \diff{x} \\
	&= \sum_{i=0}^{N_g-1} w_i e^{x_i} g(x_i) \phi^L_n(x_i)
\end{align}
\noindent where $w_i$ are the Gauss-Laguerre quadrature weights, also computed using \pyth{scipy.special.roots_laguerre}.
We synthesize the basis functions with the standard recursion relations in extended precision, which prevents underflows or overflows in problems with hundreds of modes.
Quadrature-based Matrix-Multiply Transforms (MMTs) convert between grid values and coefficients.

\subsection{Compound bases}

An arbitrary number of adjacent polynomial segments can be connected to form a \pyth{Compound} basis.
The spectral coefficients on each subinterval are concatenated to form the compound coefficient vector, and the standard transforms operate on each subinterval.
The compound basis grid is similarly the concatenation of the subinterval grids.
There are no overlapping gridpoints at the interfaces since the polynomial bases use interior grids.
Continuity is not required a priori at the interfaces, but is imposed on the solutions when solving equations (see \ref{sec.matrix_construction}).

The subintervals making up a compound basis may have different resolutions and different lengths, but must be adjacent.
Compound bases are useful for placing higher resolution (from clustering near the endpoints of polynomial grids) at fixed interior locations.
Compound expansions can also substantially reduce the number of modes needed to resolve a function that is not smooth if the positions where the function becomes non-differentiable are known.
\figref{fig.basis_grids} shows the grid of a compound basis composed of three Chebyshev segments.

\subsection{Scaled transforms \& dealiasing}

Each basis implements transforms between $N_c$ coefficients and scaled grids of size $N_g = s N_c$, where $s$ is the \emph{transform scale}.
When $s < 1$, the coefficients are truncated after the first $N_g$ modes before transforming.
Such transforms are useful for viewing compressed (i.e.\ filtered) versions of a field in grid space.
When $s > 1$, the coefficients are padded with $N_g - N_c$ zeros above the highest modes before transforming.
Padding is useful for spectral interpolation, i.e.\ to view low resolution data on a fine grid.

Transforms with $s > 1$ are necessary to avoid aliasing errors when calculating nonlinear terms, such as products of fields, in grid space.
For each basis, the \pyth{dealias} scale is set at instantiation and defines the transform scale that is used when evaluating mathematical operations on fields.
The well-known ``3/2 rule'' states that properly dealiasing quadratic nonlinearities calculated on the grid requires a transform scale of $s \ge 3/2$.
In general, an orthogonal polynomial of degree $N_g + n$ will alias down to degree $N_g - n$ when evaluated on the collocation grid of size $N_g$.
A nonlinearity of order $P$ involving expansions up to degree $N_c - 1$ will have power up to degree $P (N_c - 1)$.
For this maximum degree to not alias down into degree $N_c - 1$ of the product, we must have $2 N_g > (P + 1) (N_c - 1)$.
Picking a dealias scale of $s = (P + 1) / 2$ is therefore sufficient to evaluate the nonlinearity without aliasing errors in the first $N_c$ coefficients.
Non-polynomial nonlinearities, such as negative powers of fields, cannot be fully dealiased using this method, but the aliasing error can be reduced by increasing $s$.

\subsection{Transform plans}

To minimize code duplication and maximize extensibility, our algorithms require that each transform routine can be applied along an arbitrary axis of a multidimensional array.
Scipy transforms include this functionality, and we built Cython wrappers around the FFTW Guru interface to achieve the same.
The wrappers produce plans for FFTs along one dimension of an arbitrary dimensional array by collapsing the axes before and after the transform axis, and creating an FFTW plan for a two-dimensional loop of rank-1 transforms.
For example, to transform along the third axis of a five-dimensional array of shape $(N_1, N_2, N_3, N_4, N_5)$, the array would be viewed as a three-dimensional array of shape $(N_1 N_2, N_3, N_4 N_5)$ and a loop of $N_1 N_2 \times N_4 N_5$ transforms of size $N_3$ would occur.
This approach allows for the unified planning and evaluation of transforms along any dimension of an array of arbitrary dimension, reducing the risk of coding errors that might accompany treating different dimensions of data as separate cases.

The plans produced by FFTW are cached by the corresponding basis objects and executed using the FFTW new-array interface.
This centralized caching of transform plans reduces both precomputation time and the memory footprint necessary to plan FFTW transforms for many data fields.
The FFTW planning rigor, which determines how much precomputation should be performed to find the optimal transform algorithm, is also wrapped through the Dedalus configuration interface.

\section{Domains}
\label{sec.domains}

\pyth{Domain} objects represent physical domains, discretized by the direct product of one-dimensional spectral bases.
A Dedalus simulation will typically contain a single domain object, which functions as the overall context for fields and problems in that simulation.
A domain is instantiated with a list of basis objects forming this direct product, the data type of the variables on the domain (double precision real (64-bit) or complex (128-bit) floating point numbers), and the process mesh for distributing the domain when running Dedalus in parallel.

\subsection{Parallel data distribution}

Computations in Dedalus are parallelized by subdividing and distributing the data of each field over the available processes in a distributed-memory MPI environment.
The domain class internally constructs a \pyth{Distributor} object that directs the decomposition and communication necessary to transform the distributed fields between grid space and coefficient space.
Specifically, a domain can be distributed over any lower-dimensional array of processes, referred to as the process mesh.
The process mesh must be of lower dimension than the domain so that at least one dimension is local at all times.
Spectral transforms are performed along local dimensions and parallel data transpositions change the data locality to enable transforms across all dimensions.

To coordinate this process, the distributor constructs a series of \pyth{Layout} objects describing the necessary transform and distribution states of the data between coefficient space and grid space.
Consider a domain of dimension $D$ and shape $(N_1, N_2, ..., N_D)$ distributed over a process mesh of dimension $P < D$ and shape $(M_1, M_2, ..., M_P)$:
\begin{itemize}
    \item
    The first layout is full coefficient space, where the first $P$ array dimensions are block-distributed over the corresponding mesh axes, and the last $D-P$ dimensions are local.
    That is, the $i$-th dimension is split in adjacent blocks of size $B_{i,i} = \ceil{N_i/ M_i}$, and the process with index $(m_1, m_2, ..., m_P)$ in the mesh will contain the data block from \mbox{$m_i B_{i,i} : (m_i+1) B_{i,i}$} in the $i$-th dimension.
    \item
    The subsequent $D-P$ layouts sequentially transform each dimension to grid space starting from the last dimension and moving backwards.
    \item
    After $D-P$ transforms, the first $P$ dimensions are distributed and in coefficient space, and the last $D-P$ dimensions are local and in grid space.
    A global data transposition makes the $P$-th dimension local in the next layout.
    This transposition occurs along the $P$-th mesh axis, gathering the distributed data along the $P$-th array dimension and redistributing it along the $(P+1)$-th array dimension.
    This is an all-to-all communication within each one-dimensional subset of processes in the mesh defined by fixed $(m_1, ..., m_{P-1})$.
    \item
    The next layout results from transforming the $P$-th dimension (now local) to grid space.
    \item
  The transposition step then repeats to reach the next layout: all-to-all communication transposes the $(P-1)$ and $P$-th array dimensions over the $(P-1)$-th mesh axis.
    \item
    The next layout results from transforming the $(P-1)$-th dimension (now local) to grid space.
    \item
    This process repeats, reaching new layouts by alternately gathering and transforming sequentially lower dimensions until the first dimension becomes local and is transformed to grid space.
\end{itemize}
The final layout is full grid space.
The first dimension is local, the next $P$ dimensions are distributed in blocks of size \mbox{$B_{n,n-1} = \ceil{N_{n} / M_{n-1}}$}, and the final $D-P-1$ dimensions are local.
Moving from full coefficient space to full grid space thus requires $D$ local spectral transforms and $P$ distributed array transpositions.
This sequence defines a total of $D + P + 1$ data layouts.

\begin{figure*}
\centering
\includegraphics[width=\textwidth]{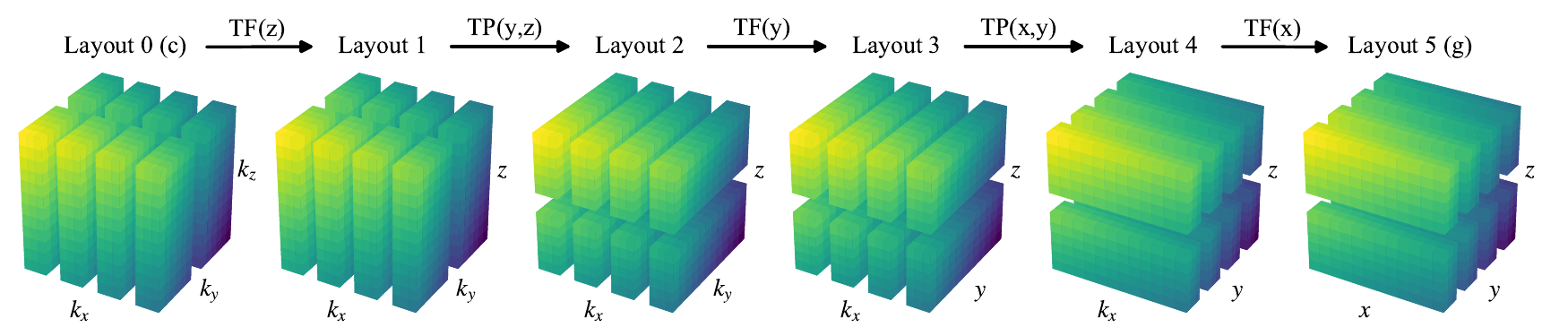}
\caption{The parallel data distribution for 3D data over a process mesh of shape $(4, 2)$.
The global data is depicted as being split into the portions that are local to each process.
The dimensions are labeled e.g.\ $k_x$ when the corresponding dimension is in coefficient space, and e.g.\ $x$ in grid space.
The transforms (TF) and transpositions (TP) stepping between layouts are indicated.}
\label{fig.layouts}
\end{figure*}

\figref{fig.layouts} shows the data distribution in each layout for 3D data distributed over a process mesh of shape $(4, 2)$.
The layout system provides a simple, well-ordered sequence of transform/distribution states that can be systematically constructed for domains and process meshes of any dimension and shape.
Conceptually, the system propagates the first local dimension down in order for each spectral transform to be performed locally.
Care must be taken to consider edge cases resulting in empty processes for certain domain and process shapes.
In particular, if $(M_i-k) B_{i,i} > N_i$ or $(M_i-k) B_{i+1,i} > N_{i+1}$ for any mesh axis $i$, then the last $k$ hyperplanes along the $i$-th axis of the mesh will be empty.
For instance, if $M_1 = 4$ and $N_1 = 9$, then the initial block size along the lowest dimension will be $B_{1,1} = 3$, and therefore processes with $m_1 = 4$ will be empty.
These cases are typically avoidable by choosing a different process mesh shape for a fixed number of processes.

For simplicity, we discussed fixed-shape global data throughout the transform process.
The implementation also handles arbitrary transform scales along each dimension, meaning $N_i = N_{c,i}$ in coefficient space, and $N_i = N_{g,i} = s_i N_{c,i}$ in grid space.
The default process mesh is one-dimensional and contains all available MPI processes.

\subsection{Transpose routines}

Consider the first transposition when moving from coefficient space to grid space, i.e.\ transposing the $P$ and $(P+1)$-th array dimensions over the $P$-th mesh axis.
This transposition does not change the data distribution over the lower mesh axes; it consists of separate all-to-all calls within each one-dimensional subset of processes defined by fixed $(m_1, ..., m_{P-1})$.

The transposition is planned by first creating separate subgroup MPI communicators consisting of each group of $M_p$ processes with the same $(m_1, ..., m_{P-1})$.
Each communicator plans for the transposition of an array with the \emph{global subgroup shape} $(B_{1,1}, ..., B_{P-1,P-1}, N_{P}, N_{P+1}, ..., N_{D})$, i.e.\ the subspace of the global data spanned by its subgroup processes.
This array is viewed as a four-dimensional array with the \emph{reduced global subgroup shape} $(B_{1,1} \times ... \times B_{P-1,P-1}, N_{P}, N_{P+1}, N_{P+2} \times ... \times N_{D})$, constructed by collapsing the pre- and post-transposition dimensions.
In this way, the general case of transposing a $D$-dimensional array distributed over a $P$-dimensional process mesh along an arbitrary mesh axis is reduced to the problem of transposing a four-dimensional array across its middle two dimensions.

When transposing the distribution along the $p$-th mesh axis between the $p$-th and $(p+1)$-th array dimensions, the global subgroup shape is given by $(n_1, ..., n_d)$ where
\begin{equation}
    n_i =
    \begin{cases}
    B_{i,i} & i < p \\
    N_i & i = p ,\, p+1 \\
    B_{i,i-1} & p+1 < i \leq P+1 \\
    N_i & i > P+1
    \end{cases}
\end{equation}
where the first $p$ array dimensions are distributed over the corresponding mesh axes, the $p$-th and $(p+1)$-th array dimensions are alternating between being local and distributed over the $p$-th mesh axis, the following $P-p$ array dimensions have already undergone a transposition and are distributed over the corresponding mesh axes less one, and the remaining array dimensions are local.
This global shape is collapsed to the reduced global subgroup shape $(G_1, G_2, G_3, G_4)$ where
\begin{equation}
    G_1 = \prod_{i=1}^{p-1} n_i, \quad
    G_2 = n_p, \quad
    G_3 = n_{p+1}, \quad
    G_4 = \prod_{i=p+2}^{D} n_i.
\end{equation}

Routines using either MPI or FFTW are available for performing the reduced data transpositions.
The MPI version begins with the local subgroup data of shape $(G_1, B_{p,p}, N_{p+1}, G_4)$ and splits this data into the blocks of shape $(G_1, B_{p,p}, B_{p+1,p}, G_4)$ to be distributed to the other processes.
These blocks are then sequentially copied into a new memory buffer so that the data for each process is contiguous.
A MPI all-to-all call is then used to redistribute the blocks from being row-local to column-local, in reference to the second and third axes of the reduced array.
Finally, the blocks are extracted from the MPI buffer to form the local subgroup data of shape $(G_1, N_p, B_{p+1,p}, G_4)$ in the subsequent layout.
The FFTW version performs a hard (memory-reordering) local transposition to rearrange the data into shape $(G_2, G_3, G_1, G_4)$, and uses FFTW's advanced distributed-transpose interface to build a plan for transposing a matrix of shape $(G_2, G_3)$ with an itemsize of $G_1 \times G_4 \times Q$, where $Q$ is the actual data itemsize.

\begin{figure}
\centering
\includegraphics[width=\linewidth]{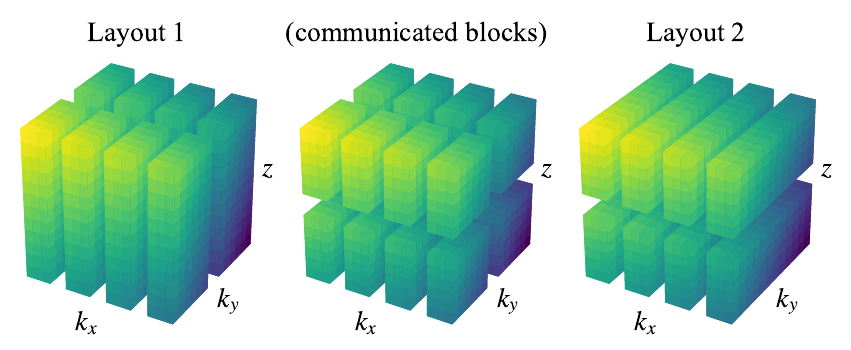}
\caption{The effective data redistribution that occurs during the distributed transposition between layouts 1 and 2 of the example shown in \figref{fig.layouts}.
This transposition is switching the second mesh axis with $M_2=2$ from distributing $k_y$ to distributing $z$.}
\label{fig.transpose_blocks}
\end{figure}

\figref{fig.transpose_blocks} shows the conceptual domain redistribution strategy for the transposition between layouts 1 and 2 of the example shown in \figref{fig.layouts}.
Both the MPI and FFTW implementations require reordering the local data in memory before communicating.
However, they provide simple and robust implementations encompassing the general transpositions required by the layout structure.
The MPI implementation serves as a low-dependency baseline, while the FFTW routines leverage FFTW's internal transpose optimization to improve performance when an MPI-linked FFTW build is available.
The FFTW planning rigor and in-place directives for the transpositions are wrapped through the Dedalus configuration interface.

These routines can also be used to group transpositions of multiple arrays simultaneously.
Transposing $S$-many arrays concatenates their local subgroup data and the reduced global subgroup shape is expanded to $(S \times G_1, G_2, G_3, G_4)$.
A plan is constructed and executed for the expanded shape.
This concatenation allows for the simultaneous transposition of multiple arrays while reducing the latency associated with initiating the transpositions.
The option to group multiple transpositions in this manner is controlled through the Dedalus configuration interface.

\subsection{Distributed data interaction}

For arbitrary transform scalings in each dimension, the layout objects contain methods providing: the global data shape, local data shape, block sizes, local data coordinates, and local data slices for \pyth{Field} objects.
These methods provide the user with the tools necessary to understand the data distribution at any stage in the transformation process.
This is useful for both analyzing distributed data and initializing distributed fields using stored global data.

The domain class contains methods for retrieving each process's local portion of the $N$-dimensional coordinate grid and spectral coefficients.
These local arrays are useful for initializing field values in either grid space or coefficient space.
Code that initializes field data using these local arrays is robust to changing parallelization scenarios, allowing scripts to be tested serially on local machines and then executed on large systems without modification.

\section{Fields}

\pyth{Field} objects represent scalar-valued fields defined over a domain.
Each field object contains a metadata dictionary specifying whether that field is constant along any axis, the scales along each axis, and any other metadata associated with specific bases (such as \pyth{'parity'} for the sine/cosine bases or \pyth{'envelope'} for Hermite and Laguerre bases).
When the transform scales are specified or changed, the field object internally allocates a buffer large enough to hold the local data in any layout for the given scales.
Each field also contains a reference to its current layout, and a data attribute viewing its memory buffer using the local data shape and type.

\subsection{Data manipulation}

The \pyth{Field} class defines a number of methods for transforming individual fields between layouts.
The most basic methods move the field towards grid or coefficient space by calling the transforms or transpositions to increment or decrement the layout by a single step.
Other methods direct the transformation to a specific layout by taking sequential steps.
These methods allow users to interact with the distributed grid data and the distributed coefficient data without needing to know the details of the distributing transform mechanism and intermediate layouts.

The \pyth{__getitem__} and \pyth{__setitem__} methods of the field class allow retrieving or setting the local field data in any layout.
Shortcuts \pyth{'c'} and \pyth{'g'} allow fast access to the full coefficient and grid data, respectively.
To complete a fully parallelized distributed transform:
\begin{python}
f = Field(domain=domain)
f['g'] = ...  # Set local grid data
f['c']  # Returns local coefficients
\end{python}
The \pyth{set_scales} method modifies the transform scales:
\begin{python}
f.set_scales(10)  # Set transform scales
f['g']  # Returns 10x spectral interpolant
\end{python}

\subsection{Field Systems}

The \pyth{FieldSystem} class groups together a set of fields.
The class provides an interface for accessing the coefficients corresponding to the same transverse mode, or pencil, of a group of fields.
A \emph{transverse mode} is a specific product of basis functions for the first $D-1$ dimensions of a domain, indexed by a multi-index of size $D-1$.
Each transverse mode has a corresponding 1D \emph{pencil} of coefficients along the last axis of a field's coefficient data.
The linear portion of a PDE that is uncoupled across tranverse dimensions splits into separate matrix systems for each transverse mode.

A \pyth{FieldSystem} of $S$ fields will build an internal buffer of size
\begin{equation}
    (B_{1,1}, ..., B_{P,P}, N_{P+1}, ..., N_{D-1}, N_D \times S).
\end{equation}
That is, the local coefficient shape with the last axis size multiplied by the number of fields.
The system methods \pyth{gather} and \pyth{scatter} copy the separate field coefficients into and out of this buffer.
Each size-$N_D \times S$ system pencil contains the corresponding field pencils, grouped along the last axis, in a contiguous block of memory for efficient access.

A \pyth{CoeffSystem} allocates and controls just the unified buffer rather than also instantiating field objects.
Coefficient systems are used as temporary arrays for all pencils, avoiding the memory overhead associated with instantiating new field objects.

\section{Operators}
\label{sec.operators}

The \pyth{Operator} classes represent mathematical operations on fields, such as arithmetic, differentiation, integration, and interpolation.
An operator instance represents a specific mathematical operation on a field or set of fields.
Operators can be composed to build complex expressions.
The operator system serves two simultaneous purposes:
\begin{enumerate}
	\item It allows the deferred and repeated evaluation of arbitrary operations.
	\item It can produce matrix forms of linear operations.
\end{enumerate}
\noindent Together, these features allow the implicit and explicit evaluation of arbitrary expressions, which is the foundation of Dedalus' ability to solve general PDEs.

\subsection{Operator classes and evaluation}

Operators accept operands (fields or operators) from the same domain, as well as other arguments such as numerical constants or strings.
Each operator class implements methods determining the metadata of the output based on the inputs; e.g.\ the output parity of a \pyth{SinCos} basis.
Operator classes also have a \pyth{check_conditions} method that checks if the operation can be executed in a given layout.
For example, spectral differentiation along some dimensions requires that dimension to be in coefficient space as well as local for derivatives that couple modes.
Finally, operators have an \pyth{operate} method which performs the operation on the local data of the inputs once they have been placed in a suitable layout.

\begin{figure}
\centering
\includegraphics[width=0.9\linewidth,trim={0cm 1cm 0cm 1cm},clip]{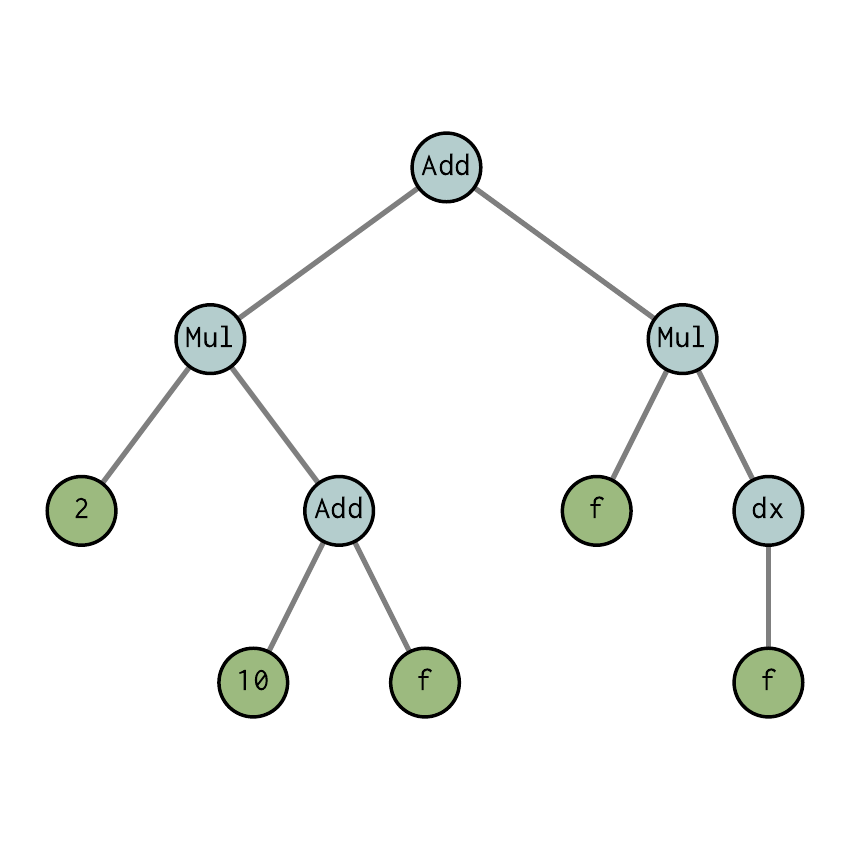}
\caption{An operator tree representing the expression \mbox{\pyth{2*(10+f) + f*dx(f)}}.}
\label{fig.operator_tree}
\end{figure}

Operators can be combined to build complex expressions.
An arbitrary expression belongs to the root operator class, with operands that belong to other operator classes, eventually with fields or input parameters forming the leaves at the end of the expression tree (see \figref{fig.operator_tree}).
The \pyth{evaluate} method computes compound operators by recursively evaluating all operands, setting the operands' transform scales to the dealias scales, transforming the operands to the proper layout, and calling the \pyth{operate} method.
Arbitrary expression trees are evaluated in a depth-first traversal.
The \pyth{evaluate} method can optionally cache its output if it may be called multiple times before the values of the leaves change.
The \pyth{attempt} method tries to evaluate a field, but will not make any layout changes while evaluating subtrees.
It therefore evaluates an expression as much as possible given the current layouts of the involved fields.
Finally, operators also implement a number of methods allowing for algebraic manipulation of expressions, described in \secref{sec.expr_manip}.

\subsection{Arithmetic operators}

The \pyth{Add}, \pyth{Multiply}, and \pyth{Power} classes implement addition, multiplication, and exponentiation respectively.
Different subclasses of these operators are invoked depending on the types of input.
A Python metaclass implements this multiple dispatch system, which examines the arguments before instantiating an operator of the proper subclass.
For example, the \pyth{AddFieldField} subclass adds two fields by adding the local data of each field.
The operation can be evaluated as long as both fields are in the same layout.
Addition between fields requires compatible metadata, e.g.\ the same \pyth{parity} or \pyth{envelope} settings.
The \pyth{AddScalarField} class likewise adds a constant to a field.
Multiplication and exponentiation of fields must occur in grid space, but otherwise have similar implementations.

The overloaded \pyth{__add__}, \pyth{__mul__}, and \pyth{__pow__} methods allow for easy arithmetic on fields and operators using Python infix operators.
For example, with a Dedalus field, \pyth{f}, the expression \pyth{f+5} will produce an \pyth{AddScalarField} instance.
We also override the \pyth{__neg__}, \pyth{__sub__}, and \pyth{__truediv__} methods for negation, subtraction, and division:
\begin{python}
-f     # Becomes Multiply(-1, f)
f - g  # Becomes Add(f, Multiply(-1, g))
f / g  # Becomes Multiply(f, Power(g, -1))
\end{python}

\subsection{Unary grid operators}

The \pyth{UnaryGridFunction} class implements common nonlinear unary functions: \pyth{np.absolute}, \pyth{np.sign}, \pyth{np.conj}, \pyth{np.exp}, \pyth{np.exp2}, \pyth{np.log}, \pyth{np.log2}, \pyth{np.log10}, \pyth{np.sqrt}, \pyth{np.square}, \pyth{np.sin}, \pyth{np.cos}, \pyth{np.tan}, \pyth{np.arcsin}, \pyth{np.arccos}, \pyth{np.arctan}, \pyth{np.sinh}, \pyth{np.cosh}, \pyth{np.tanh}, \pyth{np.arcsinh}, \pyth{np.arccosh}, and \pyth{np.arctanh}.
The operation proceeds by applying the function to the local grid space data of the operand.
The overloaded \pyth{__getattr__} method intercepts Numpy universal function calls on fields and operators and instantiates the corresponding \pyth{UnaryGridFunction}.
This allows the direct use of Numpy ufuncs to create operators on fields.
For example \pyth{np.sin(f)} on a Dedalus field \pyth{f} will return \pyth{UnaryGridFunction(np.sin, f)}.

\subsection{Linear spectral operators}

Linear operators acting on spectral coefficients are derived from the \pyth{LinearOperator} base class and the \pyth{Coupled} or \pyth{Separable} base classes if they do or do not couple different spectral modes, respectively.
These operators are instantiated by specifying the axis along which the operator is to be applied, which is used to dispatch the instantiation to a subclass implementing the operator for the corresponding basis.
These operators implement a \pyth{matrix_form} method which produces the matrix defining the action of the operator on the spectral basis functions.
For a basis ${\phi_n}$ and an operator $A$, the matrix form of $A$ is
\begin{equation}
    A_{ij} = \langle \phi_i | A \phi_j \rangle.
\end{equation}
For separable operators, this matrix is diagonal by definition, and represented with a one-dimensional array.
For coupled operators, this matrix is returned as a Scipy sparse matrix.

In general, linear operators require their corresponding axis to be in coefficient space to be evaluated.
Coupled operators further require that the corresponding axis is local.
The local data of the operand is contracted with the matrix form of the operator along this axis to produce the local output data.
Operators may override this process by implementing an \pyth{explicit_form} method if a more efficient or stable algorithm exists for forward-applying the operator.
For example, forward Chebyshev differentiation uses a recursion rather than a matrix multiplication.

\subsubsection{Differentiation}

\pyth{Differentiate} classes are implemented for each basis.
Differentiation of the \pyth{Fourier} and \pyth{SinCos} bases is a separable operator and therefore a diagonal matrix.
Differentiation of the polynomial bases is a coupled operator.
Differentiation of the Hermite polynomials and enveloped functions are both naturally banded and therefore require no conversion into a different test basis to retain sparsity.
Differentiation of the Chebyshev, Legendre, and Laguerre bases have dense upper triangular matrices, but these are never used when solving equations.
Instead, we always convert the differential equations into a test basis with banded differentiation matrices (see \secref{sec.sparse_diff_theory}).
These conversions are applied as left-preconditioners for the equation matrices.
\appref{sec.diff_conv_matrices} shows the full differentiation and conversion matrices.
For the Chebyshev, Legendre, and Laguerre bases, forward differentiation uses $\bigO(N_c)$ recurrence relations rather than dense matrices.
Template matrices are rescaled according to the affine map between the native and problem coordinates.
The differentiation matrix for the compound basis is the block-diagonal combination of the subbasis differentiation matrices.

For each basis type, a differentiation subclass is referenced from the basis-class \pyth{Differentiate} method.
These methods are aliased as e.g.\ \pyth{dx} for a basis with name \pyth{'x'} during equation parsing.
Additionally, a factory function called \pyth{differentiate} (aliased as \pyth{d}) provides an easy interface for constructing higher-order and mixed derivatives using the basis names, by composing the appropriate differentiation methods:
\begin{python}
dx(f)  #  Becomes xbasis.Differentiate(f)
d(f, x=2, y=2)  # Becomes dx(dx(dy(dy(f))))
\end{python}
The differentiation subclasses also examine the \pyth{'constant'} metadata of their operand, and return \pyth{0} instead of instantiating an operator if the operand is constant along the direction of differentiation.

\subsubsection{Integration}

Integration is a functional that returns a constant for any input basis series.
Integration operators therefore set the \pyth{'constant'} metadata of their corresponding axes to \pyth{True}.
The operator matrices are nonzero only in the first row (called the operator vector).

The native integration vectors for each basis are rescaled by the stretching of the affine map between the native and problem coordinates.
Integration for the compound basis concatenates each of the sub-basis integration vectors and places the result in the rows corresponding to the constant terms in each sub-basis.

For each basis type, an integration subclass is referenced from the basis-class \pyth{Integrate} method.
Additionally, a factory function called \pyth{integrate} (aliased as \pyth{integ}) provides an easy interface for integrating along multiple axes, listed by name, by composing the appropriate integration methods:
\begin{python}
integ(f, 'x', 'y')
# xbasis.Integrate(ybasis.Integrate(f))
\end{python}
If no bases are listed, the field will be integrated over all of its bases.
If the operand's metadata indicates that it is constant along the integration axis, the product of the constant and the interval length will be returned.

The \pyth{antidifferentiate} method of the \pyth{Field} class implements indefinite integration.
This method internally constructs and solves a simple linear boundary value problem and returns a new \pyth{Field} satisfying a user-specified boundary condition, fixing the constant of integration.

\subsubsection{Interpolation}

Interpolation operators are instantiated with an operand and the interpolation position in problem coordinates.
The operator matrices are again nonzero except in the first row (called the operator vector) and depend on the interpolation position.
The specified interpolation positions are converted to the native basis coordinates via the basis affine map.
The strings \pyth{'left'}, \pyth{'center'}, and \pyth{'right'} are also acceptable inputs indicating the left endpoint, center point, and right endpoint of the problem interval.
Specifying positions in this manner avoids potential floating-point errors when evaluating the affine map at the endpoints.

The interpolation classes construct interpolation vectors consisting of the pointwise evaluation of the respective basis functions.
Interpolation for the compound basis takes the interpolation vector of the sub-basis containing the interpolation position and places the result in the rows corresponding to the constant terms in each sub-basis.
If the interpolation position is at the interface between two sub-basis, the first sub-basis is used to break the degeneracy.

For each basis type, an interpolation subclass is referenced from the \pyth{Interpolate} basis-class method.
Additionally, a factory function called \pyth{interpolate} (aliased as \pyth{interp}) provides an easy interface for interpolating along multiple axes, specified using keyword arguments, by composing the appropriate integration methods:
\begin{python}
interp(f, x=0.5, y=1)
# xbasis.Interpolate(
#    ybasis.Interpolate(f, 1), 0.5)
\end{python}
If the operand's metadata indicates that it is constant along the interpolation axis, instantiation will be skipped and the operand itself will be returned.

\subsubsection{Hilbert transforms}

The Hilbert transform of a function $f(x)$ is the principal-value convolution with $(\pi x)^{-1}$:
\begin{equation}
    H(f)(x) = \frac{1}{\pi} \mathrm{p.v.}\int_{-\infty}^{\infty} \frac{f(x')}{x - x'} \diff{x'}.
\end{equation}
The Hilbert transform has a particularly simple action on sinusoids,
\begin{equation}
    H(\exp(ikx))(x) = - i \sgn(k) \exp(ikx)
\end{equation}
The actions on cosine/sine functions result from taking the real/imaginary parts.
The Hilbert transform is implemented as a separable operator for the \pyth{Fourier} and \pyth{SinCos} bases and referenced from the \pyth{HilbertTransform} basis-class methods.
These methods are aliased as e.g.\ \pyth{Hx} for a basis with name \pyth{'x'}.
Additionally, a factory function called \pyth{hilberttransform} (aliased as \pyth{H}) provides an easy interface for constructing higher-order and mixed Hilbert transforms using the basis names, similar to the \pyth{differentiate} factory function.
The Hilbert transform subclasses also examine the \pyth{'constant'} metadata of their operand, and return \pyth{0} instead of instantiating an operator if the operand is constant along axis to be transformed.

\subsection{User-specified functions}
\label{sec.user-spec-func}
The \pyth{GeneralFunction} class wraps and applies general Python functions to field data in any specified layout.
For example, a user-defined Python function \pyth{func(A, B)} which accepts and operates on grid-space data arrays, can be wrapped into a Dedalus \pyth{Operator} for deferred evaluation using Dedalus fields \pyth{fA} and \pyth{fB} as
\begin{python}
GeneralFunction(func, args=(fA, fB), layout='g')
\end{python}
\secref{sec.stokes_example} gives a detailed example.

\subsection{Manipulating expressions}
\label{sec.expr_manip}

The operator classes also implement a number of methods allowing the algebraic manipulation of operator expressions, i.e.\ a simple computer-algebra system.
\secref{sec.dedalus_problems} describes how this enables the construction of solvers for general partial differential equations.

These methods include:
\begin{itemize}
    \item
    \pyth{atoms}: recursively constructs the set of leaves of an expression matching a specified type.
    \item
    \pyth{has}: recursively determines whether an expression contains any specified operand or operator type.
    \item
    \pyth{expand}: recursively distributes multiplication and linear operators over sums of operands containing any specified operand or operator type.  It also distributes derivatives of products containing any specified operand or operator type using the product rule.
    \item
    \pyth{canonical_linear_form}: first determines if all the terms in an expression are linear functions of a specified set of operands, and raises an error otherwise.
    In the case of nested multiplications, it rearranges the terms so that the highest level multiplication directly contains the operand from the specified set.
    \item
    \pyth{split}: additively splits an expression into a set of terms containing specified operands and operator types, and a set of terms not containing any of them.
    \item
    \pyth{replace}: performs a depth-first search of an expression, replacing any instances of a specified operand or operator type with a specified replacement.
    \item
    \pyth{order}: recursively determines the compositional order of a specified operator type within an expression.
    \item
    \pyth{sym_diff}: produces a new expression containing the symbolic derivative with respect to a specified variable.
    The derivative is computed recursively via the chain rule.
    \item
    \pyth{as_ncc_operator}: constructs the NCC multiplication matrix associated with multiplication by the expression.
    It requires that the corresponding domain only have a single polynomial basis, and that this basis forms the last axis of the domain.
    It further requires that the expression is constant along all other (``transverse'') axes, so that multiplication by the operand does not couple the transverse modes.
	The method first evaluates the expression, then builds the NCC matrix with the resulting coefficients.
	The method allows the NCC expansion to be truncated at a maximum number of modes ($N_m < N_c$) and for terms to be excluded when the coefficient amplitudes are below some threshold ($|f_n| < \delta$) so that the matrix is sparse for well-resolved functions (see \secref{sec.sparse_ncc_theory}).
    \item
    \pyth{operator_dict}: constructs a dictionary representing an expression as a set of matrices acting on the specified pencils of a specified set of variable.
    This method requires that the expression be linear in the specified variables and contains no operators coupling any dimensions besides the last.

    The dictionary is constructed recursively, with each coupled linear operator applying its matrix form to its operand matrices, and each transverse linear operator multiplying its operand matrices by the proper element of its vector form.
    Addition operators sum the matrices produced by their operands.
    Multiplication operators build the matrices for the operand containing the specified variables.
    They then multiply these matrices by the NCC matrix form of the other operand.
\end{itemize}

\subsection{Evaluators}

An \pyth{Evaluator} object attempts to simultaneously evaluate multiple operator expressions, or tasks, as efficiently as possible, i.e.\ with the least number of spectral transforms and distributed transpositions.
Tasks are organized into \pyth{Handler} objects, each with a criterion for when to evaluate the handler.
Handlers can be evaluated on a specified cadence in terms of simulation iterations, simulation time, or real-world time (wall time) since the start of the simulation.
Handlers from the \pyth{SystemHandler} class organize their outputs into a \pyth{FieldSystem}, while handlers from the \pyth{FileHandler} class save their outputs to disk in HDF5 files via the h5py package (\secref{sec.analysis}).
The \pyth{add_task} method adds tasks to a handler and accepts operator expressions or strings (which are parsed into operator expressions using a specified namespace).

When triggered, the evaluator examines the attached handlers and builds a list of the tasks from each handler scheduled for evaluation.
The evaluator uses the \pyth{attempt} methods to evaluate the tasks as far as possible without triggering any transforms or transpositions.
If the tasks have not all completed, the evaluator merges the remaining atoms from the remaining tasks, and moves them all to full coefficient space, and reattempts evaluation.
If the tasks are still incomplete, the evaluator again merges the remaining atoms from the remaining tasks, moves them forward one layout, and reattempts evaluation.
This process repeats, with the evaluator simultaneously stepping the remaining atoms back and forth through all the layouts until all of the tasks have been fully evaluated.
Finally, the \pyth{process} method on each of the scheduled handlers is executed.

This process is more efficient than sequentially evaluating each expression.
By attempting all tasks before changing layouts, it makes sure that no transforms or transpositions are triggered when any operators are able to be evaluated.
Additionally, it groups together all the fields that need to be moved between layouts so that grouped transforms and transpositions can be performed to minimize overhead and latency.

\section{Problems}
\label{sec.dedalus_problems}

\pyth{Problem} classes construct and represent systems of PDEs.
Separate classes manage linear boundary value problems (\pyth{LBVP}), nonlinear boundary value problems (\pyth{NLBVP}), eigenvalue problems (\pyth{EVP}), and initial value problems (\pyth{IVP}).
After creating a problem, the equations and boundary conditions are entered in plain text, with linear terms on the LHS and nonlinear terms on the RHS.
The LHS is parsed into a sparse matrix formulation, while the RHS is parsed into an operator tree to be evaluated explicitly.

\subsection{Problem creation}

Each problem class is instantiated with a domain and a list of variable names.
Domains may have a maximum of one polynomial basis, which must correspond to the last axis.
The linear portion of the equations must be no higher than first-order in time and coupled derivatives.
Auxiliary variables must be added to render the system first-order.
Optionally, an amplitude threshold and a cutoff mode number can be specified for truncating the spectral expansion of non-constant coefficients on the LHS.
For eigenvalue problems, the eigenvalue name must also be specified; it cannot be \pyth{'lambda'} since this is a Python reserved word.
For initial value problems, the temporal variable name can optionally be specified, but defaults to \pyth{'t'}.

For example, to create an initial value problem for an equation involving the variables $u$ and $v$, we would write
\begin{python}
problem = de.IVP(domain, variables=['u', 'v'])
\end{python}

\subsection{Variable metadata}

Metadata for the problem variables is specified by indexing the \pyth{problem.meta} attribute by variable name, axis, and then property.

The most common metadata to set is the \pyth{'constant'} flag for any dimension, the \pyth{'parity'} of all variables for each \pyth{SinCos} basis, the \pyth{'envelope'} flag for the \pyth{Hermite} and \pyth{Laguerre} bases, and the \pyth{'dirichlet'} flag for recombining the \pyth{Chebyshev}, \pyth{Legendre}, and \pyth{Laguerre} bases (enabled by default).
Default metadata values are specified in the basis definitions.

For example, we can set the parity of variables in our problem along the $x$ axis with
\begin{python}
problem.meta['u']['x']['parity'] = 1  # cosine
problem.meta['v']['x']['parity'] = -1 # sine
\end{python}

\subsection{Parameters and non-constant coefficients}

Before adding the equations, any parameters (fields or scalars used in the equations besides the problem variables) must be added to the parsing namespace through the \pyth{problem.parameters} dictionary.
Scalar parameters are entered by value.
Non-constant coefficients (NCCs) are entered as fields with the desired data.
NCCs on the LHS can only couple polynomial dimensions; an error will be raised if the \pyth{`constant'} metadata is not set to \pyth{True} for all separable axes.

For example, we would enter scalar and NCC parameters for a 3D problem on a double-Fourier ($x$, $y$) and Chebyshev ($z$) domain as:
\begin{python}
# Scalar parameter
problem.parameters['A'] = 1e-4

# NCC parameter
z = domain.grid(2)
ncc = domain.new_field()
ncc.meta['x', 'y']['constant'] = True
ncc['g'] = z**2
problem.parameters['B'] = ncc
\end{python}

\subsection{Substitutions}

One of the most powerful features of Dedalus is the ability to define substitutions which act as string-replacement rules to be applied during the equation parsing process.
Substitutions can be used to provide short aliases to quantities computed from the problem variables and to define functions similar to Python lambda functions, but with normal mathematical-function syntax.

For example, several substitutions that might be useful in a hydrodynamical simulation are:
\begin{python}
# Substitution defining the kinetic energy
# density for a 3D fluid simulation with
# density rho and velocity (u,v,w).
problem.substitutions['KE_density'] = \
    "rho * (u*u + v*v + w*w) / 2"

# Substitution defining the Cartesian Laplacian
# of a field. Here A and Az are dummy variables
# that would be replaced by simulation variables
# in the equations. Note the system is written
# in first-order form along the z dimension.
problem.substitutions['L(A,Az)'] = \
    "dx(dx(A)) + dy(dy(A)) + dz(Az)"
\end{python}
Substitutions of the first type are created by parsing their definitions in the problem namespace, and aliasing the result to the substitution name.
Substitutions of the second type are turned into Python lambda functions producing their specified form in the problem namespace.
Substitutions are composable, and form a powerful tool for simplifying the entry of complex equation sets.

\subsection{Equation parsing}

Equations and boundary conditions are entered in plain text using the \pyth{add_equation} and \pyth{add_bc} methods.
Optionally, these methods accept a \pyth{condition} keyword, which is a string specifying which transverse modes that equation applies to.
This is necessary to close certain equation sets where, for example, the equations become degenerate for the transverse-mean mode and/or certain variables require gauge conditions.

First, the string-form equations are split into LHS and RHS strings which are evaluated over the problem namespace to build LHS and RHS operator expressions.
The problem namespace consists of:
\begin{itemize}
    \item The variables, parameters and substitutions defined in the problem.
    \item The axis names representing the individual basis grids.
    \item The derivative, integration, and interpolation operators for each basis.
    \item Time and temporal derivatives for the IVP (defaulting to \pyth{'t'} and \pyth{'dt'}).
    \item The eigenvalue name for the EVP.
    \item The universal functions wrapped through the \pyth{UnaryGridFunction} class.
\end{itemize}

A number of conditions confirming the validity of the LHS and RHS expressions are then checked.
For all problem types, the LHS expression and RHS must have compatible metadata (e.g.\ parities).
The LHS expression must be nonzero and linear in the problem variables.
The LHS must also be first-order in coupled derivatives.
The expressions entered as boundary conditions must be constant along the last axis.

For the individual problem classes, the following additional restrictions and manipulations are applied to the LHS and RHS expressions:

\subsubsection{Linear boundary value problems}

The linear boundary value problem additionally requires that the RHS is independent of the problem variables.
This allows for linear problems with inhomogeneous terms on the RHS.
Since the LHS terms are linear in the problem variables, this symbolically corresponds to systems of equations of the form
\begin{equation}
    \mathcal{L} \cdot \mathcal{X} = \mathcal{F}
\end{equation}
where $\mathcal{X}$ is the state vector of variable fields, and $\mathcal{L}$ is a matrix of operators.
The LHS expressions are expanded and transformed into canonical linear form before being stored by the problem instance.

\subsubsection{Nonlinear boundary value problems}

Nonlinear boundary value problems are systems of the form
\begin{equation}
    \mathcal{L} \cdot \mathcal{X} = \mathcal{F(X)}
\end{equation}
The RHS can be any nonlinear function of the problem variables.
In addition to the $\mathcal{L} \cdot \mathcal{X}$ and $\mathcal{F}$ expressions, the problem constructs the Frechet differential of the RHS with respect to the problem variables:
\begin{equation}
    \mathcal{F}_\mathcal{X} \cdot \Delta \mathcal{X} = \pdiff{\epsilon} \mathcal{F}(\mathcal{X} + \epsilon \Delta \mathcal{X}) |_{\epsilon=0}.
\end{equation}
This linear operator indicates the sensitivity, or directional functional derivative, of $\mathcal{F}$ with respect to changes in $\mathcal{X}$ along $\Delta \mathcal{X}$.
It is constructed symbolically using the operator methods described in \secref{sec.expr_manip} roughly as
\begin{python}
dF = 0
for var, pert in zip(vars, perts):
    dFi = F.replace(var, var + ep*pert)
    dFi = dFi.sym_diff(ep)
    dFi = dFi.replace(ep, 0)
    dF += dFi
\end{python}

In general, the Frechet derivative of an expression will contain non-constant coefficients involving the problem variables $\mathcal{X}$, which would generally couple horizontal modes.
Therefore, Dedalus only supports 1D NLBVPs.
The LHS and Frechet differential expressions are expanded and transformed into canonical linear form before being stored by the problem instance.

\subsubsection{Eigenvalue problems}

The eigenvalue problem requires that the RHS is zero, and that the LHS terms must be linear in or independent of the eigenvalue, which we refer to as $\sigma$.
This corresponds to systems of equations of the form
\begin{equation}
    \sigma \mathcal{M} \cdot \mathcal{X} + \mathcal{L} \cdot \mathcal{X} = 0
\end{equation}
which are generalized linear eigenvalue problems.
The $\mathcal{M} \cdot \mathcal{X}$ and $\mathcal{L} \cdot \mathcal{X}$ expressions are extracted by splitting the LHS expression on the presence of the eigenvalue variable, before replacing it with $1$.
These expressions are expanded and transformed into canonical linear form before being stored by the problem instance.

\subsubsection{Initial value problems}

The initial value problem requires that the LHS coefficients are time independent, the LHS is first-order in time derivatives, and the RHS has no time derivatives.
This corresponds to systems of the form
\begin{equation}
    \mathcal{M} \cdot \pdiff{t} \mathcal{X} + \mathcal{L} \cdot \mathcal{X} = \mathcal{F}(\mathcal{X},t).
\end{equation}
The $\mathcal{M} \cdot \mathcal{X}$ and $\mathcal{L} \cdot \mathcal{X}$ expressions are extracted by splitting the LHS expression on the presence of the time derivative dummy operator, before replacing it with the identity operator.
These expressions are expanded and transformed into canonical linear form before being stored by the problem instance.

\section{Solvers}
\label{sec.solvers}

Each problem type has a corresponding solver type which builds the spectral matrices for the problem equations, implements methods for computing the solution to the equations, and stores the solution state as a \pyth{FieldSystem}.

\subsection{Matrix construction}
\label{sec.matrix_construction}

The problem classes begin by building the operator matrices for the LHS expression groups ($\mathcal{M} \cdot \mathcal{X}$, $\mathcal{L} \cdot \mathcal{X}$, and/or $\mathcal{F}_\mathcal{X} \cdot \Delta \mathcal{X}$).
The matrices are constructed by first taking the set of equations and boundary conditions that apply to each pencil and calling the \pyth{operator_dict} method on each expression to build the matrices acting on the corresponding coefficients of the problem variables.
For each pencil's matrix to be solvable, the number of applicable equations must equal the number of variables in the problem, $S$.
For a given pencil, we refer to the operator matrix from the $i$-th equation acting on the $j$-th variable as e.g.\ $L^{i,j}$.
Each of these matrices is processed as follows:
\begin{itemize}
	\item
	If the $i$-th equation is constant along the coupled direction, then all rows except the first of each $L^{i,j}$ are dropped.
    \item
    If the $i$-th equation contains a coupled derivative on the LHS, each $L^{i,j}$ is left-multiplied by the basis preconditioner matrix, which renders the derivative matrix banded (e.g.\ the T-to-U conversion for Chebyshev bases; see \secref{sec.sparse_diff_theory}).
    \item
    By default, if the $i$-th equation contains a coupled derivative on the LHS, the last row of each $L^{i,j}$ is dropped and replaced with one of the boundary conditions.
    The keyword option \pyth{tau} in the \pyth{problem.add_equation} method overrides this default behavior and explicitly forces or prevents the solver from replacing the last row in that equation.
    This implements boundary conditions using the tau method (see \secref{sec.spectral_pde_methods}).
    For the system to be solvable, it typically requires the same number of boundary conditions as coupled differential equations.
    The \pyth{tau} keyword allows for non-standard cases, such as problems with singular end points.
    \item
    If the last basis is compound, the rows corresponding to the final coefficient of each sub-basis, except for the last, are dropped and replaced with internal boundary conditions matching the sub-basis values at each internal interface for each variable.
    This enforces continuity of all variables across the subsegments.
    \item
    If the $j$-th variable has been marked as constant along the coupled direction, then all columns except the first of each $L^{i,j}$ are dropped.
    \item
    If the $j$-th variable has been marked for Dirichlet preconditioning (enabled by default), then each $L^{i,j}$ is right-multiplied by the Dirichlet conversion matrix.
    This has the effect of rearranging the columns so that the matrix acts on the coefficients of the Dirichlet expansion of the corresponding variable, rendering all Dirichlet boundary conditions banded (see \secref{sec.sparse_bc_theory}).
\end{itemize}

Finally, the processed operator matrices are joined to produce the full preconditioned pencil matrix $\tilde{L}$.
The matrices are interleaved so that the columns and rows are grouped by mode rather than by field (see \secref{sec.sparse_system_theory}).
The bandwidth of the pencil matrix then becomes $S$ times the maximum bandwidth of any of the individual sub-blocks, which is roughly set by the bandwidth of the non-constant coefficient expansions.
Interleaved-block matrices $P^L_p$ and $P^R$ combining all the left and right preconditioners, projections, and reorderings are created and stored, since they will need to be applied to the RHS and solution vectors, respectively, when solving the matrix system.

The pencil matrices are stored as Scipy sparse matrices.
The matrices produced for each of the LHS expression groups ($\mathcal{M} \cdot \mathcal{X}$, $\mathcal{L} \cdot \mathcal{X}$, and/or $\mathcal{F}_\mathcal{X} \cdot \Delta \mathcal{X}$) are expanded to occupy the union of their sparsity patterns so that they can be added efficiently.

\subsection{Linear boundary value solver}

The linear boundary value solver is instantiated from a linear boundary value problem.
It first constructs the matrices $\tilde{L}_p = P^L_p L_p P^R$ for each local pencil $p$ from the stored LHS expression group $\mathcal{L} \cdot \mathcal{X}$.
Here $P^L_p$ and $P^R$ explicitly indicate that the expression matrices have been preconditioned from the left and the right, and the $p$ subscripts indicate that the left preconditioner and expression matrices vary by pencil.
The solver then constructs a system handler for evaluating the RHS equation and boundary condition expressions ($\mathcal{F}$).

The solver class contains a \pyth{solve} method, which first evaluates the RHS handler for $\mathcal{F}$.
At this point, the linear boundary value problem is fully discretized, and conceptually consists of solving an independent matrix problem for each pencil given by
\begin{equation}
    L_p X_p = F_p
\end{equation}
The equivalent preconditioned system is given by
\begin{equation}
    \underbrace{P^L_p L_p P^R}_{\textstyle \tilde{L}_p} \underbrace{(P^R)^{-1} X_p}_{\textstyle \tilde{X}_p} = \underbrace{P^L_p F_p}_{\textstyle \tilde{F}_p}
\end{equation}
For each pencil, this system is solved in the following manner:
\begin{itemize}
    \item The RHS vector $F_p$ is constructed by taking the pencil data from the RHS handler.
    \item The pencil's left-preconditioner is applied to $F_P$ to produce $\tilde{F}_p$.
    \item Since $\tilde{L}_p$ is sparse and banded, it can be efficiently solved against $\tilde{F}_p$ to produce $\tilde{X}_p$.
    \item The state-vector pencil is recovered from $\tilde{X}_p$ by reapplying the right-preconditioner as
    \begin{equation}
        P^R \tilde{X}_p = P^R (P^R)^{-1} X_p = X_p
    \end{equation}
    and the result is assigned to the state-vector \pyth{FieldSystem}.
\end{itemize}

After the RHS is evaluated, this process is trivially parallelized over pencils, with each process performing a series of local sparse matrix solves for its local pencils.
The sparsity and bandedness of the matrix $\tilde{L}_p$ makes the linear solve an efficient process, executing in $\bigO(N^c_D)$ time.
Finally, although $(P^R)^{-1}$ is a dense matrix, it never needs to be constructed, as reapplying the sparse $P^R$ matrix to the output of the linear solve reverses the implicit preconditioning of the unknowns.

\subsection{Nonlinear boundary value solver}

The nonlinear boundary value solver is instantiated from the nonlinear boundary value problem.
It constructs the matrices $\tilde{L}_p$ for each pencil and handlers for evaluating the expressions $\mathcal{F}$ and $\mathcal{L} \cdot \mathcal{X}$.

The solver class contains a \pyth{newton_iteration} method, which performs a single iteration of Newton's method to move the state vector towards the nonlinear solution.
Conceptually, the Newton step iteratively approaches the solution of the nonlinear problem
\begin{equation}
    L_p X_p = F(X)_p
\end{equation}
by solving for the update $\delta X^n$ to the state vector that will cause the future state vector $X^{n+1} = X^n + \delta X^n$ to solve the NLBVP when linearized around the current iteration $X^n$:
\begin{align}
    L_p X^{n+1}_p &= F(X^{n+1})_p \\
    L_p (X^n_p + \delta X^n_p) &= F(X^n + \delta X^n)_p \\
    &\approx F(X)_p + F_X \delta X_p \\
    \implies \underbrace{(L_p - F_X)}_{\textstyle A_p} \delta X_p &\approx \underbrace{F(X)_p - L_p X_p}_{\textstyle B_p}
\end{align}
The Newton iteration begins by evaluating the RHS handlers for $\mathcal{F}$ and $\mathcal{L} \cdot \mathcal{X}$ and building the matrices $\tilde{F}_p$, which discretize the Frechet-derivative of $\mathcal{F}$ using the current state vector.
For each pencil, the update is then determined in the following manner:
\begin{itemize}
    \item The RHS vector $B_p$ is constructed by combining the pencil data from the RHS handlers.
    \item The left-preconditioner is applied to produce $\tilde{B}_p$.
    \item The LHS matrices are combined to produce $\tilde{A}_p$, which is solved against $\tilde{B}_p$ to produce $\delta \tilde{X}_p$.
    \item The right-preconditioner is applied to recover $\delta X_p$.
    \item The state vector is updated as $X_p \rightarrow X_p + \delta X_p$.
\end{itemize}

We note that the sparse matrix being solved changes at each iteration, since it depends on the evaluation of the Frechet derivative at the current state vector.
The magnitude of the perturbations can be monitored to determine when the solver has converged.
Convergence can depend sensitively on the initial values of the state vector, but the iterations converge rapidly (quadratically) for sufficiently good starting positions.
The initial conditions are set by modifying the fields in the \pyth{solver.state} \pyth{FieldSystem}.

\subsection{Eigenvalue solver}

The eigenvalue solver is instantiated from the eigenvalue problem.
It constructs the $\tilde{M}_p$ and $\tilde{L}_p$ matrices and solves the eigenvalue problem for a single pencil at a time, storing the resulting eigenvalues and eigenvectors.
The class contains a \pyth{set_state} method which will set the solver's state vector to the specified eigenmode for visualization or further computation.

The solver class contains two methods for solving the generalized eigenvalue problem for a specified pencil, which conceptually takes the form
\begin{equation}
    \sigma M_p X_p + L_p X_p = 0.
\end{equation}

\subsubsection{Dense solver}

The first is the \pyth{solve_dense} method, which converts the LHS matrices to dense arrays and uses the \pyth{scipy.linalg.eig} routine to directly solve the full generalized eigenvalue problem.
This has the advantage of solving for all of the $S N^c_D$ eigenmodes of the discretized system.
However, the computational cost scales as $\bigO((S N^c_D)^3)$, which becomes prohibitive at large resolutions.

\subsubsection{Sparse solver}

The second is the \pyth{solve_sparse} method, which solves for a subset of the eigenmodes near a specified target eigenvalue $\sigma_T$.
The generalized problem for the preconditioned matrices is first rearranged as a regular eigenvalue problem using a shift and inversion:
\begin{equation}
    \underbrace{(\tilde{L}_p + \sigma_T \tilde{M}_p)^{-1}  \tilde{M}_p}_{\tilde{A}_p} \tilde{X}_p = - \frac{1}{(\sigma - \sigma_T)} \tilde{X}_p = \lambda_p  \tilde{X}_p
\end{equation}

A Scipy sparse linear operator is constructed to represent the left-side operator $\tilde{A}_p$.
This is applied to a vector by first applying $\tilde{M}_p$, and then solving the result against $(\tilde{L}_p + \sigma_T \tilde{M}_p)$.
This generalized linear operator is then passed to the \pyth{scipy.sparse.linalg.eig} routine, which uses the implicitly-restarted Arnoldi method in ARPACK to iteratively compute a specified number of eigenmodes with the largest magnitude $\lambda$.
The right-preconditioner is applied to the resulting eigenmodes to recover $X_p$, and the computed values of $\lambda$ are inverted and shifted to recover the corresponding $\sigma$.

This shift-and-invert formulation allows using sparse regular eigenvalue solvers for the generalized eigenvalue problem, with the requirement that $(\tilde{L}_p + \sigma_T \tilde{M}_p)$ is full rank.

\subsection{Initial value solver}

The initial value solver is instantiated from the initial value problem with one of the timestepping classes as an argument, defining the integrator to be used to step the problem forward in time.
The solver constructs the $\tilde{M}_p$ and $\tilde{L}_p$ matrices for each pencil and a handler for evaluating the RHS expressions $\mathcal{F}$.

Conceptually, the discretized problem takes the form
\begin{equation}\label{eq.ivp}
    M_p \pdiff{t} X_p + L_p X_p = F(X,t)_p
\end{equation}
where the systems for different pencils are only coupled through the RHS terms.
In general, $M_p$ may not be a full-rank matrix, due to the presence of constraint equations and boundary conditions.
This system is integrated using mixed implicit-explicit schemes, where the LHS terms are integrated implicitly, and the RHS terms are integrated explicitly.
The timestepping loop is written by the user, allowing for detailed control of and interaction with the model as timestepping progresses.

\subsubsection{Initial conditions}

The solver's initial state must be set before beginning a simulation.
The solver state is stored in the \pyth{solver.state} field system, and initial conditions are set by directly modifying the variables in this system before beginning integration, e.g.:
\begin{python}
# Set the initial u field
x = domain.grid(0)
u = solver.state['u']
u['g'] = np.cos(x)
\end{python}
When possible, it is best to begin a simulation with \emph{consistent} initial conditions that satisfy the constraint equations and boundary conditions; see \secref{sec.qg}.
Initial conditions that are inconsistent may introduce persistent errors or stability problems with some timestepping schemes.

Initial conditions can also be loaded from the analysis files produced by Dedalus (\secref{sec.analysis}) via the \pyth{solver.load_state} method.
This is particularly useful for restarting simulations from a checkpoint saved by a previous simulation.

\subsubsection{Time evolution}

The \pyth{step} method of the initial value solver advances the state by one timestep, producing $X^{n+1}$ from $X^n$, where the superscripts denote the temporal iteration of the state vector.
The method accepts the timestep \pyth{dt} as an argument.
The method then gathers the state system, calls the specified integration routine to update the system, and scatters the updated state system back to the field objects.
In general, the integration step will evaluate the RHS handler to perform the temporal integration and will simultaneously evaluate any other scheduled handlers (e.g.\ analysis tasks) attached to the evaluator.

\subsubsection{Timestep determination}

The implemented timestepping schemes accommodate changing the timestep between iterations during a simulation.
While the user can implement any desired algorithm for determining the timestep, the \pyth{CFL} class in the \pyth{dedalus.extras.flow_tools} module can help determine what timestep might adequately resolve physical timescales in the evolving solution.
The \pyth{add_frequencies} and \pyth{add_velocities} methods allow users to enter expressions corresponding to state-dependent frequencies and velocities of processes in their simulation, using the same string-based parsing system that is used to enter equations.
CFL frequencies are derived from the entered velocities by dividing their values by the grid spacing.
Internally, the CFL class builds an auxiliary handler to evaluate these frequencies at a specified cadence, and as the simulation runs the suggested timestep is determined via the \pyth{compute_dt} method as follows:
\begin{itemize}
    \item At each point on the grid, all of the specified frequencies are added.
    \item The maximum total frequency from the entire grid is taken and inverted to determine the CFL timestep.
    \item This timestep is then multiplied by a \emph{safety factor} (typically $0.1-0.5$), specified at the CFL instantiation with the \pyth{safety} keyword.
    \item The resulting timestep is then bounded to lie with absolute levels set by the \pyth{min_dt} and \pyth{max_dt} keywords, and a within relative factors of the previous timestep set by the \pyth{min_change} and \pyth{max_change} keywords.
    \item If the fractional change from the previous timestep to the newly determined timestep is smaller than the \pyth{threshold} parameter, the previous timestep is returned.
    Otherwise, the newly determined timestep is returned.
\end{itemize}
The absolute limits can be useful to prevent the timestep from vastly overstepping relevant dynamics or grinding to a halt due to a spurious feature of the solution.
The relative limits help prevent ill-conditioning that may occur for some schemes when the timestep varies too suddenly.
The thresholding option allows the timestep to be frequently reevaluated but avoids modifying it by inconsequential amounts.
This can have significant performance advantages since factorizations of the IVP matrices are stored and reused when the timestep remains the same between iterations.

\subsubsection{Termination}

To help determine when a simulations should terminate, the initial value solver implements the \pyth{proceed} property, which determines whether any of the following three criteria apply:
\begin{itemize}
    \item The simulation time has exceeded the value assigned to the \pyth{solver.stop_sim_time} attribute.
    \item The wall time (in seconds) since the solver was instantiated has exceeded the value assigned to the \pyth{solver.stop_wall_time} attribute.
    \item The iteration count has exceeded the value assigned to the \pyth{solver.stop_iteration} attribute.
\end{itemize}
The wall-time stop is particularly useful for stopping simulations before hard time-limits on cluster job submissions have been reached, allowing for clean termination and potential post-processing of the data before a job is terminated by the system.

A simple timestepping loop in an IVP can take the form:
\begin{python}
while solver.proceed:
    dt = CFL.compute_dt()
    solver.step(dt)
\end{python}
This will continue timestepping until any of the specified stopping criteria have been reached, adjusting the timestep along the way via the CFL handler.

\subsection{Timesteppers}

Rather than implementing a single specific timestepping scheme, Dedalus implements general algorithms for applying mixed implicit-explicit (IMEX) multistep and Runge-Kutta integrators along with a range of specific integrators of each type.
These IMEX schemes implicitly integrate the LHS terms and explicitly integrate the RHS terms.
This provides temporal stability for linearly stiff equations without requiring iterative algorithms for integrating the nonlinear terms.

\subsubsection{Multistep IMEX integrators}

A general multistep IMEX scheme with $s$ steps temporally discretizes preconditioned systems of the form of \eqref{eq.ivp} into the general form
\begin{equation}
    \sum_{j=0}^{s} a_j \tilde{M}_p \tilde{X}^{n-j}_p + \sum_{j=0}^{s} b_j \tilde{L}_p \tilde{X}^{n-j}_p = \sum_{j=1}^{s} c_j \tilde{F}^{n-j}_p
\end{equation}
where in general the coefficients $a_j$, $b_j$, and $c_j$ depend on the timesteps separating the steps, $\diff{t}^{n-1} = t^n - t^{n-1}$.
This expansion is rearranged to solve for the new state $X^n_p$ as
\begin{equation}
    \underbrace{(a_0 \tilde{M}_p + b_0 \tilde{L}_p)}_{\textstyle \tilde{A}^n_p} \tilde{X}^{n}_p = \underbrace{\sum_{j=1}^{s} c_j \tilde{F}^{n-j}_p - a_j \tilde{M}_p \tilde{X}^{n-j}_p - b_j \tilde{L}_p \tilde{X}^{n-j}_p}_{\textstyle \tilde{B}^n_p}
\end{equation}

The \pyth{MultistepIMEX} class implements this structure using double-ended queues to store \pyth{CoeffSystems} containing $\tilde{M} \tilde{X}$, $\tilde{L} \tilde{X}$, $\tilde{F}$, and $\diff{t}$ for the $s$ most recent steps.
The class implements a \pyth{step} method, called with the latest timestep $\diff{t}^{n-1}$, which produces $X^n$ as follows:
\begin{itemize}
    \item The timestep queue is rotated with the newest value replacing the oldest and the scheme coefficients $a_j$, $b_j$, $c_j$ are evaluated using the timestep history.
    \item $\tilde{M} \tilde{X}^{n-1}$ and $\tilde{L} \tilde{X}^{n-1}$ are evaluated for all local pencils without building the dense inverse of the right preconditioner as e.g.
    \begin{equation}
        \tilde{L}_p \tilde{X}_p = P^L_p L_p P^R (P^R)^{-1} X_p = P^L_p L_p X_p
    \end{equation}
    \item The RHS handler is evaluated and the data for each pencil is left-preconditioned and stored in the $\tilde{F}^{n-1}$ coefficient system.
    \item For each pencil, $\tilde{A}^n_p$ is solved against $\tilde{B}^n_p$ to produce $\tilde{X}^n_p$.
    A matrix solver which stores and reuses factorizations of each $\tilde{A}^n_p$ can reduce the solve time if the coefficients $a_0$ and $b_0$ remain unchanged from the previous iteration.
    \item Applying the right-preconditioner recovers the state vector $X^n_p$.
\end{itemize}

Specific multistep schemes are implemented as subclasses of the \pyth{MultistepIMEX} and define the scheme coefficients $a_j$, $b_j$, and $c_j$ via the \pyth{compute_coefficients} method.
Dedalus currently implements a number of Crank-Nicolson leap-frog, Crank-Nicolson Adams-Bashforth, and semi-implicit BDF methods from \citet{Wang:2008vk_fixed}, ranging from first to fourth order schemes.
The multistep methods only require a single evaluation of the RHS per iteration.
However, since they depend on previous iterations, they cannot run full-order when beginning a simulation.
Instead, each scheme falls back on lower-order schemes for the first $s$ iterations of a simulation.
Multistep schemes may also become ill-conditioned if the timestep is varied abruptly.

\subsubsection{Runge-Kutta IMEX integrators}

A general Runge-Kutta IMEX scheme temporally discretizes preconditioned systems of the form of \eqref{eq.ivp} by constructing stages indexed by $i=1,...,s$ as
\begin{equation}
    \tilde{M}_p \tilde{X}^{n,i}_p - \tilde{M}_p \tilde{X}^{n,0}_p + \diff{t} \sum_{j=0}^{i} H_{i,j} \tilde{L}_p \tilde{X}^{n,j}_p = \diff{t} \sum_{j=0}^{i-1} A_{i,j} \tilde{F}^{n,j}_p
\end{equation}
where $\tilde{F}^{n,j}$ is evaluated at time $t^{n,j} = t^{n,0} + \diff{t} c_j$, $X^{n,0} = X^n$, and $t^{n,0} = t^n$.
The $H$, $A$, and $c$ tableaus define a specific scheme.
This expansion is rearranged to sequentially solve for the stages as
\begin{multline}
    \underbrace{(\tilde{M}_p + \diff{t} H_{i,i} \tilde{L}_p)}_{\textstyle \tilde{A}^i_p} \tilde{X}^{n,i}_p = \\
    \underbrace{\tilde{M}_p \tilde{X}^{n,0}_p + \diff{t} \sum_{j=0}^{i-1} A_{i,j} \tilde{F}^{n,j}_p - \diff{t} \sum_{j=0}^{i-1} H_{i,j} \tilde{L}_p \tilde{X}^{n,j}_p}_{\textstyle \tilde{B}^i_p}
\end{multline}
We implement ``globally stiffly accurate'' methods where the final stage is the advanced solution, i.e.\ $X^{n+1} = X^{n,s}$ and $t^{n+1} = t^{n,s} = t^{n} + \diff{t}$.
These schemes do not require $\tilde{M}_p$ to be full rank, which it generally is not for Dedalus problems with algebraic constraints and/or boundary conditions.

The \pyth{RungeKuttaIMEX} class implements this structure using \pyth{CoeffSystem}s to store $\tilde{M} \tilde{X}^{n,0}$ as well as $\tilde{L} \tilde{X}^{n,i}$ and $\tilde{F}^{n,i}$ for all of the stages.
The class implements a \pyth{step} method, called with the timestep $\diff{t}$, which produces $X^{n+1}$ as follows:
\begin{itemize}
    \item $\tilde{M}_p \tilde{X}^{n,0}$ is evaluated for all local pencils.
    \item Then for each stage $i = 1, ..., s$:
        \begin{itemize}
        \item The RHS handler is evaluated and the data for each pencil is left-preconditioned and stored in the $\tilde{F}^{n,i-1}$ coefficient system.
        For each pencil, $\tilde{L}_p \tilde{X}^{n,i-1}$ is also evaluated.
        \item For each pencil, $\tilde{A}^i_p$ is solved against $\tilde{B}^i_p$ to produce $\tilde{X}^{n,i}_p$.
        A matrix solver which stores and reuses the factorizations of each $\tilde{A}^i_p$ can be used to reduce the solve time if the timestep $\diff{t}$ has remained unchanged from the previous iteration.
        \item The right-preconditioner is applied to recover $X^{n,i}_p$, which is assigned to the state-vector.
        The solver simulation time is set to $t^{n,0} + \diff{t} c_i$.
    \end{itemize}
\end{itemize}

Specific multistep schemes are implemented as subclasses of the \pyth{RungeKuttaIMEX} base class and define the $H$, $A$, and $c$ tableaus.
Dedalus currently implements a number of first, second, and third-order methods from \citet{Ascher:1997dy} and \citet{Sprague:2006js}.
A particular advantage of the Runge-Kutta methods is that they do not depend on any previous iterations of the state variables, so they can take full-order timesteps at the beginning of a simulation and trivially accommodate adaptive timestepping.
The cost is that the higher-order schemes perform multiple evaluations of the RHS per iteration, but they tend to run stably with larger CFL safety factors than the multistep schemes.

The ease of switching integrators allows users to easily test a variety of schemes to find the best option for their particular problem.
In addition, the multistep and Runge-Kutta base classes make it straightforward to implement new timestepping schemes.

\subsection{Matrix solvers}

Dedalus implements a generic interface for matrix solvers in the \pyth{dedalus.libraries.matsolvers} module which simplifies the implementation and comparison of different routines.
The primary routines are direct sparse matrix solvers from the SuperLU and UMFPACK libraries, wrapped through \pyth{scipy.sparse} package.
For each library, we implement fast single-solve routines as well as routines that store and apply the factorized form of a matrix.
These routines enable fast solves against multiple right-hand sides at the expense of initially computing the factorization.
This is particularly useful for initial value problems where the timestep is fixed or varying slowly.

Additional routines implement a variety of algorithms specialized to banded matrices and block-diagonal matrices, which result from full-Fourier problems and can be efficiently inverted.
The solver interface is designed to be easily extensible, allowing users to simply wrap and test new routines for specific problems.

The matrix solver routine can be specified when instantiating a \pyth{Solver} object, and the default can be set using the Dedalus configuration interface.

\section{Analysis and post-processing}
\label{sec.analysis}

The Dedalus handler system enables saving arbitrary analysis tasks while an initial value problem is running.
This system utilizes the same symbolic parsing system as is used to specify equations and efficiently evaluates the analysis tasks alongside the RHS terms on a specified cadence.
Post-processing tools simplify merging and interacting with the resulting analysis files.

\subsection{File handlers}

After building a initial value solver, instances of the \pyth{FileHandler} class can be attached to the solver's evaluator object to coordinate the periodic output of some simulation data to HDF5 files using the h5py library.
Each file handler is instantiated with an output directory path and the cadence at which handler's tasks will be evaluated.
This cadence can be in terms of any combination of simulation time (specified with \pyth{sim_dt}), wall time (specified with \pyth{wall_dt}), and iteration (specified with \pyth{iter}).
Simulation time cadences are often useful for data analysis; wall time cadences are often useful for checkpointing, e.g.\ saving the full state of a simulation every hour.
To limit the file sizes produced by the handler, the outputs are split up into different \emph{sets} over time, each containing some number of \emph{writes} that can be limited with the \pyth{max_writes} keyword.
For example, to setup a file handler to be evaluated every few iterations:
\begin{python}
output = solver.evaluator.add_file_handler(
    'output', iter=5, max_writes=100)
\end{python}

Multiple file handlers can compute and save different sets of tasks at different cadences.
For example, you may want to occasionally save full copies of the state variables for checkpointing, more frequently save snapshots of some variables for visualization, and very frequently save scalar quantities such as the total energy in the simulation.

\subsection{Analysis tasks}

Tasks, or expressions to be computed and saved by the file handler, are added to a given handler using the \pyth{add_task} method.
Tasks are entered in plain text and parsed using the same namespace that is used for equation entry.
For each task the output layout, scaling factors, and a name can also be specified.
For example, creating a task to evaluate the kinetic energy density of a flow might look like:
\begin{python}
 output.add_task("0.5*rho*(u**2+v**2+w**2)",
    layout='g', name='KE')
\end{python}
For checkpointing, you can also simply specify that all of the state variables should be saved:
\begin{python}
output.add_system(solver.state, layout='g')
\end{python}

\subsection{Post-processing}

By default, the output files for each file handler are arranged hierarchically as follows:
\begin{enumerate}
    \item At the top level is output directory that was specified when the handler was constructed, e.g.\ \pyth{'./output/'}.
    \item Within this directory are subdirectories for each set of outputs, with the same name plus a set number, e.g.\ \pyth{'output_s1/'}.
    \item Within each subdirectory are HDF5 files containing the local data for each process, with the same name plus a process number, e.g.\ \pyth{output_s1_p0.h5}.
\end{enumerate}

Often it is preferable to deal with the global dataset when performing analysis or visualization in post-processing.
The distributed process files can be merged into global files for each set using the \pyth{merge_process_files} function from the \pyth{dedalus.tools.post} module.
For some analyses, it is additionally convenient to merge the output sets together into a single file that is global in space and time, which can be done with the \pyth{merge_sets} function.
However, this can generate very large files, and is not usually necessary for analyses that are local in time, e.g.\ individually plotting each output of a task.
To assist with performing such tasks in parallel, the \pyth{visit_writes} function will coordinate all available processes to apply a given function to each output across all sets from a handler.

Together, the symbolic specification of analysis tasks and helper functions for merging and interacting with the output files can dramatically simplify user interactions with simulation products.
High-level plotting functions for plotting slices of fields and tasks are implemented in the \pyth{dedalus.extras.plot_tools} module, and example scripts utilizing these tools to construct visualizations in parallel are available.
The HDF5 output file format was chosen because it is widely used in the scientific community, and allows users to easily examine and visualize simulation outputs using a wide variety of tools and languages.

\section{Benchmarks and examples}
\label{sec.examples}

We demonstrate the features and performance of the Dedalus codebase with a variety of examples involving different types of PDEs from a variety of fields.
The examples and some of the unique features that they demonstrate are:
\begin{itemize}
\item \textbf{Parallel scaling}: strong scaling test of an incompressible hydrodynamics IVP across many nodes.
\item \textbf{Kelvin-Helmholtz}: accuracy benchmark of compressible hydrodynamics with a finite-volume code.
\item \textbf{Nonlinear Schr{\"o}dinger Network}: complex-valued PDE on a network of 1D segments, programatic extension of Dedalus interface to form spectral element method.
\item \textbf{Orszag-Tang Vortex}: moderate Mach-number vortices in compressible magnetohydrodynamics, regularized and resolved shocks.
\item \textbf{Quasigeostrophic Flow}: asymptotically reduced equations for rotating incompressible flow, LBVP to balance initial conditions.
\item \textbf{Cylindrical Stokes Flow}: low Reynolds-number flow in an annulus using polar coordinates and non-constant coefficients.
\item \textbf{Atmospheric Waves}: NLBVP to solve for the structure of an atmosphere with radiative diffusion, EVP to examine normal modes.
\item \textbf{Diamagnetic Levitation}: electrodynamics and rigid body mechanics, immersed boundaries, ODE integration, non-local boundary conditions.
\end{itemize}

Example scripts for these problems are available \chref{https://github.com/DedalusProject/methods_paper_examples}{online}.
The \chref{http://dedalus-project.org/gallery/}{project gallery} is updated with user contributed examples on an ongoing basis.

\subsection{Parallel scaling}

\begin{figure}
\centering
\includegraphics[width=\linewidth]{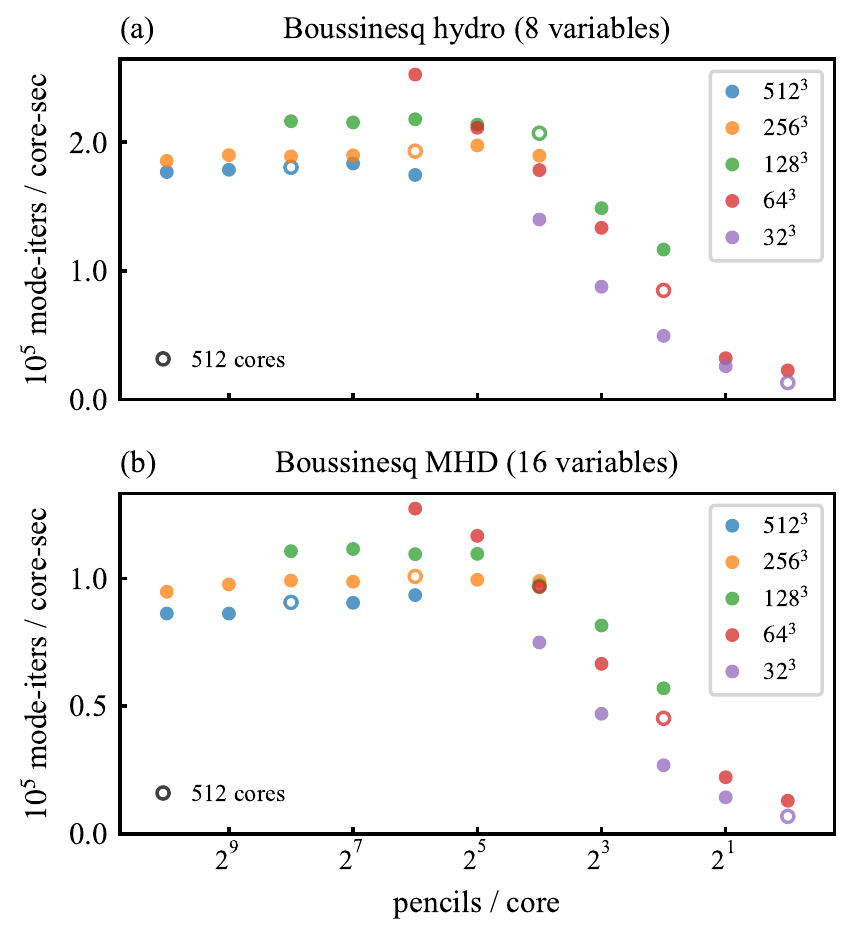}
\caption{
Dedalus performance (mode-iterations per core-second) vs parallel scaling (pencils per core) from 32 to 2048 cores for 3D Boussinesq (a) hydrodynamics and (b) magnetohydrodynamics.
Colors correspond to different 3D (Fourier-Fourier-Chebyshev) resolutions.
For each resolution, the open circle indicates the run on 512 cores, with the core count increasing by factors of two to the right.
Efficient strong scaling is seen down to roughly 8 pencils per core.
Weak scaling shows the expected $\propto 1 / \log N$ efficiency for FFT-based computations.
Execution time scales linearly with the number of problem variables.
}
\label{fig.scaling}
\end{figure}

A \chref{https://github.com/DedalusProject/scaling}{parallel scaling suite} for Dedalus is publicly available.
\figref{fig.scaling} shows performance and parallel scaling results from 32 to 2048 cores for 3D Boussinesq hydrodynamics and magnetohydrodynamics simulations in Fourier-Fourier-Chebyshev domains.
The tests were run on the \chref{https://www.sdsc.edu/support/user_guides/popeye-simons.html}{Flatiron Institute's Popeye cluster} using 32 Intel Ivy Bridge cores per node.

The code's speed, as measured in mode-iterations per core-second, is plotted against the number of cores, as measured in pencils per core.
The plateaus for each resolution indicate the regions of ideal strong scaling; large 3D problems are able to scale efficiently to thousands of cores.
The roll off at high core counts indicates the end of efficient strong scaling: the parallel efficiency of Dedalus typically remains above 50\% down to 8 pencils per core.
The ratios of the plateau values for different resolutions indicate that the weak scaling efficiency is proportional to $1 / \log N$, as expected for FFT-based computations.
Boussinesq hydro with 8 variables is twice as fast as Boussinesq MHD with 16 variables, indicating that execution time scales linearly with the number of problem variables.

\subsection{Kelvin-Helmholtz accuracy benchmark}

\begin{figure}
\centering
\includegraphics[width=\linewidth]{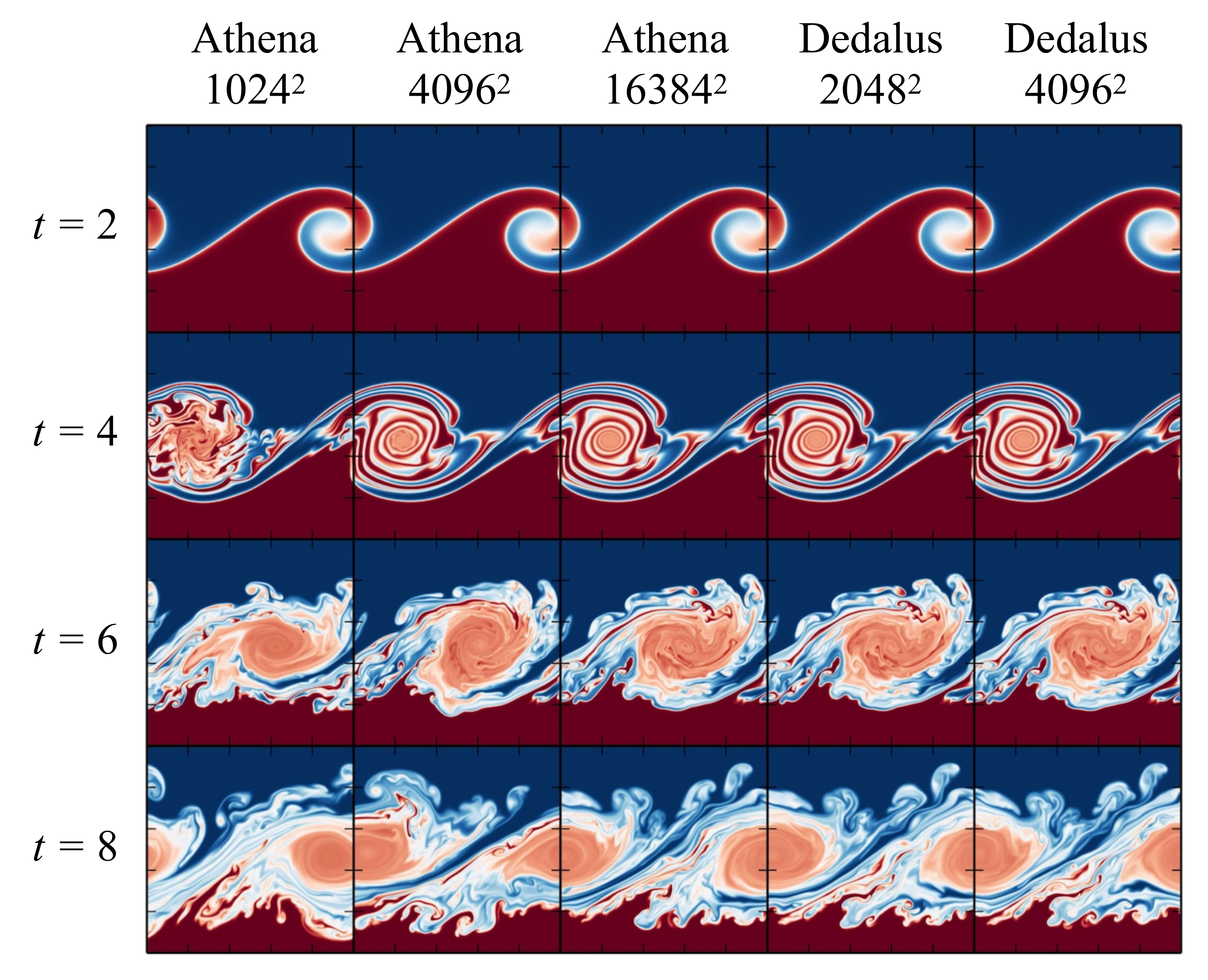}
\caption{Snapshots of a moderate Mach-number Kelvin-Helmholtz instability test problem simulated at various resolutions with a finite-volume code (Athena) and Dedalus.
The finite-volume method introduces small errors which trigger unphysical secondary instabilities in the vortex rolls.
These spurious instabilities disappear as the simulation resolution is increased.
Quantitative comparisons show comparable accuracy between the finite-volume method with $16384^2$ degrees of freedom and Dedalus with $2048^2$ degrees of freedom.
Figure adapted from \citet{Lecoanet:2015jx}}
\label{fig.kh}
\end{figure}

\citet{Lecoanet:2015jx} performed an accuracy benchmark comparing the finite-volume code \chref{https://github.com/PrincetonUniversity/Athena-Cversion}{Athena} and Dedalus.
Both codes simulated the Kelvin-Helmholtz instability in a moderate Mach-number compressible flow.
At low-to-moderate resolution, numerical errors from the finite-volume method can cause unphysical secondary instabilities to develop within the rolls created by the flow.
By directly comparing the nonlinear evolution of the flows at late times, the authors found that the finite-volume method requires a resolution of $16384^2$ cells to avoid these spurious instabilities and achieve the same accuracy as Dedalus at a resolution of $2048^2$ modes (\figref{fig.kh}).

This test demonstrates the power of high-order methods for solving PDEs with smooth solutions.
At low-to-moderate Mach numbers with finite dissipation, the flow solution lacks strong shocks and its spectral expansion converges rapidly.
Generally, for incompressible and low-Mach-number flows in simple geometries, the rapid convergence of spectral methods outweighs their larger per-iteration computation cost, making them the ideal method for simulating a broad range of astrophysical and geophysical flows.

\subsection{Nonlinear Schr{\"o}dinger Network}

The nonlinear Schr{\"o}dinger equation (NLS) is classical field equation describing the dispersive behavior of wavepackets in a weakly nonlinear medium \citep{Ablowitz:2008ev}.
The \textit{focusing} NLS is a PDE for the complex-valued field $\psi$ given by
\begin{equation}
    i \pderiv{\psi}{t} + \frac{1}{2} \npderiv{\psi}{x}{2} = -\psi |\psi|^2.
\end{equation}
\noindent In this example, we simulate the 1D NSE on a quantum graph, i.e.\ a network of connected segments with differential equations (see \citet{berkolaiko2013introduction} and \citet{Noja:2013gg} for applications).
This is achieved using a Dedalus domain with a single Chebyshev segment but separate variables for the solution on each segment, each governed by the NSE and coupled through their boundary conditions to mimic the network.
This demonstrates how the parsing system in Dedalus can be combined with a simple Python workflow to mimic a spectral element method by ``connecting'' distinct domains via boundary conditions.

\begin{figure}
\centering
\includegraphics[width=\linewidth]{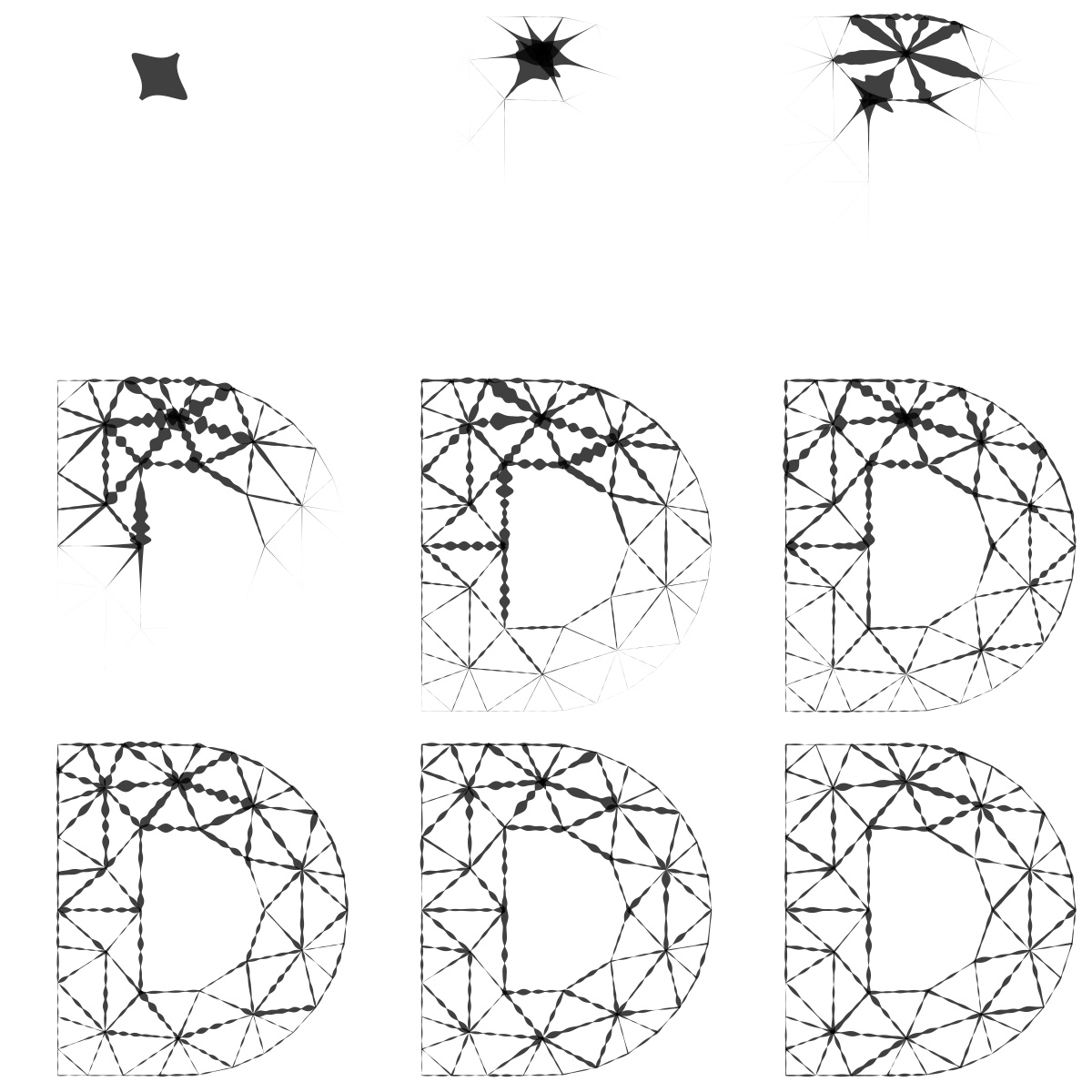}
\caption{Evolution of the nonlinear Schr{\"o}dinger equation on a network, simulated by coupling the boundaries of different fields on a 1D Chebyshev segment.
A soliton initially isolated to one segment scatters at the vertices and fills the network over time.}\label{fig.ex_network_1}
\end{figure}

We begin by loading any user-defined network defined by the planar positions of $N_v$-many vertices and a list of $N_e$-many directed edges connecting two vertices.
We then build a domain with a single 1D, 64-mode Chebyshev basis, which serves as the underlying elemental basis for the spectral element method.
We define an IVP by constructing variables and adding equations for each edge.
This is done by looping over edges to create variable names encoding the edge index and using Python string substitution to insert these names into the equations.
This programatically adds $N_e$ copies of the NLS to the problem, one for each edge, with the derivative operators weighted by the inverse of the corresponding edge length ($L_{e}$) to account for the different true lengths of the graph segments:
\begin{equation}
    i \pderiv{\psi_e}{t} + \frac{1}{2 L_e^2} \npderiv{\psi_e}{x}{2} = -\psi_e |\psi_e|^2, \quad  e = 1, ..., N_e.
\end{equation}
We encode edge-neighbor information using the graph incidence matrix.
Continuity of the solution at each vertex is imposed by iteratively matching the solution at each incident edge:
\begin{equation}
    \psi_e(v_n) = \psi_{e'}(v_n), \quad  \forall \,e, \, e' \text{ incident to vertex } v_n.
\end{equation}
\noindent Finally, Kirchoff's law for conservation of flux is imposed at each vertex by requiring the sum of the gradients of the solution on each incident edge, weighted by the edge lengths and signed by whether the edge is incoming or outgoing to that vertex ($\sigma_{e} = \pm 1$), to be equal to zero:
\begin{equation}
    \sum_{e\in v_{n}} \left. \frac{\sigma_{e}}{L_e} \pderiv{\psi_e}{x} \right|_{v_n} = 0.
\end{equation}

A bright soliton is placed on a single edge as the initial condition, and the problem is integrated forward in time using the SBDF2 timestepper.
The soliton translates to the end of the segment then scatters and disperses into the other segments and eventually fills the graph.
Snapshots of the evolution are shown in \figref{fig.ex_network_1}.
This example demonstrates the composability of the Dedalus API and the advantage of working within a high-level Python environment.

\subsection{Orszag-Tang Vortex}

Spectral methods can also correctly solve for high-Mach-number flows which develop shocks, provided that diffusion is introduced in the simulation to regularize the shocks.
The Orszag-Tang vortex problem \citep{Orszag:1979kv} is a standard compressible magnetohydrodynamics (MHD) test problem in astrophysics (e.g. for the fixed-grid Godunov code Athena \citep{Stone:2008bo, Felker:2018gz}, the finite difference code Flash \citep{Lee:2009kq}, smoothed-particle hydrodynamic codes \citep{Tricco:2016gy}, moving mesh codes \citep{Mocz:2017kl}, etc.).
The problem can be simulated with shock-capturing algorithms (e.g. Riemann solvers) or those with numerical diffusivity which can regularize the shocks (e.g. SPH).
Here, we explicitly add diffusion of momentum, heat, and magnetic fields to the model to regularize the shocks.

We note that a diffusive shock creates entropy at a rate that is independent of (but mediated by) the microphysical diffusion. 
For example, in Burger's equation, $\partial_{t} u + u \partial_{x} u = \partial_{x} ( \nu \partial_{x} u)$, the dissipation rate across a shock is 
\begin{equation}
\frac{d}{dt} \int \frac{u^{2}}{2} \text{d} x = -\int \nu |\partial_{x} u|^{2} \text{d} x \approx - \frac{\Delta u^{3}}{12}.
\end{equation}
The leading order dissipation is \emph{independent of the viscosity}, and the correction is on the order of the inverse Reynolds number with logarithmic corrections.
Numerical simulations of shocks should therefore give results that are mostly independent of the viscosity, provided the Reynolds number is large enough.
This principle underlies shock-capturing algorithms, but can also be leveraged in spectral simulations. 
Because of the weak dependence on the Reynolds number, diffusivities much larger than the natural values can be used to regularize shocks while still respecting important global balances. 
As long as the resulting diffusive-shock length scale $\Delta \ell \sim  \nu / \Delta u$ is resolved, a spectral computation will be free of Gibbs ringing and produce accurate results.

We simulate the Orszag-Tang vortex in a 2D domain which is periodic in both $x$ and $y$ directions, and comprises $[0,1]^2$.
A vortex with velocity $\vec{u} = (-\sin(2\pi y),\sin(2\pi x) )$ is initialized in an ideal gas with constant density and pressure, $\rho=25/(36\pi)$ and $p=5/(12\pi)$, and ratio of specific heats $\gamma=5/3$.
The initial magnetic field is specified by a vector potential $A_z = B_0 (\cos(4\pi x)/(4\pi) + \cos(2\pi y)/(2\pi) )$, with $\vec{B} = \vec{\nabla}\vec{\times}\vec{A}$, and $B_0=1/\sqrt{4\pi}$.
The adiabatic sound speed $\gamma p/\rho=1$, so the flow is supersonic in parts of the domain, which leads to the formation of MHD shocks.
We solve the equations
\begin{align}
&\partial_t\vec{u} + \vec{\nabla} T' + T_0\vec{\nabla}\Upsilon - \nu \left(\nabla^2 \vec{u} + (1/3)\vec{\nabla}\vec{\nabla}\vec{\cdot}\vec{u}\right) = \nonumber \\
& -\vec{u}\vec{\cdot}\vec{\nabla}\vec{u}  - T'\vec{\nabla}\Upsilon  + \nu \vec{\nabla}\Upsilon\vec{\cdot} \mathbf{S}  \nonumber \\
& + (\vec{B}\vec{\cdot}\vec{\nabla}\vec{B} - (1/2)\vec{\nabla}|\vec{B}|^2)e^{-\Upsilon}, \\
&\partial_t\Upsilon + \vec{\nabla}\vec{\cdot}\vec{u} = - \vec{u}\vec{\cdot}\vec{\nabla}\Upsilon, \\
&\partial_t T' + (\gamma-1)T_0\vec{\nabla}\vec{\cdot}\vec{u} - (\chi/c_v)\nabla^2 T' = \nonumber \\
& \vec{u}\vec{\cdot}\vec{\nabla}T' - (\gamma-1)T'\vec{\nabla}\vec{\cdot}\vec{u} + (\chi/c_v)\vec{\nabla} T' \vec{\cdot}\vec{\nabla}\Upsilon \nonumber \\
&  + (\nu/2c_v)(\text{Tr}(\mathbf{S}^{2}) + \text{Tr}(\mathbf{S})^{2}) + (\eta/c_v)|\vec{\nabla}\vec{\times}\vec{B}|^2 e^{-\Upsilon}, \\
&\partial_t A_z - \eta \nabla^2 A_z = \vec{e}_z\vec{\cdot}(\vec{u}\vec{\times}\vec{B}).
\end{align}
Here, $\nu$, $\chi$, and $\eta$ are the viscosity, thermal diffusivity, and magnetic diffusivity, $T_0=1/\gamma$ is the background temperature, $T'$ is the temperature perturbation, and $\Upsilon=\log\rho$.
The heat capacity at constant volume $c_v=(\gamma-1)^{-1}=3/2$ normalizes the conduction and heating in the energy equation for $T'$.
In the nonlinear viscous terms, the symmetric stress tensor $\mathrm{S} = \vec{\nabla}\vec{u} + (\vec{\nabla} \vec{u})^{T} - (2/3) \vec{\nabla}\vec{\cdot}\vec{u} \, \mathrm{I}$.
This is an MHD-analog of the equations introduced in \citet{Lecoanet:2014ij}.

We run the simulation at rather high resolution, using $4096$ modes in the $x$ and $y$ directions.
We use $3/2$ dealiasing in each direction, although this does not eliminate aliasing errors from converting from $\Upsilon$ to $\rho$.
We set all the diffusivities equal to $10^{-4}$.
For timestepping, we use a third-order, four-stage DIRK/ERK method (RK443 of \citealt{Ascher:1997dy}), with an initial timestep size of $2.5\times 10^{-5}$.
We use adaptive timestepping based on the CFL condition associated with both flow and Alfv\'{e}n speeds (but not the sound speed), with a safety factor of $0.6$.
The simulation is run to $t=1$.

\begin{figure}
\centering
\includegraphics[width=\linewidth]{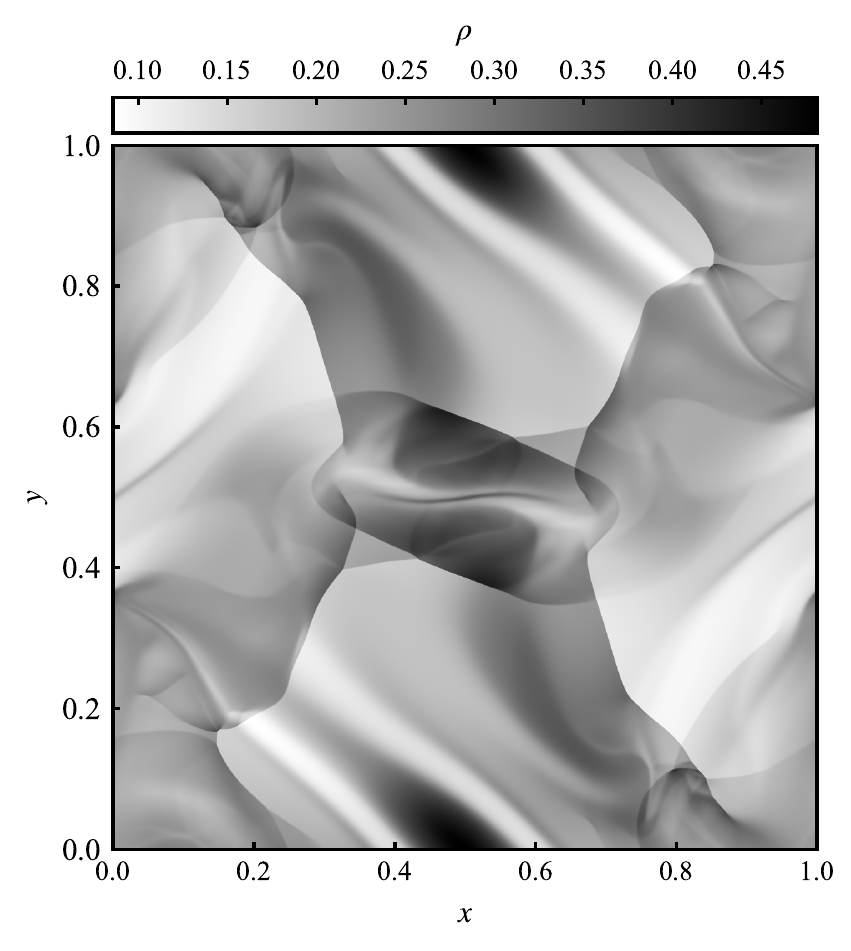}
\caption{Density at $t=0.5$ in the Orszag-Tang vortex test.
The sharp density jumps visualize the MHD shocks.
The structures in the upper left and lower right corners are due to shock-shock interactions, which are correctly treated in Dedalus.}
\label{fig.OT.full}
\end{figure}

\figref{fig.OT.full} shows the density at $t=0.5$.
There are very sharp density gradients due to the formation of shocks. Even more impressive, Dedalus is able to correctly follow the interaction between multiple shocks, which produces the features in the upper left and lower right corners of the figure.
These are similar to what is found in high resolution simulations with shock-capturing codes \citep[e.g.][]{Felker:2018gz}.

\begin{figure}
\centering
\includegraphics[width=\linewidth]{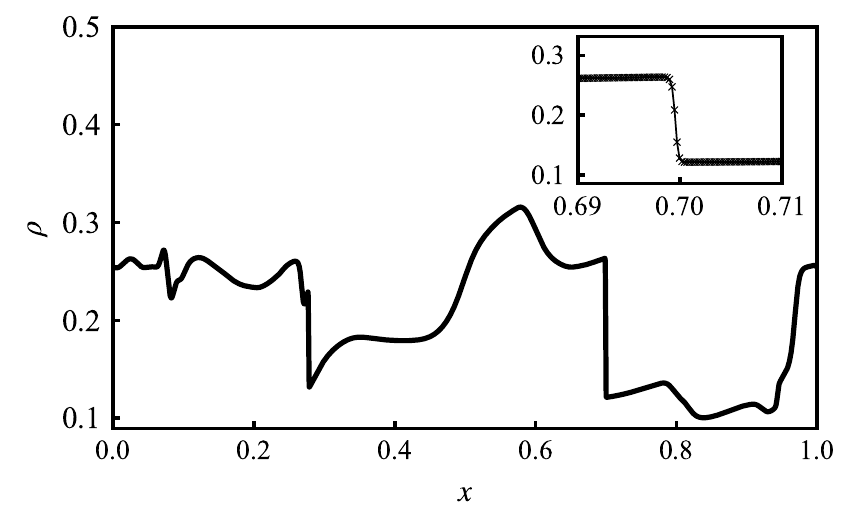}
\caption{A profile of the density at $t=0.5$ and $y=0.3125$ in the Orszag-Tang vortex test.
The inset shows the density profile around $x=0.7$, with crosses denoting grid points (using the non-dealiased grid).
The shock is resolved by several grid points.}
\label{fig.OT.line}
\end{figure}

To better visualize the shocks, we plot a horizontal profile of the density at $y=0.3125$ in \figref{fig.OT.line}.
At this height, there are shocks at $x\approx 0.3$ and $\approx 0.7$.
The inset shows the solution very close to $x=0.7$.
The shock is well-resolved by 4--5 grid points.
Note that the density stays smooth despite no diffusion in the density equation.
Rather, it appears that the combination of viscosity and thermal diffusivity cause the temperature and pressure to regularize, which in turn regularizes the density via the ideal gas equation of state, $p=\rho T$.

\subsection{Quasigeostrophic Flow}
\label{sec.qg}

One of the original motivations behind Dedalus was to allow the straightforward computation of non-standard equation sets.
It is a common practice for modelers to start with something universal, such as Navier-Stokes, or Maxwell Equations, make a series of asymptotic reductions, and arrive at a new set of equations that can capture an interesting physical regime.
It is often not clear how useful these equations are until they are simulated.
But it is risky to write ``off-the-shelf'' solvers for new equations before there is good evidence that they are useful to at least a few people.

The Quasigeostrophic (QG) model is a classic approximation used in geophysical and astrophysical fluid dynamics.
Initially, the QG model was intended to reduce the overall cost of simulations.
The idea is to filter sub-dominant fast-timescale waves, while retaining essential nonlinear dynamics.
Current state-of-the art simulations no longer require the QG assumption on the basis of cost.
However the equations are still used widely because of their relative simplicity and explanatory power.
QG is a non-standard model that became widely adopted.
Since the advent of QG, other strongly nonlinear reduced models appear in different fields occasionally, but computational results can lag by several years.
For example, in stratified fluids \cite{Durran:1989dl,Vasil:2013ij,Lecoanet:2014ij}, in magnetized fluids \cite{Zank:1992fm,Ng:2012fj}, in Langmuir turbulence \cite{Chini:2009em,Malecha:2014gv}, in near-inertial wave dynamics \cite{Wagner:2017dc,Wagner:2016iy,Wagner:2015iw}, and in rapidly rotating fluids \cite{Julien:1998iv,Julien:2006dz,Julien:2016js} (theory) \cite{Sprague:2006js,Julien:2012dc,Plumley:2016eo} (simulation).

The solution of the QG equation has a long history starting with numerical weather prediction in the 1950s \cite{Charney:1953dt,Phillips:1956du}.
Under certain restrictive assumptions, the QG problem can be reduced to 2D equations on the surface \cite[e.g.][]{Holland:1980fd,Salmon:2006gi}, or three-dimensional equations in a triply periodic domain \cite[e.g.][]{McWilliams:1994ey,Vallgren:2010gp}.
However, the most general case requires a three-dimensional layer with boundaries \cite[e.g.,][]{Tulloch:2009ji}, and non-constant coefficients.
In our current work, we pick QG as an interesting example problem because it highlights many features unique to Dedalus.
We choose the most difficult and general case of a finite 3D layer because it illustrates the most design features within a single model.
Like previous work, it would be straightforward to simplify our example script to model a triply periodic domain, or a two-dimensional layer, if one wanted.

The traditional QG equations collapse to a single, second-order equation for the potential vorticity (PV).
This is equivalent to a first-order formulation with two dynamical variables and two first-order equations.
A first-order formulation does not increase the computational cost because the second-derivative matrix has double the bandwidth of the first-derivative matrix.
Dedalus can use the traditional PV formulation in terms of the streamfunction and its vertical derivative.
Here, we describe an alternative formulation choosing more physically meaningful variables, which makes the problem more straightforward to pose and generalize; at no additional cost. 

We solve for two variables: $w$ (upwelling) and $p$ (pressure).
All other physical quantities (e.g. buoyancy and velocity) are diagnostic in terms of $w$ and $p$.
The upwelling field is not part of the traditional formulation.
Although we solve for two variables instead of a single variable (PV) in the traditional formulation, using the vertical velocity makes the boundary conditions much simpler; we do not need to solve a nonlinear buoyancy advection equation on the boundary.

We use Dedalus substitutions to define several of the important physical variables in terms of the pressure, $p$,
\begin{eqnarray}
u &=& -\partial_{y}p \\
v &=& \ \ \partial_{x} p , \\
\theta &=& \ \  \partial_{z} p , \\
\zeta &=&  \ \  \partial_{x}v - \partial_{y}u = (\partial_{x}^2+\partial_{y}^2)p.
\end{eqnarray}
Dedalus substitutions can be recursive, e.g., to define $\zeta$.
The substitutions reflect the well-known geostrophic and hydrostatic diagnostic balances inherent in the theory.
Pressure assumes the role of the stream function, $\psi$.
We subsequently define the advection substitution with an argument,
\begin{eqnarray}
D(q) \ \equiv \ u \, \partial_{x} q \ + \ v\, \partial_{y}q.
\end{eqnarray}
We also use substitutions to prescribe horizontal 4th-order hyperdiffusion,
\begin{eqnarray}
L(q) \ \equiv \  \partial_{x}^{2} q \ + \ \partial_{y}^{2}q, \quad \Delta_{4}(q) = L^{4}(q).
\end{eqnarray}
The problem parameters are the Coriolis variation $\beta$, the Ekman friction $\gamma$, the stratification profile $\Gamma(z)$, the thermal-wind profile $U(z)$, and the hyper-diffusivities $\nu_4$ and $\kappa_4$.
We solve the coupled prognostic equations,
\begin{eqnarray}
\partial_{t} \zeta  + U  \partial_{x}\zeta + \beta v -  \partial_{z}w + \nu_{4} \Delta_{4}(\zeta)= -D(\zeta)
\\
\partial_{t} \theta  + U  \partial_{x}\theta - U'  v + \Gamma  w + \kappa_{4} \Delta_{4}(\theta)= -D(\theta)
\end{eqnarray}

The traditional formulation solves the entire buoyancy evolution equation on the boundary.
This is a complicated way to impose no vertical flow.
Because the upwelling is part of our formulation, we impose the boundary condition directly.
More precisely, we impose Ekman flux boundary conditions.
\begin{eqnarray}
w \ \mp \ \gamma_{\pm} \, \zeta \ = \ 0  \quad \quad \text{at} \quad z = 0,1
\end{eqnarray}
Ekman pumping conditions result from a closure model of an asymptotically thin viscous boundary layer \cite{Stellmach:2014kpa}.
We could easily include boundary stress or topography on the right-hand side.

The $-U'$ shear term in the bulk equation comes from a large-scale background thermal-wind profile
\begin{eqnarray}
\partial_{z} U = -\partial_{y} \Theta,
\end{eqnarray}
where $\Theta(y,z) = y \, S(z)$ is a linear north-south-varying buoyancy profile.
Unstable modes extract energy from this thermal variation.
The evolution equations are a coupled pair of 1st-order equations in $\partial_{z}$ for  $p$ and $w$.
There is only one independent time derivative as $\zeta$ and $\theta$ are both related to $p$, as is PV
\begin{eqnarray}
\text{PV}  =  \zeta + \partial_{z} (\Gamma^{-1} \, \theta)  =  \left(\partial_{x}^{2} + \partial_{y}^{2} + \partial_{z} \Gamma^{-1}  \partial_{z}\right)p
\end{eqnarray}
Looking at the diagnostic balances, we can see that
\begin{eqnarray}
\partial_{z} \zeta - (\partial_{x}^{2} + \partial_{y}^{2})\theta \ = \ 0
\end{eqnarray}
This is the ``thermal-wind" balance that follows from geostrophy and hydrostatic balance.
The vertical velocity field, $w$, satisfies its own diagnostic balance at each time step, which removes non-balanced forcing terms.
\textit{The vertical velocity here is exactly analogous to the pressure field for incompressible hydrodynamics}; it is a Lagrange multiplier enforcing the consistent evolution of $\zeta$ and $\theta$, which are both related to the pressure.

\textit{Balanced initial conditions --- } We initialize the pressure with filtered random noise on the grid; vorticity and buoyancy follow directly.
In practice, this is likely to be sufficient to initialize the simulation.
Some timesteppers may correct for unbalanced initial conditions.
For completeness, we illustrate how to solve a Linear Boundary-Value Problem (\pyth{LBVP}) for the upwelling given an initial pressure field.

Knowing $p(t=0,x,y,z)$, we set up the coupled problem for $w(t=0,x,y,z)$:
\begin{eqnarray}
& L(\varpi_{t}) \!-\!  \partial_{z}w  \!=\! -D(\zeta) \!-\! U  \partial_{x}\zeta \!-\! \beta v \!-\!  \nu_{4} \Delta_{4}(\zeta) \\
& \partial_{z} \varpi_{t}  \!+\! \Gamma  w \! = \! - D(\theta)  \!-\! U \partial_{x}\theta \!+\! U' v \!-\! \kappa_{4} \Delta_{4}(\theta)
\\
& w = \pm \gamma_{\pm}  \zeta  \ \ \text{at} \ \ z = 0,1
\end{eqnarray}
We could collapse these to a single system for $w$.
However, we can mostly reuse the original system if we introduce a \textit{slack variable}, $\varpi_{t}$.
The slack variable takes the place of the time evolution terms in the full system.
Solving this \pyth{LBVP} gives the vertical velocity at $t=0$ and $\varpi_t=\partial_t p|_{t=0}$.
This general approach can be used to derive balanced initial conditions for other equations with time-independent constraints, e.g., incompressible hydrodynamics.

 \begin{figure}
  \centering
  \includegraphics[width=\linewidth]{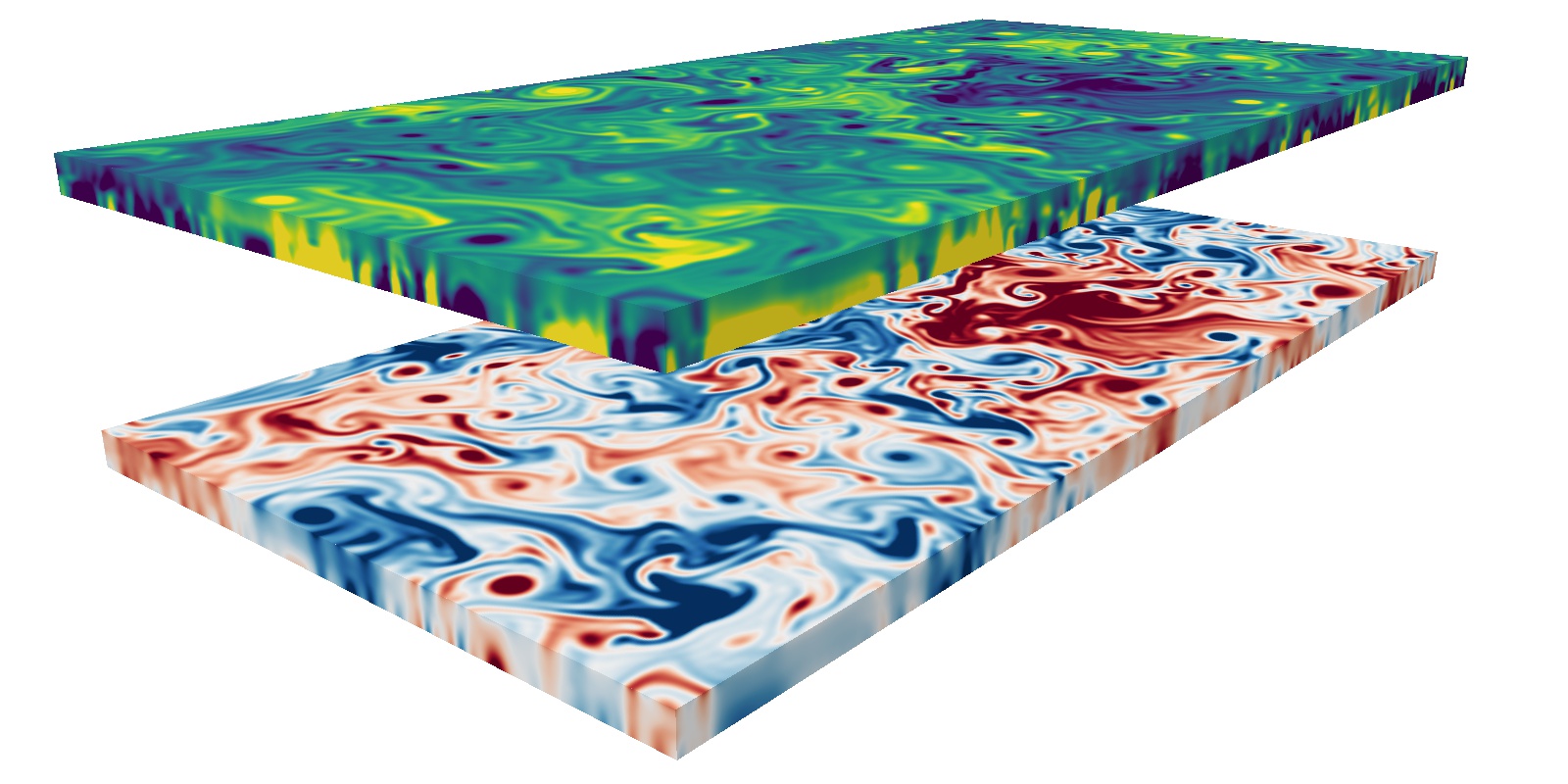}
  \caption{Surface and lateral slices of the PV (top) and buoyancy perturbation $\theta$ (bottom) in a 3D quasigeostrophic flow. Both images are at $t=200$ in the statistically saturated state.}
  \label{fig:QG}
\end{figure}

 \textit{The simulation --- } We non-dimensionalize $z$ with the domain depth, $H$.
 We non-dimensionalize $x,y$ with the Rossby radius of deformation at the top of the box, $L \ = \ N\, H/f$, where $N$ is the characteristic background buoyancy frequency at $z=H$, and $f$ is twice the background planetary rotation rate.
We solve for the relative nonlinear fluctuations around a linearly unstable thermal wind profile; $U(z) \ = \ z$, $U'(z) = 1$.
We non-dimensionalize time using the background shear rate.
Here we instead use a depth-dependent non-dimensional stratification parameter, which necessitates a fully three-dimensional approach.
\begin{eqnarray}
\Gamma(z) = e^{2z-2} 
\end{eqnarray}
Dedalus expands this profile in a Chebyshev series up to a cutoff tolerance of $10^{-8}$.

The presence of the shear implies a large-scale background buoyancy gradient.
We include a background planetary vorticity gradient ($\beta$-effect) with $\beta=0.1$.
To saturate the simulation we use hyper-diffusivities $\nu_4=\kappa_4=10^{-6}$, bottom friction $\gamma_{-}=0.16$, but no top friction.
The horizontal domain size is $40\times20$ Rossby radii in the $x,y$ directions respectively.
We use $256\times128$ Fourier modes in the $x,y$ directions, and $32$ Chebyshev modes in the $z$ direction, all using 3/2 dealiasing.
The Chebyshev tau method balances energy to exponentially high accuracy; exact energy conservation could be still be imposed, however, by formulating a self-adjoint system with an alternative orthogonal polynomial basis \cite{Watwood:2019kv,Vasil:2019ir,Aurentz:2019vt}.

The whole configuration is baroclinically unstable.
The motion is a kind of side-ways convective heat transport.
For background, we highly recommend the book ``Atmospheric and Oceanic Fluid Dynamics'' by Geoffrey K. Vallis \cite{Vallis2006}. To summarize the phenomenology: \textit{
``It is the instability that gives rise to the large- and mesoscale motion in the atmosphere and ocean --- it produces atmospheric weather systems, for example --- and so is, perhaps, the form of hydrodynamic instability that most affects the human condition.''}

We compute the nonlinear evolution until the system reaches a statistically stationary state, which takes roughly 75 dynamical time units.
\figref{fig:QG} shows the PV and vertical vorticity $\zeta$ at $t=200$.
The solution contains a sea of compact eddies that form, merge, and breakup over several dynamical times.
For these parameters, the system is mostly two-dimensional.
There are, however, nontrivial variations in the depth-dependence of PV.

\newcommand{\Reyn}{\ensuremath{\mathrm{Re}}}
\newcommand{\Pecl}{\ensuremath{\mathrm{Pe}}}

\subsection{Stokes Flow}
\label{sec.stokes_example}

Many important biophysical and industrial fluid problems occur in a high-drag (or low Reynolds number) limit ($u_{0} \ell_{0} / \nu = \Reyn \ll 1$) where the momentum equation reduces to Stokes flow.
Here, we present and solve a simple test problem demonstrating boundary-driven Stokes flow in curvilinear geometry with time-dependent tracer fields.

We simulate the classic ``unmixing'' demonstration \citep[e.g.][]{Fonda:2017ev}, where a Taylor-Couette device with an inner-cylinder radius, $R_{\mathrm{in}}$, turns $n$ times and then reverses back to the start.
For low $\Reyn$, the flow is reversible, and $m$-many dye tracers will appear to mix and then unmix.
We non-dimensionalize lengths with the gap width, $\ell_{0} = R_{\mathrm{out}} - R_{\mathrm{in}}$, and velocities with the maximum inner-cylinder speed,  $u_0$.
The Peclet number $u_{0} \ell_{0} / \kappa = \Pecl$ controls the tracer mass diffusion. Taking $\Reyn \to 0$,
\begin{align}
  &\nabla^2 \vec{u} = \vec{\nabla} p\\
  &\vec{\nabla \cdot u} = 0\\
  &\partial_t c_m + \vec{u} \cdot \vec{\nabla} c_m = \frac{\nabla^2 c_m}{\Pecl}, \quad m = 1,2,3.
\end{align}
The domain is a no-slip two-dimensional cylindrical annulus with radius $1 \leq r \leq 2$ and angle $0 \leq \theta < 2\pi$.
The diffusive dye flux $\mathcal{F}_m = -\vec{\nabla} c_m/\Pecl = 0$ at the boundaries.
The flow vanishes at the outer wall; $\vec{u}(t, R_\mathrm{out}, \theta) = 0$.
The tangential velocity at the inner wall is a time-dependent regularized square wave
\begin{equation}
  \label{eq:stokes_1}
  u_{\theta}(t, R_\mathrm{in},\theta) = \frac{\arctan(50\sin(t/n_{\mathrm{rot}}))}{\arctan(50)}.
\end{equation}
The time-dependent boundary condition uses the \pyth{GeneralFunction} mechanism described in \secref{sec.user-spec-func}.
The initial dye fields are circular Gaussians with widths $\delta=1/4$ centered at $r_{\text{dye}}=3/2$, $\theta_m = 2\pi m /3$.

Since the domain is a cylindrical annulus, which excludes the origin, it can be accurately discretized using a direct product of Fourier and Chebyshev bases in $\theta$ and $r$, respectively.
For simulating the full disk, non-direct product bases imposing certain regularity conditions at the origin are necessary \citep{Vasil:2016kb} (these will be implemented in future versions of Dedalus).

In polar coordinates, the differential operators $\nabla$, $\nabla^{2}$ contain non-constant coefficients proportional to $1/r$ and $1/r^{2}$ multiplying the radial and azimuthal derivatives, which can simply be added when entering the equations.
Naively (for example) one might implement the angular part of the Laplacian as \pyth{'d(u,th=2)/r**2'} where \pyth{th} is the $\theta$ spatial variable.
However, the Chebyshev expansion $r^{-2}$ converges very slowly, leading to dense non-constant coefficient matrices.
To avoid this, we multiply all equations by $r$ or $r^2$ (depending on the order of derivatives).
The resulting non-constant coefficient matrices only contain bandwidth one or two.

\begin{figure*}
  \centering
  \includegraphics[width=\textwidth]{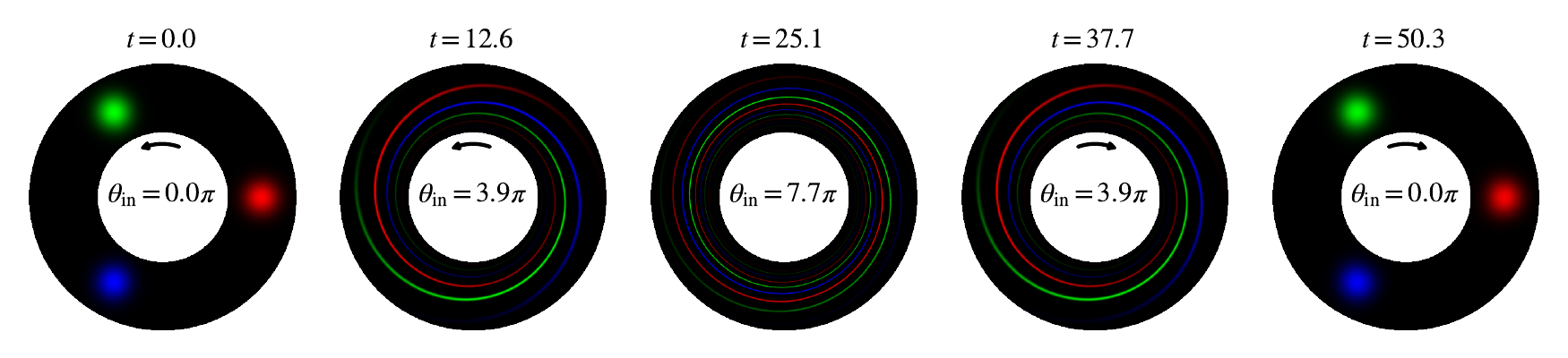}
  \caption{Evolution of the tracer fields in a reversible Taylor-Couette Stokes flow.
  The three separate tracer fields are visualized as red, green, and blue in a single RGB image.
  As time progresses, we rotate the inner cylinder first counter clockwise and then clockwise.
  By the end of the simulation, the flow has returned to its initial conditions, modified only by mass diffusion of the dye.}
  \label{fig:stokes_flow}
\end{figure*}

We solve this system with $512 \times 512$ modes using the Runga-Kutta 443 time stepper and a fixed $dt = 0.005$.
The simulation runs approximately four rotations forwards and backwards; $n_{\mathrm{rot}}=8$.
\figref{fig:stokes_flow} shows snapshots of the flow at five times, symmetric about the middle of the simulation.
The dye spreads out into thin sheets as the inner cylinder rotates in the counter-clockwise direction; the sheets congregate back, differing from the original Gaussian blobs due to the action of the dye.
The final shapes are not Gaussian.
Shear dispersion leads to thin structure (sharp gradients) in the radial direction, and hence the diffusion is not isotropic.

This simulation uses high $\Pecl = 10^7$, causing the sheets to become very thin.
The dye-patches area stays constant in the absence of diffusion.
Equating the initial and final area $A_0 \sim \pi \delta^2/4  \approx A_{\mathrm{f}} \sim n_\mathrm{rot} 2\pi r_{\text{dye}} \delta r$, implies the sheet width $\delta r \sim 1/200$.
For the diffusion time across the sheets to be greater than the total simulation time, $ \tau_D \gg 2 n_{\mathrm{rot}} \simeq 10$, we require $\Pecl \gtrsim 10/\delta r^2 \sim 4 \times 10^5$.
The example code also works with $\Pecl = \infty$, disabling the $\nabla^2 c_m/\Pecl$ terms and flux boundary conditions.
In this case, the well-resolved sheets return to the original Gaussian with only slight time-stepping errors.

\subsection{Atmospheric waves}

Stratified (non-rotating, neutral) atmospheres support two classes of linear waves: acoustic waves and gravity waves.
The properties of these waves are easy to compute in constant coefficient atmospheres (like isothermal atmospheres), but require numerical solutions when the atmosphere structure introduces non-constant coefficients.
Solving for the nonlinear atmospheric background structure is challenging.
Here we use a nonlinear boundary value problem (\pyth{NLBVP}) to solve for an atmosphere profile, and then solve an eigenvalue problem (\pyth{EVP}) for the wave modes.

We study an optically thin plane-parallel atmosphere coupled to an underlying, optically thick adiabatic layer.
This models the Sun's photosphere and lower atmosphere, or Jupiter's atmosphere spanning the radiative-convective boundary and the lower stratosphere.
For the background structure, we solve hydrostatic and thermal equilibrium, including radiation transport under the Eddington tensor approximation \citep[e.g.,][]{Mihalas&Mihalas_1984, Castor_2007, Jiang:2012kp, Jiang:2015ky}:
\begin{align}
 	\pderiv{P_g}{z} + \pderiv{P_r}{z} &= - \rho g,\\
 	\pderiv{P_r}{z} &= - \frac{\rho \kappa F_r}{c},
\label{eq: eddington P_r}
\end{align}
with $P_g$ the gas pressure, $P_r$ the radiation pressure, $\rho$ the density, $g$ the constant gravity, $\kappa$ the opacity, $F_r$ the constant radiative flux, and $c$ the speed of light.
We use an ideal gas equation of state $P_g = \rho R T$ with $R$ the gas constant and $T$ the temperature, the Eddington tensor closure $P_r = f a T^4$ with $a$ the radiation constant and $f = 1/3$, and a Kramer-like opacity law for $\kappa$:
\begin{equation}
  \kappa(\rho, T) = \kappa_0 \left(\frac{\rho}{\rho_0}\right)^a \left(\frac{T}{T_0}\right)^b,
\end{equation}

With these substitutions, and non-dimensionalizing by $\rho_0$, $T_0$, and a lengthscale $L$, the equations become
\begin{equation}
	\frac{1}{\rho} \pderiv{\rho}{z} + \frac{1}{T} \pderiv{T}{z} = - \frac{g^*}{T} \left[1 - \left(\frac{F}{F_\mathrm{Edd}}\right) \rho^a T^{b}\right]
\end{equation}
\begin{equation}
\frac{1}{T} \pderiv{T}{z} = - Q \rho^{a+1} T^{b-4},
\label{eq: eddington thermal balance}
\end{equation}
where $g^* = g L / R T_0$, $F_\mathrm{Edd} = g c / \kappa_0$, and \mbox{$Q = \rho_0 \kappa_0 F_r L / 4 f a c T_0^4$}.

For $F \ll F_\mathrm{Edd}$, the optically thick deep solution will be nearly polytropic with $\rho \propto T^m$ where $m = (3 - b) / (1 + a)$.
We fix $a = 1$, $b= 0$ and take $g^* = m+1$ to define the lengthscale $L$.
$Q$ is taken to be $1 - 1.68 \times 10^{-4}$, corresponding to $\ln (T_\mathrm{eff}/T_0) = -2$ in a gray atmosphere.
We solve this nonlinear boundary value problem in Dedalus using $\ln \rho$ and $\ln T$ as dynamical variables with 128 Legendre modes.
A dealiasing factor of 2 is used, which reduces but cannot eliminate aliasing errors since exponential nonlinearities (formally including infinite polynomial orders) are present in the formulation.
Solutions converge to a relative tolerance of $10^{-8}$ in $\bigO(10)$ Newton iterations.
When $F=0$ there is an analytic solution due to \citet{Brandenburg:2016bu} which our numerical solution matches to 10 digits in the $L_2$ norm, verifying the correctness of the implementation.
We proceed here with $F/F_\mathrm{Edd}=10^{-5}$, for which there is no analytic solution.

\begin{figure}
  \includegraphics[width=\linewidth]{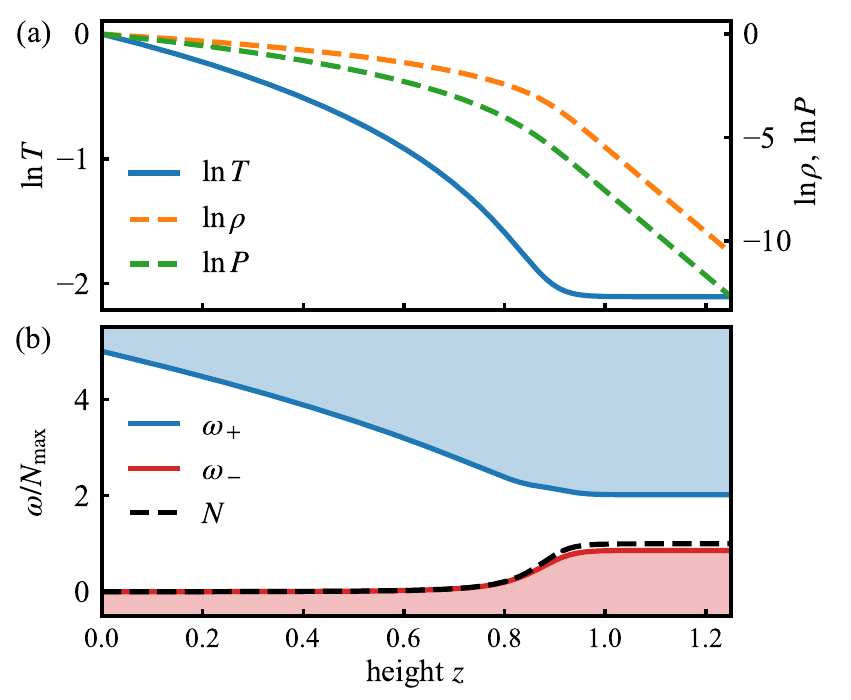}
  \caption{(a) Structure of a balanced atmosphere with an Eddington closure for radiation transport and a Kramer-like opacity law with $a=1$, $b=0$.
  At depth, the atmosphere is nearly an adiabatic polytrope with constant $s/C_P$ and $N^2 \approx 0$.
  The upper atmosphere is nearly isothermal with exponentially decaying $\rho$ and $P$.
  (b) The upper atmosphere is a resonant cavity for internal gravity waves with $N^2 > 0$.
  Frequencies are normalized by the Brunt-V\"{a}is\"{a}l\"{a} frequency $N$ in the isothermal layer.}
  \label{fig:eddington_atm}
\end{figure}

We plot the temperature, density, and pressure in \figref{fig:eddington_atm}.
The bottom, optically thick, part of the atmosphere is nearly polytropic with $\rho\propto T^{(3/2)}$, as expected for this choice of $a$ and $b$ \citep{Barekat:2014kx}.
The top, optically thin, part of the atmosphere is nearly isothermal while the density and pressure drop exponentially.
This region has a constant Brunt-V\"{a}is\"{a}l\"{a} (buoyancy) frequency
\begin{equation}
  N^2 = -g \partial_z s/C_P.
  \label{eq:eddington Brunt squared}
\end{equation}

Next we consider the oscillation modes of this atmosphere.
The character of waves modes depends on their frequency relative to $\omega_\pm$, which are defined as
\begin{align}
  \omega_+^2 &= \omega_L^2 + \omega_\mathrm{ac}^2,\\
  \omega_-^2 &= \frac{\omega_L^2}{\omega_L^2 + \omega_\mathrm{ac}^2} N^2,
\end{align}
where the Lamb frequency $\omega_L$ and acoustic frequency $\omega_\mathrm{ac}$ are related to the properties of the background atmosphere \citep[e.g.,][]{Hindman:1994eh}.
Acoustic waves have frequencies greater than $\omega_+$, while gravity waves have frequencies less than $\omega_-$.
We plot $\omega_\pm$ for our atmosphere in \figref{fig:eddington_atm}.
High-frequency sound waves can propagate all the way to the bottom of the atmosphere, whereas gravity waves mostly stay in the optically thin isothermal layer.

We write the ideal, linearized fully compressible equations as
\begin{align}
\partial_t w + \partial_z T_1 + T_0 \partial_z \ln \rho_1 + T_1 \partial_z \ln \rho_0 & = 0, \\
\partial_t u + \partial_x T_1 + T_0 \partial_x \ln \rho_1 & = 0, \\
\partial_t \ln \rho_1 + w \partial_z \ln \rho_0 + \left(\partial_x u + \partial_z w\right)  & = 0, \\
\partial_t T_1 + w \partial_z T_0 + (\gamma - 1) T_0 \left(\partial_x u + \partial_z w\right) & = 0,
\end{align}
for the perturbation velocities $w$ and $u$, and thermal and log density perturbations $T_1$ and $\ln \rho_1$.
The non-constant coefficients of the atmosphere are $\partial_z \ln \rho_0$, $T_0$ and $\partial_z T_0$ and we take $\gamma =5/3$.
For simplicity, we impose impenetrable top and bottom boundaries ($w = 0$).

We formulate this as an eigenvalue problem in Dedalus by replacing $\partial_t = i \omega$ and $\partial_x = - i k_x$ and solving for complex eigenvalue $\omega$ given specified horizontal wavenumbers $k_x$.
We use the \chref{https://github.com/DedalusProject/eigentools}{eigentools} package to automatically test whether eigenvalues are numerically converged using the techniques described in \citet[][ch 7.5]{Boyd:2001wu}.
As expected, we find about 50\% of the eigenvalues are converged for sufficiently high resolution.
Here we solve with 256 Legendre modes for twenty distinct $k_x$, themselves logrithmically spaced.
Legendre expansions produce optimal polynomial approximations in the $L^2$ norm and may be preferable to Chebyshev expansions in problems where the lack of a fast transform is unimportant, such as dense eigenvalue problems.

\begin{figure}
  \includegraphics[width=\linewidth]{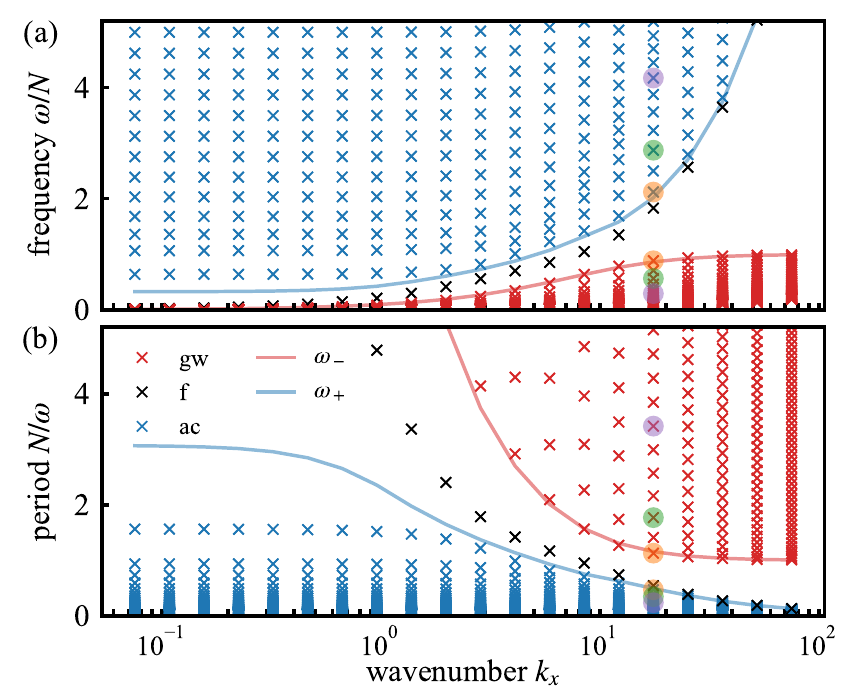}
  \caption{Eigenmode (a) frequencies and (b) periods for a radiative atmosphere.
  The mode frequencies $\omega$ are normalized by the Brunt-V\"{a}is\"{a}l\"{a} frequency $N$ in the isothermal layer.
  Acoustic modes ("ac", blue), gravity modes ("gw", red), and f-modes ("f", black) are identified by their frequencies.
  Acoustic waves are expected to have nearly equal frequency spacing, while gravity waves have nearly equal period spacing, as is observed.
  Colored circles highlight modes shown in \figref{fig:wave eigenfunctions}.
  \label{fig:wave_eigenvalues}}
\end{figure}

\figref{fig:wave_eigenvalues} shows the frequencies (or equivalently periods) of the wave modes calculated with the Dedalus eigenvalue solver.
Because sound waves have frequencies greater than $\omega_+$ and gravity waves have frequencies less than $\omega_-$, we also plot the minimum of $\omega_-$ and maximum of $\omega_+$ over the vertical extent of the atmosphere (both of which occur in the isothermal layer).
This allows us to easily distinguish sound waves from gravity waves.
We also find that, at fixed horizontal wavenumber, the frequency spacing between sound waves is about constant, whereas the period spacing between gravity waves is about constant.
This well-known property follows from the dispersion relation in the large-vertical-wavenumber limit.

\begin{figure}
  \includegraphics[width=\linewidth]{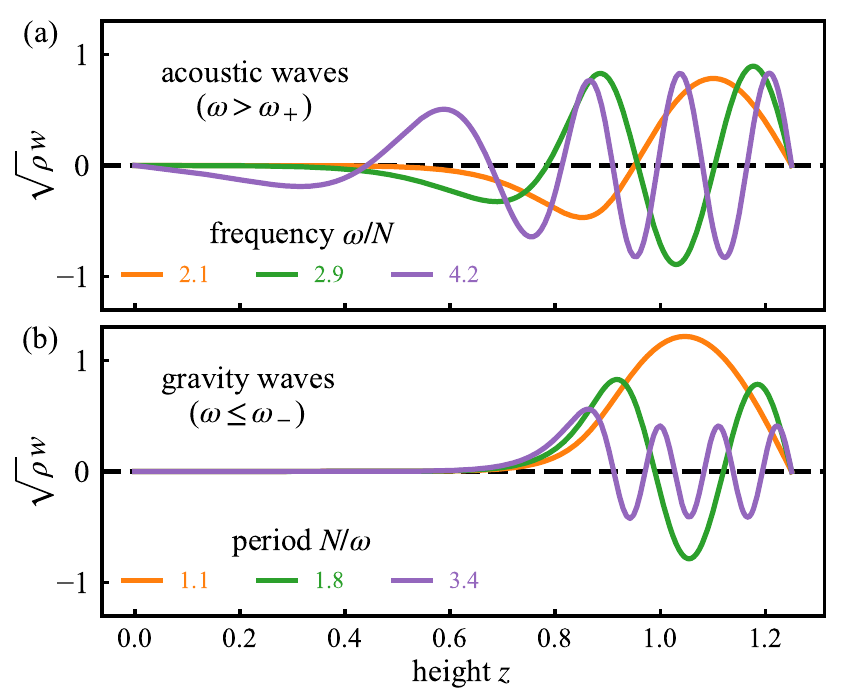}
  \caption{Vertical velocity eigenfunctions for a radiative atmosphere, scaled by $\sqrt{\rho}$, including (a) acoustic modes and (b) gravity modes.
  Gravity modes are almost entirely confined to the isothermal portion of the atmosphere and are evanescent in the adiabatically stratified deep interior.
  Acoustic modes propagate in both regions, with high frequency waves propagating deeper.
  \label{fig:wave eigenfunctions}}
\end{figure}

\figref{fig:wave eigenfunctions} shows vertical velocity eigenfunctions $w$, scaled by $\sqrt{\rho}$.
These modes correspond to the marked modes in \figref{fig:wave_eigenvalues}.
All eigenfunctions have been normalized by the mode kinetic energy:
\begin{equation}
  w = \frac{w_\mathrm{EVP}}{W}, \quad W = \sqrt{\frac{\int \rho \left(u_\mathrm{EVP}^2+w_\mathrm{EVP}^2\right) \mathrm{d}z}{\int \rho \mathrm{d}z}}.
\end{equation}
We find that the eigenfunctions follow the intuition of the propagation diagram in \figref{fig:eddington_atm}.
Gravity modes are trapped in the upper atmosphere, while acoustic modes span a larger region of the full atmosphere with the highest frequency modes reaching the bottom of the domain.
The effects of the non-constant coefficients are visible in the acoustic modes which have a varying vertical wavelength in the deep adiabatic interior and a nearly constant vertical wavelength in the upper isothermal region.

This example of wave eigenfunctions and eigenvalues demonstrates the ability to link a complex atmosphere with detailed solves.
This example considered ideal waves in a bounded atmosphere, but can be easily extended to non-adiabatic waves with thermal damping, to magnetohydrodynamic waves propagating through a background magnetic field, and to systems with open boundary conditions.

\subsection{Diamagnetic Levitation}

This example computes the motion of a levitating rigid body under the competing influence of gravity and an imposed diffuse background magnetic field.
There are three general classes of magnetic levitation:
(I) \textit{AC Electromagnetic suspension} of a conductor relying on an alternating background field;
(II) \textit{Motion-induced suspension} such as Maglev trains and spinning tops;
and (III) \textit{diamagnetic levitation}, which results from magnetically induced eddy currents from a DC field within special types of macroscopic media.
Diamagnetic levitation is usually the weakest form, but superconductivity provides the important exception.
Past experimental work on diamagnetic levitation used an approximately 16 Tesla solenoid to suspend a frog quasi-stably (and mostly safe for the frog) \cite{Simon:2000jk}.

For context, magnetic levitation of any kind is actually a rigid-body version of magnetohydrodynamic buoyancy (or Parker instability), which are well-know in astrophysics. \cite{Acheson:1979fp}.
Maglev train physics is very similar to differential-rotation induced magnetic buoyancy; which is important in stellar and planetary interiors \cite{Vasil:2008bz}.
All types of levitation would be amenable simulation in Dedalus; including magnetohydrodynamic buoyancy.
Here we focus exclusively on solid-body diamagnetic levitation.

\textit{Electrodynamics ---}
We use the (non-relativistic) Maxwell's equations for the magnetic flux density, $B$; electric field, $E$; magnetic field, $H$; and current density, $J$.
\begin{eqnarray}
\partial_{t} B \ + \ \nabla \times E \ = \ 0\\
\nabla \cdot B\ = \  0\\
\nabla \times H \ = \ J
\end{eqnarray}
Closing the system requires a constitutive relation between $H$ and $B$, and Ohm's Law between $E$ and $J$ (in a frame moving with velocity $v$),
\begin{eqnarray}
H \ =\  \frac{B}{\mu},
\quad
E \ = \ \rho\,J \ - \ v \times B
\end{eqnarray}
The constitutive parameters are the resistivity $\rho$ (inverse electrical conductivity), the magnetic permeability $\mu$, and the diffusivity $\eta=\rho/\mu$.

We use a magnetic potential $B \ =\ \nabla \times A$ to enforce the magnetic Gauss's law $\nabla\cdot B = 0$.
In two dimensions we can use a scalar magnetic potential (three dimensions would require a gauge choice).
Then Faraday's law of induction is
\begin{eqnarray}
\partial_{t} A = v_{x} B_{y} - v_{y} B_{x} + \rho \left[ \partial_{y}\frac{B_{x}}{\mu}- \partial_{x} \frac{B_{y}}{\mu} \right],
\end{eqnarray}
where $B_{x} = \partial_{y} A$ and $B_{y} = - \partial_{x} A$.

\textit{Eulerian velocity ---}
We embed a freely movable (yet coupled) rigid body within the neutral background.
First, the rigid-body translation and rotation produce the local Eulerian ``fluid'' (continuum) velocity
\begin{eqnarray}
& \text{if} \quad (x,y) \ \in \ \mathcal{M}(x_{0},y_{0},\theta),\quad \text{then}  \nonumber \\
&v_x(t,x,y) \ = \ \dot{x}_{0}(t) -  (y - y_{0}(t)) \, \dot{\theta}(t) \\
&v_y(t,x,y) \ = \  \dot{y}_{0}(t) +  (x - x_{0}(t) )\,\dot{\theta}(t)
\end{eqnarray}
The set $\mathcal{M}(x_{0},y_{0},\theta)$ represents the points inside the body with center of mass $x_{0}(t),y_{0}(t)$ and orientation angle $\theta(t)$.
The Eulerian velocity vanishes outside the body.

\textit{Mask function ---}
We use a \textit{smooth} mask function (rather than strictly binary) to indicate the solid-object region.
We model our solid as an ellipse with 2:1 semi-major/semi-minor axis ratio.
We use an adjustable error function profile for the mask function
\begin{equation}
M_{\varepsilon}(x,y) \ = \ \tfrac{1}{2} \left\{1-\text{Erf}\left[ \tfrac{\sqrt{\pi}}{2\varepsilon} \left(\tfrac{x^{2}}{a^{2}}+\tfrac{x^{2}}{b^{2}}-1\right)\right]\right\}. \label{Meps}
\end{equation}
The smoothed-mask function approach is called the ``smoothed-volume-penalty method'' (SPV).
Recent detailed analysis shows this method is quite competitive with all other techniques for treating moving solid objects.
Also, the smooth transition region aids in controlling error compared to an abrupt transition \cite{Hester:2019vk}.

\textit{Coupled ODEs ---}
Translation and rotation about the center of mass follow a set of ordinary differential equations describing Newton's laws of conservation of momentum and angular momentum.
\begin{eqnarray}
\frac{d}{dt}  \left[\!\!
\begin{array}{c}
 x_{0}(t)\\ y_{0}(t)
\end{array}\!\!
\right] &=& \ \quad \left[\!\!
\begin{array}{c}
 \overline{v}_x(t)\\
  \overline{v}_y(t)
\end{array}\!\!
\right],
\\
\frac{d}{dt}  \left[\!\!
\begin{array}{c}
 \overline{v}_{x}(t)\\ \overline{v}_{y}(t)
\end{array}\!\!
\right]  &=&  \frac{1}{m}\left[\!\!
\begin{array}{c}
 \bar{F}_x(t) \\
  \bar{F}_y(t)
\end{array}\!\!
\right] - \left[\!\!
\begin{array}{c}
 0 \\
  g
\end{array}\!\!
\right],
\\
 \frac{d}{dt}\left[\!\!
\begin{array}{c}
 \theta(t)\\ I\,\omega(t)
\end{array}\!\!
\right]  &=& \ \quad \left[\!\!
\begin{array}{c}
 \omega(t) \\
  \bar{\tau}_{z}(t)
\end{array}\!\!
\right]
\end{eqnarray}
The parameters $m$ and $I$ are the object's mass and moment of inertial; $g$ represents gravitational acceleration.
For the ellipse $I/m = \pi \, a\, b\,(a^{2} + b^{2})/4$.
Solving this system of 6 ODEs requires computing the total force, $\bar{F}_{x},\bar{F}_{y}$, and torque, $\bar{\tau}_{z}$.  Each depends on the background magnetic field and material properties of the rigid body.

\textit{Lorentz force ---}
We compute the total force on the object by integrating the Lorentz force density
\begin{eqnarray}
\mathcal{L} = J \times B =
\nabla \cdot \left[ H \, B - \tfrac{H \cdot B}{2} I \right] +  \tfrac{H \cdot B}{2} \nabla \log \mu
\end{eqnarray}
The force \textit{on} the rigid body is
\begin{eqnarray}
\bar{F} \ = \ - \int \mathcal{L}\, \text{d}x  \ = \ -\int \frac{H \cdot B}{2} \nabla \log \mu \, \text{d}x.
\end{eqnarray}
The gradient of the permeability makes sense because the material is highly localised, yet infinitely differential via \eqref{Meps}.

\textit{Magnetic susceptibility --- }
For most everyday substances (e.g., frogs) the magnetic permeability is extremely close to that of free space, $\mu_{0}$.
Defining
\begin{eqnarray}
\chi \ = \ \frac{\mu}{\mu_{0}} - 1.
\end{eqnarray}
we assume $|\chi| \ll 1$ (e.g., graphite has $\chi  \ \approx \ - 2 \times 10^{-5}$),
\begin{eqnarray}
\frac{1}{\mu}  \ \approx \ \frac{1-\chi \,M_{\varepsilon}}{\mu_{0}},
\quad
\nabla \log \mu \ \approx \ \chi\, \nabla M_{\varepsilon}
\end{eqnarray}
This simplifies our equations considerably.
This assumption would not be justified for a partial superconductor where $\chi \approx -1$.
In the case of the force and torque
\begin{eqnarray}
\bar{F} & \approx & -\chi \int \frac{|B|^{2}}{2\mu_{0}} \, \nabla M_{\varepsilon} \, \mathrm{d} x,
\\
\bar{\tau} & \approx &  -\chi \int \frac{|B|^{2}}{2\mu_{0}} \, (r-r_{0})\times \nabla M_{\varepsilon} \, \mathrm{d} x
\end{eqnarray}
Considering a negative upward magnetic pressure gradient, it follows that levitation requires $\chi < 1$.
The proportionality of the force to $\chi$ shows the need for very strong magnetic fields.

\begin{figure}
\centering
\includegraphics[width=\linewidth]{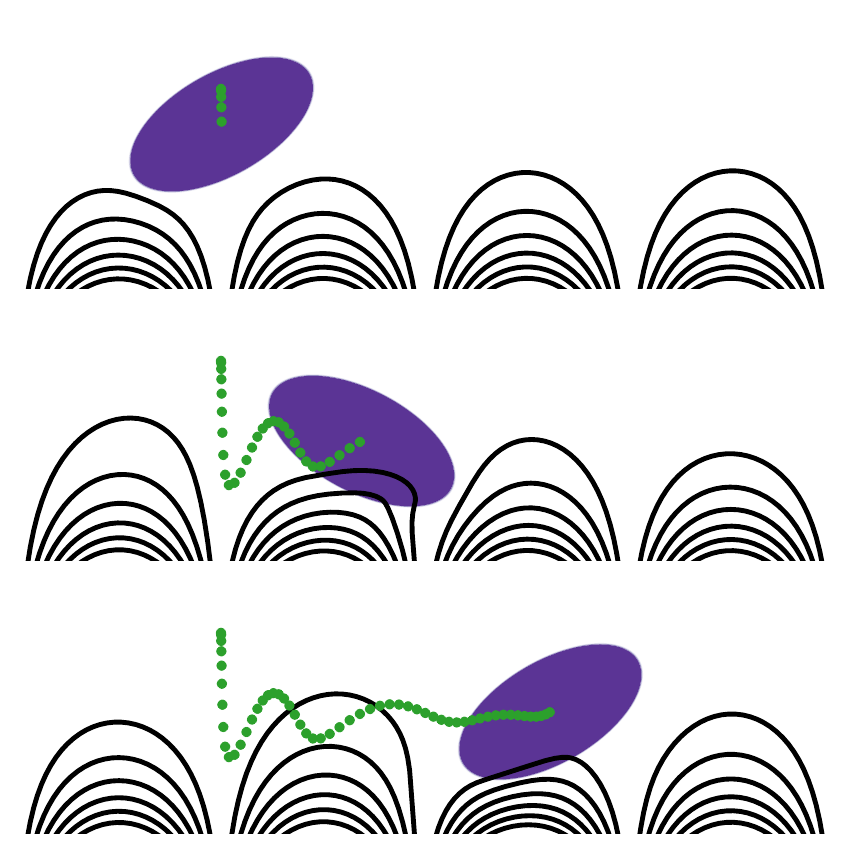}
\caption{Evolution of a diamagnetic object levitating in a magnetic field.
The purple ellipse shows the solid-body region.
The black lines show magnetic field lines the same way iron filings trace the magnetic field from a strong bar magnet.
Green dots show the time-dependent trajectory of the ellipse's center of mass at equally-spaced time intervals.}\label{fig:mag_lev}
\end{figure}

\textit{Simulation ---}
We solve the magnetic induction equation in non-moving electrically neutral background.
Most of the domain is filled with a harmonic magnetic field to a good approximation; like the air around us.
We non-dimensionalize the kinematics of the simulation based gravity and the semi-minor axis of our elliptical rigid body.
Therefore $g=1$, $a=2$, and $b=1$.
We choose a mask smoothing parameter $\varepsilon = 0.025$.
We solve the magnetic induction equation under the above approximations in a rectangular $16\times8$ domain.
We use a Fourier series with 1024 modes for the $x$ direction and a Chebyshev series with 512 modes for the $y$ direction.
In both directions we use the 3/2 dealiasing rule.
We fix the vertical magnetic field at the bottom of the domain, and allow a free vacuum field at the top.
We enforce the top boundary condition via the Hilbert transform in the $x$ direction
\begin{eqnarray}
A \ = \  -\frac{B_{0}}{k} \, \sin(kx)  & \quad \text{at} \quad & y=y_{0}\\
B_{x} - \mathcal{H}_{x}(B_{y}) \ = \   0 & \quad \text{at} \quad & y=y_{1}
\end{eqnarray}
The simulation starts with a harmonic field satisfying the boundary conditions.

We release the solid ellipse from rest with an initial 30-degree tilt.
\figref{fig:mag_lev} shows the ellipse and magnetic field at three representative times in its evolution.
The object free falls until it encounters a strong enough magnetic pressure gradient to rebound.
Along the way, some of the background field diffuses into the object.
As the object bounces upward, the captured field resists and eventually halts the rebound.
The process continues until the object is mostly sliding and rotating along the top of the magnetic arches.

\section{Conclusion \& outlook}

The integrity and reproducibility of computational science relies critically on robust, open-source, and well-supported code.
We have introduced Dedalus, a public Python framework for solving PDEs using spectral methods with an interdisciplinary community of users and developers.

Dedalus enables users to construct custom domains using the direct product of spectral series, to symbolically specify systems of equations and boundary conditions, and to perform custom data analysis.
Dedalus supports initial value, eigenvalue, and linear and nonlinear boundary value problems.
The solution of these problems is automatically parallelized using MPI, allowing for seamless scaling from individual laptops to supercomputers with tens-of-thousands of cores.
The Dedalus distribution includes simple example scripts to help users become familiar with the code's features.
In this paper, we have included a diverse range of example problems demonstrating more advanced features and the code's adaptability to many different physical models.

We are committed to continually enhancing and optimizing Dedalus and supporting its users.
Substantial extensions to the codebase are currently underway.
These include support for multiple coupled dimensions, coordinate-free equation entry, enhanced stand-alone data analysis, and non-direct-product bases for tensorial quantities in curvilinear coordinates (particularly full disks, spheres, and balls \citep{Vasil:2016kb,Vasil:2019ir,Lecoanet:2019jn}).
We actively help users to troubleshoot problems and formulate Dedalus-compatible models on our public, searchable mailing list.
Our goal for the future is to continue growing through community development and to provide robust tools for a wide range of scientific applications.

\begin{acknowledgments}

We thank Eliot Quataert; the Dedalus collaboration began under the encouragement and support of Prof. Quataert when the authors resided at or near the Berkeley Theoretical Astrophysics Center in 2011-2014.
We thank Keith Julien for many years of encouragement and freely sharing ideas about spectral methods and mathematical modeling; many key parts of this work trace back to his original insights.
We also thank Sheehan Olver and Alex Townsend for many conversations relating deep technical knowledge of spectral numerical methods.
We also greatly appreciate the sustained support for the project and its developers offered by Tom Abel, Lars Bildsten, Glenn Flierl, Mordecai-Mark Mac Low, and Nevin Weinberg.
We thank Shane Keating, K.\ Shafer Smith, and Gregory Wagner for helping to clarify the physics and nomenclature of the quasigeostrophic model.
We thank Eric Hester for helping to develop the general volume penalization methods used for magnetic levitation.

We thank the Kavli Institute for Theoretical Physics (KITP); all authors of this paper participated in the "Wave-Flow Interaction in Geophysics, Climate, Astrophysics, and Plasmas" program in Spring of 2014, when substantial early work on this project was undertaken.
Through KITP's support, this research was supported in part by the NSF under Grant No.\ NSF PHY-1748958.
We thank the Woods Hole Oceanographic Institute Geophysical Fluid Dynamics Summer Program, which has hosted most of the authors together at some point during the development of this project.
Burns and Lecoanet acknowledge the University of Sydney School of Mathematics \& Statistics Research Committee Fund for supporting multiple visits to the University of Sydney to collaborate on this project.

Burns acknowledges support from a Flatiron Research Fellowship, an NSF Graduate Research Fellowship under Grant No. 1122374, a Woods Hole GFD Fellowship, an MIT Kavli Graduate Fellowship, and a DOE SULI Internship.
Vasil acknowledges support from the Australian Research Council under project number DE140101960, and the Woods Hole GFD Program.
Oishi acknowledges support from Bates College startup funding, NASA LWS grant No.\ NNX16AC92G, Research Corporation Scialog Collaborative Award (TDA) ID\# 24231, and NSF grant AST10-09802.
Lecoanet acknowledges support from a Hertz Foundation Fellowship, an NSF Graduate Research Fellowship under Grant No.\ DGE 1106400, a PCTS Fellowship, a Lyman Spitzer Jr. Fellowship, and a Woods Hole GFD Fellowship.
Brown acknowledges support from a KITP postdoc position, University of Colorado Boulder startup funding, and NASA LWS grant No.\ NNX16AC92G.

Computations were conducted with support from the NASA High End Computing (HEC) Program through the NASA Advanced Supercomputing (NAS) Division at Ames Research Center on Pleiades with allocation GID s1647.

\end{acknowledgments}
\bibliography{methods,extras}
\appendix

\section{Spectral operator matrices}

\subsection{Differentiation and conversion matrices}
\label{sec.diff_conv_matrices}

\subsubsection{Fourier}

Fourier differentiation is a separable operator with a matrix form
\begin{equation}
    \spderiv{}{x} \phi^F_k(x) = ik \phi^F_k(x) \quad \implies \quad D^F_{k,k'} = i k \delta_{k,k'},
\end{equation}
where the $k$ in the matrix entry expression is the signed wavenumber of the corresponding mode.

\subsubsection{Sine/Cosine}

Sine/cosine differentiation is a separable operator which flips the parity of its operand and has matrix forms
\begin{align}
    \spderiv{}{x} \phi^c_k(x) = - k \phi^s_k(x) \quad &\implies \quad D^C_{k,k'} = -k \delta_{k,k'}, \\
	\spderiv{}{x} \phi^s_k(x)= k \phi^c_k(x) \quad &\implies \quad D^S_{k,k'} = k \delta_{k,k'}.
\end{align}

\subsubsection{Chebyshev}

The derivatives of the Chebyshev polynomials satisfy the recurrence relation
\begin{equation}
    \frac{\pdiff{x} T_n(x)}{n} = 2 T_{n-1}(x) + \frac{\pdiff{x} T_{n-2}(x)}{n-2}.
\end{equation}
Therefore,
\begin{align}
    D_{i,j} &= \langle T_i | \spderiv{T_j}{x} \rangle \\
    &= \frac{2 j (\modulo{(j-i)}{2})}{1 + \delta_{i,0}} [i<j].
\end{align}

The derivatives of the Chebyshev-T polynomials are sparse in the Chebyshev-U polynomials:
\begin{equation}
    \spderiv{T_n(x)}{x} = n U_{n-1}(x).
\end{equation}
Furthermore, the polynomials satisfy the identity
\begin{equation}
    2 T_n(x) = U_n(x) - U_{n-2}(x),
\end{equation}
which provides a sparse T-to-U conversion operator for left-preconditioning Chebyshev differential equations:
\begin{align}
    P^L_{i,j} &= \langle U_i | T_j \rangle \\
    &= \frac{\delta_{i,j} - \delta_{i,j-2}}{2 - \delta_{j,0}}.
\end{align}

\subsubsection{Legendre}

The derivatives of the Legendre polynomials satisfy the recurrence relation
\begin{equation}
    \pdiff{x} P_n(x) = (2 n - 1) P_{n-1}(x) + \pdiff{x} P_{n-2}(x).
\end{equation}
Therefore,
\begin{align}
    D_{i,j} &= \langle P_i | \spderiv{P_j}{x} \rangle \\
    &= (2 i + 1) (\modulo{(j-i)}{2}) [i<j].
\end{align}

The Legendre polynomials are equal to the Jacobi polynomials $J^{\alpha, \beta}$ with $\alpha = \beta = 0$.
Their derivatives are therefore sparse in the Jacobi polynomials with $\alpha = \beta = 1$:
\begin{equation}
    \spderiv{P_n}{x}(x) = \spderiv{J^{0,0}_n}{x}(x) = \frac{n + 1}{2} J^{1,1}_{n-1}(x).
\end{equation}
Furthermore, the Jacobi polynomials satisfy the identity
\begin{equation}
    2 (2 n + 1) J^{0,0}_n(x) = (n + 2) J^{1,1}_n(x) - n J^{1,1}_{n-2}(x),
\end{equation}
which provides a sparse conversion operator for left-preconditioning Legendre differential equations:
\begin{align}
    P^L_{i,j} &= \langle J^{1,1}_i | P_j \rangle \\
    &= \frac{j+2}{2 (2 j + 1)} \delta_{i,j} - \frac{j}{2 (2 j + 1)} \delta_{i,j-2}.
\end{align}

\subsubsection{Hermite}

The derivatives of Hermite polynomials are naturally sparse in the Hermite polynomials themselves:
\begin{equation}
	\spderiv{H_n(x)}{x} = 2 n H_{n-1}(x).
\end{equation}
The Hermite polynomial differentiation matrix is therefore banded and is given by
\begin{equation}
	D_{i,j} = \langle H_i | \spderiv{H_j}{x} \rangle = 2 j \delta_{i,j-1}.
\end{equation}

The enveloped Hermite functions satisfy
\begin{equation}
	\spderiv{\phi^H_n(x)}{x} = \sqrt{\frac{n}{2}} \phi^H_{n-1}(x) - \sqrt{\frac{n+1}{2}} \phi^H_{n+1}(x).
\end{equation}
The differentiation matrix for the enveloped Hermite functions is therefore also banded and given by
\begin{equation}
	D_{i,j} = \langle \phi^H_i | \spderiv{\phi^H_j}{x} \rangle = \sqrt{\frac{j}{2}} \delta_{i,j-1} - \sqrt{\frac{j+1}{2}} \delta_{i,j+1}.
\end{equation}

Since these matrices are both banded, no conversion operators / left-preconditioners are necessary for the \pyth{Hermite} basis.

\subsubsection{Laguerre}

The derivatives of Laguerre polynomials are sparse in the generalized Laguerre polynomials as
\begin{equation}
	\spderiv{L_n(x)}{x} = -L^{(1)}_{n-1}(x),
\end{equation}
where
\begin{equation}
	L^{(1)}_n(x) = \sum_{i=0}^{n} L_i(x).
\end{equation}
The Laguerre polynomial differentiation matrix is therefore
\begin{equation}
	D_{i,j} = \langle L_i | \spderiv{L_j}{x} \rangle = -1 [i<j].
\end{equation}

The enveloped Laguerre functions satisfy
\begin{equation}
	\spderiv{\phi^L_n(x)}{x} = -\frac{1}{2} \phi^L_n(x) - \sum_{i=0}^{n-1} \phi^L_i(x).
\end{equation}
The differentiation matrix for enveloped Laguerre functions is therefore
\begin{equation}
	D_{i,j} = \langle \phi^L_i | \spderiv{\phi^L_j}{x} \rangle = - \frac{1}{1 + \delta_{i,j}} [i \le j].
\end{equation}

Both differentiation matrices are rendered sparse by conversion from Laguerre polynomials to the first generalized Laguerre polynomials.
This is applied to left-precondition all Laguerre differential equations and is given by
\begin{equation}
	P^L_{i,j} = \langle L^{(1)}_i | L_j \rangle = \delta_{i,j} - \delta_{i,j-1}.
\end{equation}

\subsection{Dirichlet recombination matrices}
\label{sec.dirichlet_matrices}

\subsubsection{Chebyshev \& Legendre}

The Chebyshev and Legendre polynomials satisfy the endpoint conditions $T_n(\pm 1) = P_n(\pm 1) = (\pm 1)^n$.
These polynomials recombine to isolate the boundary support to the first two modes as
\begin{equation}
    D^T_n(x) =
    \begin{cases}
    T_n(x) & n = 0,1 \\
     T_n(x) - T_{n-2}(x) & n \geq 2
    \end{cases}.
\end{equation}
\noindent and likewise for the Legendre polynomials.
These Dirichlet polynomials therefore satisfy
\begin{equation}
    D^T_n(\pm 1) =
    D^P_n(\pm 1) =
    \begin{cases}
    (\pm 1)^{n} & n = 0,1 \\
    0 & n \geq 2
    \end{cases}.
\end{equation}

The sparse conversion matrix from the Dirichlet polynomials to the Chebyshev/Legendre polynomials is used as a right-preconditioner to compress the boundary rows corresponding to Dirichlet boundary conditions while maintaining sparsity of the equation matrices.
This matrix is given by
\begin{equation}
	P^R_{i,j} = \langle T_i | D^T_j \rangle = \langle P_i | D^P_j \rangle = \delta_{i,j} - \delta_{i,j-2} [j > 1].
\end{equation}

\subsubsection{Laguerre}

The Laguerre polynomials satisfy the endpoint condition $L_n(0) = 1$.
These polynomials recombine to isolate the boundary support to the first mode as
\begin{equation}
    D_n(x) =
    \begin{cases}
    L_0(x) & n = 0 \\
    L_n(x) - L_{n-1}(x) & n \geq 1
    \end{cases}.
\end{equation}
These Dirichlet polynomials therefore satisfy
\begin{equation}
    D_n(0) =
    \begin{cases}
    1 & n = 0 \\
    0 & n \geq 1
    \end{cases}.
\end{equation}

The sparse conversion matrix from the Dirichlet polynomials to the Laguerre polynomials is used as a right-preconditioner to compress the boundary rows corresponding to Dirichlet boundary conditions while maintaining sparsity of the equation matrices.
This matrix is given by
\begin{equation}
	P^R_{i,j} = \langle L_i | D_j \rangle = \delta_{i,j} - \delta_{i,j-1} [j \ge 1].
\end{equation}
This matrix is used to right-precondition problems using both the Laguerre polynomials and the enveloped Laguerre functions.

\subsection{NCC multiplication matrices}
\label{sec.ncc_matrices}

\subsubsection{Chebyshev}

The Chebyshev polynomials satisfy the multiplicative identity
\begin{equation}
	T_n(x) T_j(x) = \frac{T_{j+n}(x) + T_{|j-n|}(x)}{2}.
\end{equation}
The single-mode Chebyshev multiplication matrices are therefore
\begin{equation}
	\langle T_i | T_n T_j \rangle = \frac{\delta_{i,j+n} + \delta_{i,|j-n|}}{2}.
\end{equation}

\subsubsection{Legendre}

The Legendre polynomials satisfy the three-term recurrence relation
\begin{equation}
	(n+1) P_{n+1}(x) = (2n+1) x P_n(x) - n P_{n-1}(x),
\end{equation}
\noindent which is encoded in the Legendre Jacobi matrix $J$ such that
\begin{equation}
	x P_i(x) = J_{i,j} P_j(x).
\end{equation}

Since the Legendre Jacobi matrix encodes the action of multiplication by $x$, the Legendre polynomial multiplication matrices are given by the corresponding Legendre polynomials of the Jacobi matrix:
\begin{equation}
	\langle P_i | P_n P_j \rangle = (P_n(J))_{i,j} \equiv (M^P_n)_{i,j}.
\end{equation}
\noindent The Legendre polynomial multiplication matrices can therefore be constructed by applying the Legendre recurrence relation to the Legendre Jacobi matrix:
\begin{align}
	M^P_0 &= I, \\
	M^P_1 &= J, \\
	(n + 1) M^P_{n+1} &= (2n+1) J M^P_n - n M^P_{n-1}.
\end{align}

\subsubsection{Hermite}

The Hermite polynomials satisfy the three-term recurrence relation
\begin{equation}
    H_{n+1}(x) = 2 x H_n(x) - 2 n H_{n-1}(x).
\end{equation}
\noindent The Hermite polynomial multiplication matrices can therefore be constructed by applying this recurrence relation to the corresponding Hermite Jacobi matrix, as detailed above for Legendre polynomials.

Matrices for multiplication between Hermite polynomials and enveloped Hermite functions are implemented by reweighting the Hermite multiplication matrices as
\begin{equation}
	\langle \phi^H_i | H_n \phi^H_j \rangle = \frac{N_i}{N_j} \langle H_i | H_n H_j \rangle.
\end{equation}
Multiplication between enveloped Hermite functions is not currently implemented as it is not band-limited when expanded in the enveloped functions.
Including such terms would be possible by expanding the basis to include different powers of the Gaussian envelope.

\subsubsection{Laguerre}

The Laguerre polynomials satisfy the three-term recurrence relation
\begin{equation}
    (n + 1) L_{n+1}(x) = (2n + 1 - x)  L_n(x) - n L_{n-1}(x).
\end{equation}
\noindent The Laguerre polynomial multiplication matrices can therefore be constructed by applying this recurrence relation to the corresponding Laguerre Jacobi matrix, as detailed above for Legendre polynomials.

Matrices for multiplication between Laguerre polynomials and enveloped Laguerre functions utilizes the same matrices since
\begin{equation}
	\langle \phi^L_i | L_n \phi^L_j \rangle = \langle L_i | L_n L_j \rangle.
\end{equation}
Multiplication between enveloped Laguerre functions is not currently implemented as it is not band-limited when expanded in the enveloped functions.
Including such terms would be possible by expanding the basis to include different powers of the exponential envelope, i.e.~the generalized Laguerre polynomials.

\end{document}